\documentclass[letterpaper,12pt]{article}

\usepackage{tabularx} 
\usepackage{amsmath}  
\usepackage{graphicx} 
\usepackage[margin=1in,letterpaper]{geometry} 
\usepackage{cite} 
\usepackage[final]{hyperref} 
\hypersetup{
	colorlinks=true,       
	linkcolor=black,        
	citecolor=black,        
	filecolor=magenta,     
	urlcolor=black         
}
\usepackage{blindtext}
\usepackage{enumitem}
\usepackage{pbox}
\usepackage{amssymb}
\usepackage{amsmath}
\usepackage{txfonts}
\usepackage{mathdots}
\usepackage{mathptmx}
\usepackage{multirow}
\usepackage[classicReIm]{kpfonts}
\usepackage{color, colortbl}
\usepackage{makecell}
\usepackage{graphicx} 
\definecolor{tbcolor}{rgb}{0.92,0.96,1}
\newcommand{\RomanNumeralCaps}[1]
    {\MakeUppercase{\romannumeral #1}}

\numberwithin{equation}{section}
\begin{document}
\title{AME Blockchain: An Architecture Design for Closed-Loop Fluid Economy Token System}


\author{\small Lanny Z.N. Yuan, \small Huaibing Jian, \small Peng Liu, \small Pengxin Zhu, \small ShanYang Fu
\thanks{\{lannyyuan, huaibingjian, liupengbcm, zhupxs, sunyoungfu\}@gmail.com}
}

\date{\small \today, v1.0}
\maketitle

\begin{abstract}
\noindent In this white paper, we propose a blockchain-based system, named AME, which is a decentralized infrastructure and application platform with enhanced security and self-management properties. The AME blockchain technology aims to increase the transaction throughput by adopting various optimizations in network transport and storage layers, and to enhance smart contracts with AI algorithm support. We introduce all major technologies adopted in our system, including blockchain, distributed storage, P2P network, service application framework, and data encryption. To properly provide a cohesive, concise, yet comprehensive introduction to the AME system, we mainly focus on describing the unique definitions and features that guide the system implementation.
\end{abstract}
\clearpage
\section*{Introduction}

Blockchain technology is the base component in the AME technology stack. Blockchain technology combines peer-to-peer network computing and cryptography to create an immutable decentralized public ledger, which is considered as the fifth innovative computing paradigm. It has brought dramatic changes to the way of storing and exchanging data, allowing the Internet enter the era of trust economy. Blockchain technology is promising to make the world more secure and autonomous. However, the existing blockchain-based applications are still in their early stage of development, lacking a favorable environment for technical advancement. There are few blockchain applications that are mature or attractive to the public, i.e., most existing application scenarios have not significantly benefited from this technology. This indicates that there are many aspects need be further optimized for both blockchain technology and applications.

The current blockchain projects on the market mainly have the following limitations and disadvantages:

\begin{enumerate}
\item  Low transaction processing throughput, low data storage capacity, no parallelism support: In Bitcoin, for example, due to the constraints on block size and block time, only 7 transactions can be processed per second and achieving a high confidence that a transaction has been confirmed requires about an hour long wait until the transaction is 6 blocks deep into the Bitcoin blockchain. Such a low transaction rate is far from satisfying the application demand. Moreover, the long transaction latency will reduce the willingness of new users to join the network, and hence lower the competitiveness of blockchain-based products against those from traditional industries. Although there are some research advancements, such as the lightning-network, we still lack a comprehensive approach to solving the efficiency issue.

\item  ``User unfriendly'', hard to use: The blockchain is indeed a synthesis of many existing technologies, which results in high learning cost and implementation difficulty. The current blockchain applications are mostly designed for those technical experts who know how to use them, but not for mainstream consumers. In particular, almost all blockchain applications require users to install and run blockchain full nodes or ``lightweight nodes''. Its long learning curve and massive efforts lower the willingness of participation for ordinary users. For example, the game CryptoKitties might be the most user-friendly distributed application on Ethereum, but users still have to install the Metamask Wallet browser extension by themselves. Besides, users also need to know how to securely purchase cryptocurrencies from others and use them in Metamask. To attract more users, blockchain applications should be designed as simple as the modern web or mobile applications. In addition, the blockchain technology also needs to lower the learning and usage costs, support rapid deployment, and provide close-to-business interfaces.

\item  Poor extensibility, incomplete functional support: Most of current blockchain applications or services provide only limited functionalities, and are short of features that encourage code contribution from open source communities. The flexibility and extensibility of these applications need to be improved. We hope that, by providing a solid technical infrastructure and a complete set of corresponding API/SDK components, the third-party application developers can easily develop and extend their own applications. Hence, we allow developers to build customized blockchain applications based on available sub-components in the AME ecosystem.
\end{enumerate}

This project aims to address the problems and challenges discussed above arising from the development of blockchain technology and applications, including: the support of efficient data storage and concurrent transaction processing; the improvement of application usability, functional completeness, robustness and scalability. The goal of the AME system is to build an open, secure and autonomous application service platform based on the trustful relationships among all participates. By integrating advanced technologies in areas of blockchain, artificial intelligence (AI), big data, and Internet of thing (IoT), the system is able to provide secure and flexible communication channels for all participates and thus establish a security system based on trust economy. The AME system can be viewed as a distributed super-computer or a value conversion platform, on which anyone can communicate and trade with others freely. Besides, the AME system can provide native identity authentication and DNS services, which are of great significance for building the next generation Internet infrastructure. Overall, the vision of the AME system is to serve all the rational applications that have currently been proposed and conceived.

\let\thefootnote\relax\footnote{This paper is for informational purposes only. All contained technical details are not finalized, and may change during the future development and testing phases.}
\clearpage
\tableofcontents{}
\clearpage
\section{AME Blockchain Native Network -- ABNN}

In this chapter, we will introduce one of the core designs in the AME system -- AME Blockchain Native (Slim Dragonfly) Network -- ABNN. In order to achieve a flexible, extensible and decentralized P2P solution, we propose a ``Double  Ring'' topology called the AME Blockchain Native (Slim Dragonfly) Network --- ABNN, which consists of a Manager Ring, a Worker Ring, and an Origin (starting point). In \textbf{Section 1.1} we will present the system topology of ABNN. Then we will are going to introduce of Worker Ring in \textbf{Section 1.2}.
\subsection{System Topology - ABNN}
In this section, we introduce the network topology design in our AME system -- ``Double Ring''. The main design principle is to separate the data management logic from the application business logic layer, such that it is easy to achieve a flexible and extensible decentralized P2P solution. Moreover, our system adopts the ``multi-layer DHT ring'' technology for distributed data lookup.

\textbf{Figure 1.1} shows the overall structure of the ``Double Ring'' topology, which consists of two rings. We call this topology the AME Blockchain Native Network --- ABNN. The Ring 0 (inner ring) is the \textit{Manager Ring} composed of management server nodes, while the Ring 1 (outer ring) is the \textit{Worker Ring} composed of worker nodes. In the ABNN Ring topology, each node maintains the information of its sibling nodes and the corresponding head node in each group. With an optimized routing strategy, ABNN provides fast data lookup and delivery within 2-3 hops, which effectively improves the efficiency of message broadcasting in the network. ABNN is an improved Slim Fly topology\cite{1_3} based on the Hoffman-Singleton Graph\cite{1_2}.

\begin{figure}[htbp]
\centering
\includegraphics*[width=3.8in, height=3.1in, keepaspectratio=true]{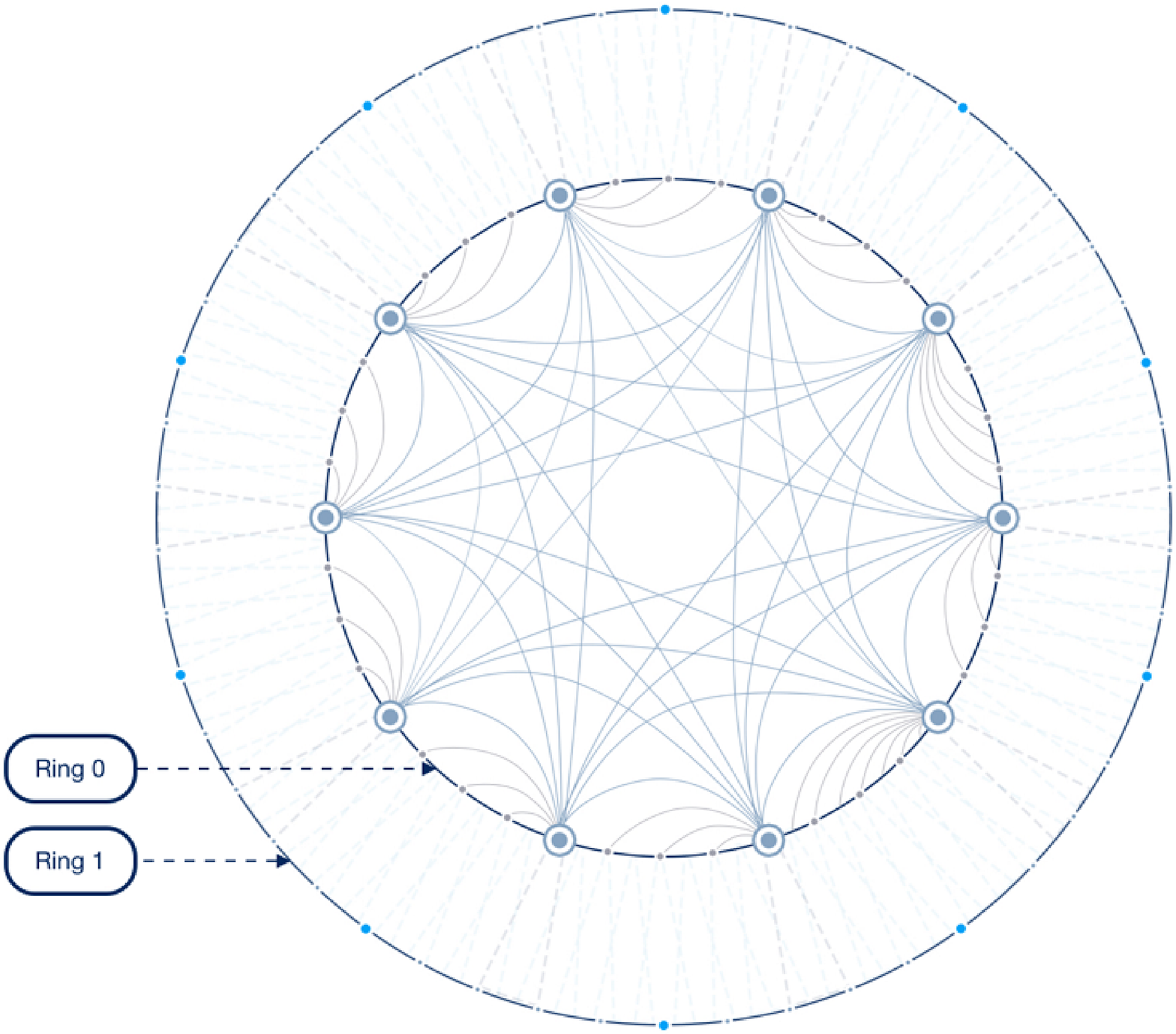}
\caption{AME Blockchain Native Network --- ABNN}
\end{figure}

\textbf{Figure 1.1} also demonstrates the data flow of the Ring topology and how the Manager Ring manages the groups of the Worker Ring. In the following, we discuss the design and implementation of each sub-components in detail.

\subsubsection{Manager Ring -- Ring 0}

As shown in \textbf{Figure 1.2}, each server node in the Manager Ring is a full node, which stores the complete information of the system (including blockchain data, Worker Ring shard information, terminal node information, user account list information, etc.). In addition, the Manager Ring also embeds distributed databases with fast query support. A query request from a client will be submitted to the Manager Ring first, and the client has to pay a certain amount of tokens (e.g. 0.01 ABC, where ABC is short for AME Blockchain Coin). Here we propose the concept of ``Unity'': each Manager Ring node (i.e., a Unity) actually consists of $N$ server nodes ($N\leq 6$) stacked vertically, which are named from server 0 to 6. The server 0 is responsible for message management, i.e., send management commands (CMD) and check message integrity (Error\_Check). The servers 1 and 2 are responsible for sending messages within the ring, while the servers 3 and 4 are responsible for computation and transmission of request responses. The server 5 is reserved and will take over others' work when they are down. Overall, the entire Unity structure consists of 6 servers, each of which guarantees strong data consistency with others. The \textbf{Figure 1.2} shows the message flow direction inside each layer: counterclockwise for server 0-2, and clockwise for server 3-4.

\begin{figure}[htbp]
\centering
\includegraphics*[width=4.4in, height=3.2in, keepaspectratio=true]{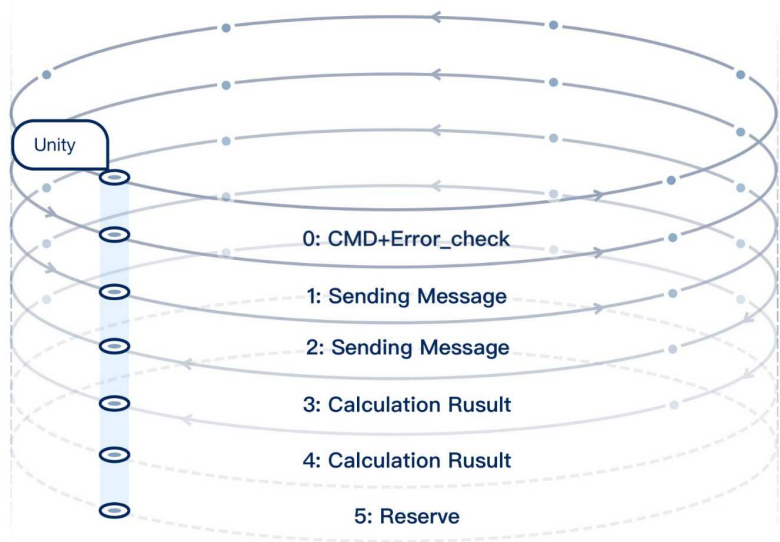}
\caption{Manager Ring Structure}
\end{figure}

As shown in \textbf{Figure 1.3}, multiple Unities logically form a ring structure, which is divided into multiple Groups. Assume that there are $N$ Unities on the ring, which is then divided into $\sqrt{N}$ Groups with $\sqrt{N}$ Unities each. For instance, the outer Worker Ring consists of 256 ``Unities'', named as $S_{1}$, $S_{2}$ ... $S_{256}$. It will be divided into 16 Groups, namely $GROUP_{1}$, $GROUP_{2}$,..., $GROUP_{16}$, each with 16 Unities.

\begin{figure}[!htbp]
\centering
\includegraphics*[width=4.65in, height=3.3in, keepaspectratio=true]{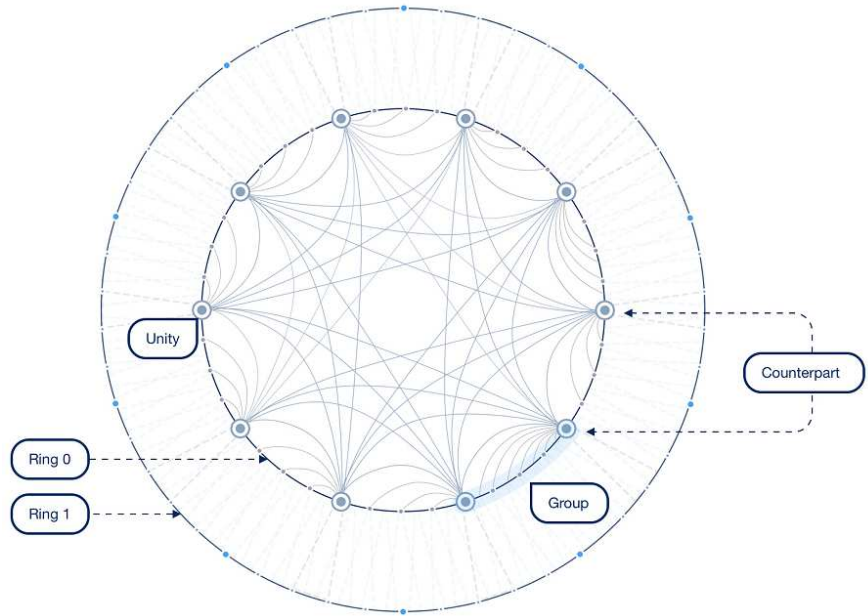}
\caption{Unity, Group, and Counterpart with Annotation}
\end{figure}

Under this partition scheme, the header node of each group (or shard) is called a ``Counterpart'' of other headers. Note that the roles of all servers/nodes on the ring are identical, i.e., any Unity can be a header node. We propose a \textit{Dynamic Sharding Paradigm}, which means that each Unity treats itself as a group header and its sibling node as the succeeding node. Besides, the corresponding group headers are treated as Counterparts. As a result, the partition scheme viewed from Unities are different and may be dynamically changed.

Whenever a terminal (or client) issues a service request, it needs to first submit a transaction request to the Manager Ring, and then the Manger Ring assigns this request to the corresponding Worker server with respect to the requesting terminal. Such a service can support, but not limited to, IM (Instant Messaging) or storage services. During the process of a request, the terminal can apply for a server replacement to ensure the service qualify, by issuing an appeal request. Upon the completion of the request, the Manager Ring distributes the tokens spent by the client to each participating server, according to the service time and quality. To protect the system security, the IP address of a Manager Ring Server is not publicly broadcasted. Instead, only some public email addresses are available, and the Manager Ring server sends messages to outer Worker Ring server in a loosely coupled manner, i.e., by sending emails.

\begin{figure}[htbp]
\centering
\includegraphics*[width=4.29in, height=3.13in, keepaspectratio=true]{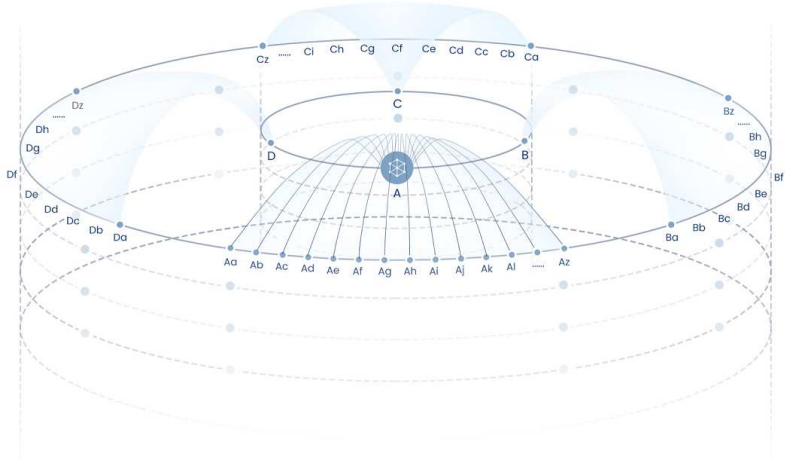}
\caption{Group Management in Manager Ring}
\end{figure}

At the same time, a server on the Manager Ring correspondingly manages one or more Groups on the Worker Ring, and maintains all index information of these Groups. As shown in \textbf{Figure 1.4}, the Manager Ring shards and manages the Groups on the Worker Ring. Each Group synchronizes its group information at a fixed frequency (to the Manager Ring).

In principle, the Manager Ring servers only operate the management logics, but are not involved in specific application services. In this manner, the scalability of the AME system can be significantly improved. We encourage developers to develop more distributed applications on top of AME systems, such as instant messaging services, cloud storage services, App Engine services, and CDN services. The servers on the Manager Ring are responsible for executing blockchain bytecode, and store transaction data in the system. The transactions are generated by the Worker Ring or terminals, and then submitted to the Manager Ring. The Manger Ring servers commit transaction data into the AME blockchain in the form of mining.

\begin{figure}[htbp]
\centering
\includegraphics*[width=4.17in, height=2.32in, keepaspectratio=true]{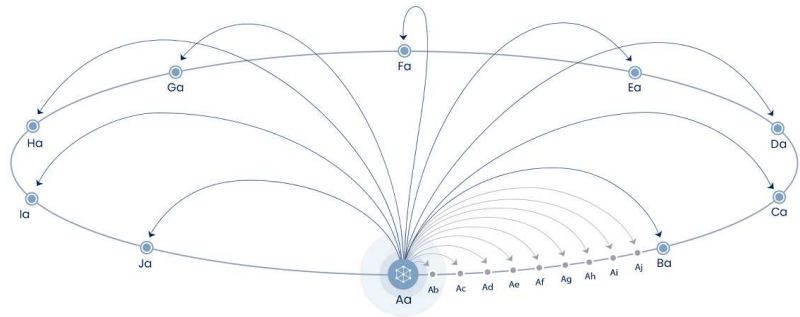}
\caption{Data Flow on the Ring}
\end{figure}

\textbf{Figure 1.5} shows the data flow on the Ring. Whenever a node needs to broadcast a message (e.g., block data), it sends the message to all neighboring nodes in the same group, as well as the header nodes (i.e., counterpart) in other groups. The message transmission pattern is actually identical to a Hamming graph (\textbf{Figure 1.6}) if we view the \textbf{Figure 1.5} from another dimension. All nodes in a same Group are fully connected, while a node and its corresponding counterparts are also fully connected.

\begin{figure}[htbp]
\centering
\includegraphics*[width=4.06in, height=2.96in, keepaspectratio=true]{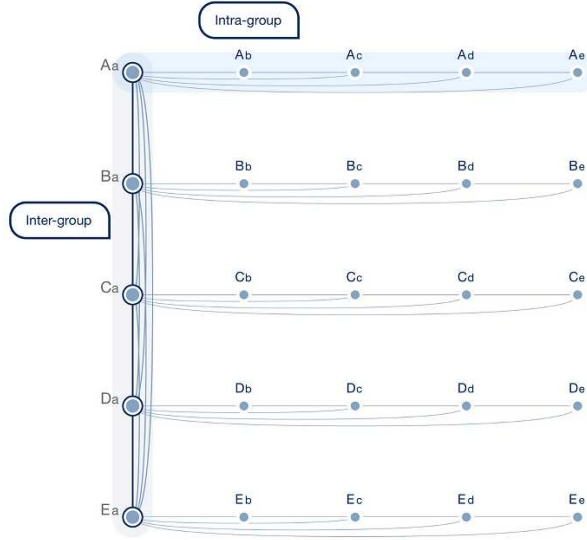}
\caption{Hamming Graph}
\end{figure}

\subsubsection{Worker Ring -- Ring 1}

The outer ring in the system is the ``Worker Ring'' composed of worker servers. We use the ``DHT Ring'' approach to manage the sharding and naming for the Worker Ring servers (the DHT in this context is just a rule for naming).

Similar to the Manager Ring Unity, the Unity of the Worker Ring is also a stereoscopic concept, which means that each Unity contains multiple servers stacked vertically (e.g., $S_{1}$ consists of $S_{1a},{S}_{1b},\ldots,S_{1e}$). Most data storages commonly apply 3-way data replication. Since our system only maintains the naming information requiring small storage space, we are able to use 5-way replication instead. All these replicated servers maintain the identical set of data, which are used as the backups to improve the system robustness. If one of the servers (e.g., ${S}_{1a}$) crashes or goes offline, the system can still operate normally, as other backup servers are able to provide the same services.
To improve the system security, such as prevent external malicious traversal search, interception or spam delivery to the AME system, the server on the Worker Ring does not keep any information related to the entire network. In our DHT algorithm, each Unity maintains the information about all other servers in the same Group (e.g., $S_{1}$ maintains the information of GROUP1 servers ${S}_{1},S_{2},\ldots,S_{16}$), as well as the information about the header servers (i.e., counterparts) in other Groups (e.g., $S_{17}$, $S_{33}$, $S_{49},\ldots$). In this manner, we guarantee that any query in the system can be handled within ``2 to 3 hops'', hence significantly improving system efficiency and reducing query overhead. The account information of the end user is maintained in the server that matches the first two characters of the hash derived from the account name.

Worker Ring servers earn tokens from providing application services to users, and the transactions are recorded on the blockchain of the Manager Ring. In addition, the Worker Ring servers are required to keep the log of the provided services, which might be audited by the Manager Ring periodically.

\subsubsection{AI Self-Management Research and Strategy Scheduling}

We deploy an automatic service scoring system on each client node, so that the client is able to evaluate the services it received. The Manager Ring rates the Worker Ring and the provided services, by recording the service scores of the outer Ring and periodically reading the service logs. With the adoption of AI reinforcement learning algorithms (including service node upgrade/downgrade, and suspicious behavior tracking, etc.), the AME system can support intelligent self-management and eventually achieve system autonomy. The designs of related AI algorithms will be introduced in \textbf{Chapter 6} in detail.

Besides, the system also provides a management mechanism based on user threat assessment: The Manager Ring identifies potential malicious behaviors (e.g., sniffing nodes of firewalls) via user threat assessment (UTA) provided by the AI component. If a potential threat is detected in the network, the Manager Ring will kick that user out of the network, in order to protect the system security.

Recall that the Worker Ring is composed of worker servers, and the number of Groups can be dynamically adjusted according to the capacity of the actual services. In each Group, we can add a number of index servers, each of which acts as a backup for others and maintains the available server information in that Group. The Manager Ring maintains all index server addresses for each Group. When a client issues a service request, it first sends a transaction request to the Manager Ring, which returns the address of an index server. After that, the client connects to that server for obtaining the address of the corresponding worker server. Finally, the worker server can provide the user specific services, such as message forwarding or storage. The worker server internally provides group-based encrypted data backup mechanism that guarantees the reliability and availability of the stored data. When a sniffing client threatens the network, it has to first obtain the index server address before further affecting the index or worker server (e.g., blocking IPs). Once the AI confirms the detection from the UTA, the malicious user is immediately kicked out from the network, which ensures the network security. Meanwhile, the supplemental index servers will be re-elected within the affected Group, and the data on the lost worker servers will be replicated on new backup servers.

\subsubsection{The Bootstrap}

The Bootstrap server is a backup server for disaster recovery. When the Manager Ring fails to operate or is out of control due to internal errors, the origin server will launch the recovery process for taking over and reorganizing the entire system. This is to ensure that the system is able to work normally even in the occurrences of accidental failures or catastrophic errors.

\subsubsection{Temporary Worker Server and Small Hardware Device}

Apart from the ``Double Ring'' and the Worker Ring 1, our system also allows ``temporary workers'' to join the network. Temporary workers can be mounted in any Group on the Worker Ring to provide additional computation power, bandwidth and storage service, such that they can absorb workloads from the overloaded worker servers. A temporary worker is able to apply for promotion to join the Worker Ring as a normal worker node, after it serves a certain amount of service requests. In addition, we will release a ``temporary worker'' hardware, and users can also earn tokens by running this hardware and providing services.

The node expansion strategy in AME is that the newly promoted temporary worker is assigned to the corresponding Group as a vertically parallel server. Once the number of servers in a group reaches the threshold, this Group splits into two Groups, then broadcasts this Ring expansion event to each Worker Ring Group and reports to the Manager Ring. When the Manager Ring wants to remove an offending or frequently failed worker server from the network, the temporary worker has the priority to take over the role of the removed server.

\begin{figure}[htbp]
\centering
\includegraphics*[width=4.3in, height=3.45in, keepaspectratio=true]{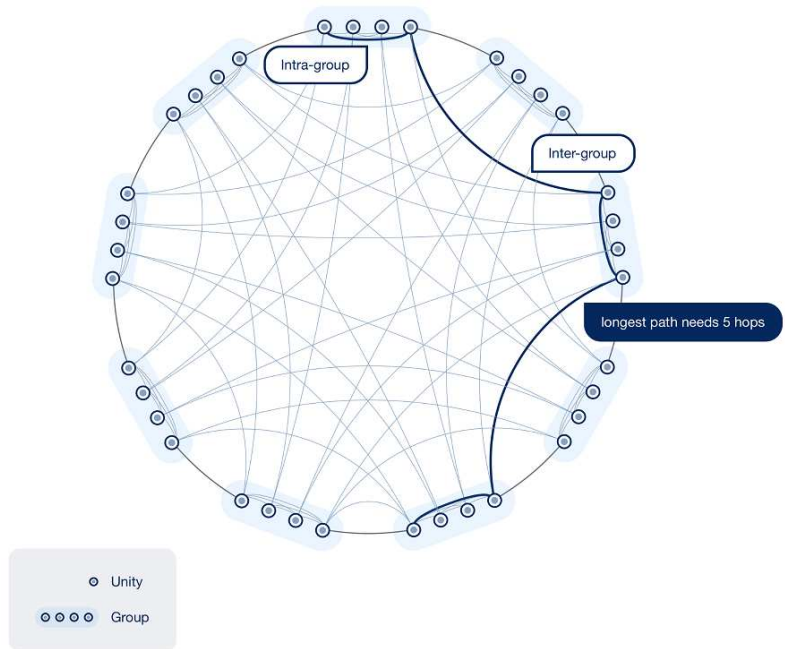}
\caption{Dragonfly Networks}
\end{figure}

\subsubsection{Comparison of ABNN Network Topology with Dragonfly Network}

We compare the ABNN network topology with the Dragonfly Network Topology. It uses an optimized topology for the blockchain application scenarios. \textbf{Figure 1.7} illustrates the Dragonfly Topology, in which the Dragonfly router randomly selects an intermediate group to transmit a datagram each time, such that the network links can be evenly utilized. Therefore, it takes up to 5 hops to complete a data transmission. With an optimized routing strategy, ABNN provides fast data retrieval and delivery within 2-3 hops, which significantly improves the efficiency of message broadcast in the network. Besides, ABNN adjusts the tradeoff between transmission efficiency and routing storage: in the ring topology of ABNN, each node maintains the information of its sibling nodes in the same Group and of the corresponding counterpart nodes in other groups. In this manner, the system increases the storage consumption slightly in exchange for the improvement of transmission efficiency. We believe that ABNN is by far the most suitable technological application scenarios for the Dragonfly topology.

\subsection{Worker Ring DHT Network}

The Worker Ring (Ring 1) is the actual provider of various resources in the system. These servers provide services to clients who use the network, and charge a certain amount of service fee. As a decentralized network, the Worker Ring network cannot rely on a group of permanent central servers as in traditional C/S architectures, such as centralized e-commerce and IM networks. Instead, in the AME system, even though a client has to connect to the Worker Ring server similar to the client-server manner, it actually can connect to an alternative worker server to be served. In this decentralized Worker Ring, there are several major issues that need to be addressed, as follows:

\begin{itemize}
\item  How to deploy servers on the Worker Ring for network construction?

\item  How to efficiently lookup the required resources on the Ring?

\item  How does an application use these servers on the Worker Ring to provide its service?
\end{itemize}
 
\subsubsection{Worker Ring Service Sub-System}

The Worker Ring Service Sub-system is the service provider in the network. It resides in the outer Ring (i.e., Worker Ring), which is composed of N adjacent Unities connected end to end as a ring. The Unity of Worker Ring is a stereoscopic concept, which consists of multiple servers and internally provides high availability and load balancing. There are a number of temporary worker servers around the Worker Ring, which are assigned by the Manager Ring and can help serve a part of the services/businesses. The Worker Ring Service Sub-system provides the concept of ``permanent'' virtual Unity, which can ``permanently'' exist and work on the Ring. The Worker Ring adopts a fully distributed structural topology design. Each Unity maintains only part of the Worker Ring network resources and provides a DHT-based resource lookup functionality.

\begin{figure}
\centering
\includegraphics*[width=4.75in, height=2.19in, keepaspectratio=true]{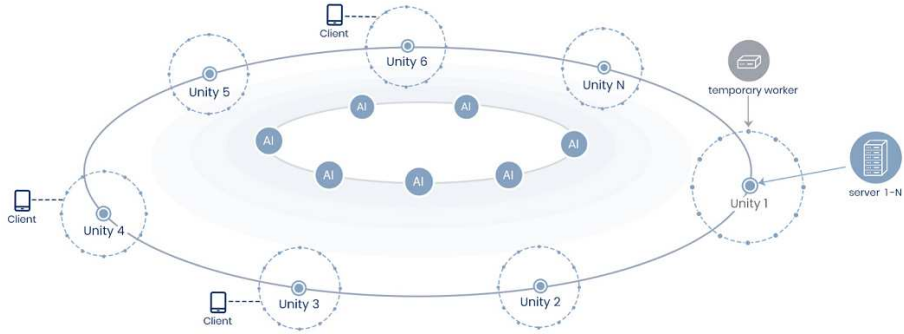}
\caption{The Structure of Worker Ring (Ring 1) Sub-System}
\end{figure}

A resource on the Worker Ring is just an abstract concept, which could be a single server, a stored file, or anything else. These resources are uniformly mapped to abstract addresses on the Worker Ring, and stored in the DHT routing table on the Unity server. Any server on the Worker Ring can locate the corresponding resource in the network by querying the table.

The service quality of the Worker Ring is monitored by the Manager Ring. The AI component in the Manager Ring will monitor the service status of each Worker Ring server, in order to change the role of these servers when necessary. A server can be either a temporary worker on the Worker Ring or a server in a Unity.

The Worker Ring servers provide bandwidth, computation and storage resources to the clients. When a client requests for a service, it issues a request to the Manager Ring, which then returns an available server according to the client request.\textbf{ }Once the request is completed, the Manager Ring sends the AME coins paid by the client to the account of the participating server and commits a transaction.

\subsubsection{Abstract Address Network on Worker Ring}

The fundament for establishing the Worker Ring is the abstract datagram in the network layer. It allows each Unity or even each server in these Unities to represent its exact ``network identity'' as a 256-bit ``abstract network datagram'', and communicate with this 256-bit network address to identify the sender and receiver. In particular, we do not need to be aware of the underlying protocols, such as IPv4 or IPv6 addresses, UDP port signals, etc., which are completely hidden behind the abstract network layer.

In the AME Worker Ring, there exists an overlay network address encapsulation (i.e., Abstract Resource Network Layer), which can (unreliably) send datagram from one abstract address to another. In principle, the Abstract Resource Network Layer (ARNL) can be implemented atop different network technologies. Therefore, we implements it over UDP protocol on the IPv4/IPv6 network. We can also choose TCP protocol, when UDP is not an option.

Option 1: ARNL is an unreliable (small) datagram protocol based on a 256-bit resource address abstraction, and it can be used as the base for more complex network protocols. For example, we can use ARNL as an alternative to the IP abstraction and build a TCP-streaming-like protocol.

Option 2: A reliable variable-sized large datagram protocol (called RLDP) can be built on ARNL to replace the TCP-like protocols. For example, we can use this reliable protocol to send RPC requests to remote hosts and receive responses.

\subsubsection{Worker Ring Unity and Group}

The Worker Ring network consists of a large number of servers, which are all acting as resource providers and data forwarders in the network. They are allocated in an address space, and mapped to different Unities on the Worker Ring. Multiple Unities logically form a ring structure.

Assume that there are $N$ Unities on the Worker Ring; then each Unity should have following properties:

\begin{itemize}
\item  Each Unity consists of 6 to 32 servers;

\item  Each Unity on the ring is of the same importance;

\item  All Unities on the ring are divided into $\sqrt{N}$ partitions, each of which is called a Group and contains $\sqrt{N}$ Unities;

\item  Each Unity maintains the index server information of $\sqrt{N}$ neighboring Unities starting from itself;

\item  Each Unity maintains all Group server information that are at the distance of $N$ Unities;

\item  Each Unity server acts as a Group server, an index server, and a resource server at the same time.
\end{itemize}

The Group server maintains the partition table, which can provide the index server address closest to the target resource; the index server maintains an intra-group lookup table, which maintains the addresses of resource servers; the resource server provides resources, such as computation power and storage space.

\subsubsection{DHT Routing Table in Unity}

Any Unity server in the Worker Ring DHT usually maintains a DHT routing table. The Worker Ring routing table consists of $n$ ($n = 2$) buckets, numbered from 0 to 1. The first table (i.e., Intra-group Routing Table) contains information about known Unities, whose distances to the Unity address \textit{a} are from 1 to $\sqrt{N}$ Unities. The second table (i.e., Inter-group Routing Table) contains information about counterparts, whose distances to the current Unity are of an integer multiple of the Group size $\sqrt{N}$. The Unity information includes the address, IP, UDP port and other useful information, such as the response latency of the last ping.

The Worker Ring Unity follows a stereoscopic design, in which multiple servers collaboratively work in the same Unity and hence can also conduct timely data synchronization (such as user online and offline events).

When the Manager Ring assigns a server to a user, multiple servers can be provided simultaneously, and the user selects the ``optimal'' server considering its actual scenario requirements. Note that all provided servers are from the same Unity on the Worker Ring.

\begin{figure}
\centering
\includegraphics*[width=4.43in, height=2.24in, keepaspectratio=true]{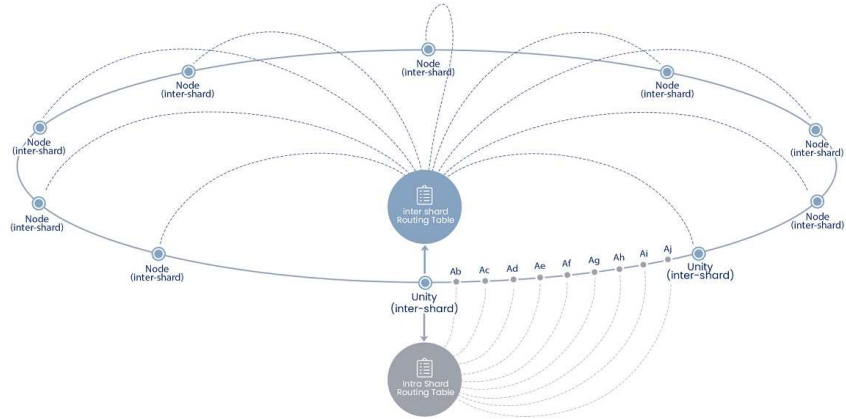}
\caption{Unify Group and Inter-Group Connection}
\end{figure}

\subsubsection{Resource Indexing in Unity DHT -- Key and Value}

The resources on the Worker Ring DHT are all indexed as key-values. The keys are all of 256-bit bytes, i.e., the SHA256 hash of the serialized resource object in most cases. The serialized object is also called the key description. In some cases, the abstract address of a Worker Ring Unity server is also used as the key on the Worker Ring DHT, since such an abstract address is also 256-bit hash from serialized objects. For example, if a Unity server does not want to publish its own IP address, others can locate it as long as they know the key of that server.

The values associated with these 256-bit keys are arbitrary byte strings with bounded lengths. The content of a byte string is determined by the corresponding application requirements.  It can be either the account login information of an IM server, or the digital fingerprint of a stored file. The value of a Worker Ring network resource is usually acquired by locating and querying the Unity that maintains the target key.

The key-value mappings of the Worker Ring DHT are maintained on DHT Unity servers, and are essentially resource address mappings in the network. Therefore, any resource in the Worker Ring network (e.g., account login information in the IM service, and the stored file) has at least one address to ensure that it is an accessible resource. Such a DHT address should not be frequently changed, otherwise other Unities will not be able to locate the keys they are looking for. If a Unity server does not want to reveal its ``real'' identity, it can generate an independent abstract address only for participating in the DHT lookups. However, the abstract address of the communication channel between Worker Ring servers have to be publically available, as it is associated with Unity server information, such as server id, IP address, and port.

\subsubsection{DHT Network Lookup Algorithm with O($1$) Complexity}

The Worker Ring DHT adopts a more efficient lookup algorithm compared to classic DHT algorithms, such as Chord and Kad. In our algorithm, only two hops (i.e., O(1) complexity) are required to lookup the corresponding resource, which improves resource lookup efficiency on the Worker Ring.

The Worker Ring DHT resources mentioned above are indexed by their keys. A resource first generates a key, and hashes it to a Unity on the Worker Ring, which then maintains the resource.

When a query is issued to the network, the resource server where the client is located becomes a query proxy, and first looks up for local Group information. If it hits in the table, the server further locates the target resource server in one step. Otherwise, the server checks inter-group information and looks-up those resource servers near the target server Group. After that, we further look up for local Group information on that server, and search for the target resource server nearby. In this manner, the lookup can be completed within two hops. The object or resource stored on the target server is just abstract, which can be either the login information of a user (i.e., AME IM user login), or a file.

Compared to Kad, the Worker Ring DHT lookup algorithm optimizes the search path in the lookup process, making it more efficient.

\subsubsection{Dynamic High Availability and Load Balancing in Unity}

In order to ensure the high availability of Unities on the Worker Ring, servers in the same Unity adopt a backup mechanism that supports hot backup composed of multiple physical servers. When a Unity server goes down or leaves, other servers can quickly switch their roles and take over the work belonging to the faulty Unity server.

When a Unity contains a relatively small number of servers, it takes over some Unity servers from neighboring Unity (considering its available resources) as backup servers. This is to ensure that each Unity has enough physical servers to support its workloads.

The servers inside the Worker Ring Unity has a load balancing mechanism, which helps achieve load balancing within the Unity by distributing the workloads properly based on the business capacity of each server.

\subsubsection{Capacity Assumption for Worker Ring}

In the Worker Ring network, we configure the minimum number of Unity to be 64. Consider that there are normally 6 servers in each stereoscopic Unity. The total number of concurrent connections that the entire AME network can accept in the initial phase is approximately 24,960,000. It can accommodate up to 24 million online users, meeting the initial capacity requirements for the AME IM system.

Now we discuss the upper limit of the AME network service, where a Unity consists of up to 32 servers. According to this design, the number of total workers in these Unities can reach 323 (32,768), and the total number of system servers is around 1 million. Consequently, the maximum number of concurrent connections that the entire AME network can accept is approximately 68,157,440,000, which can accommodate about 67 billion online users at any moment!

\subsubsection{High Frequency Transaction Processing in Worker Ring}

Since almost all events in the AME IM system have to be paid, such as logging in and sending a message, we can foresee that the transaction frequency will be very high, resulting in heavy workloads on the chain. In order to solve this problem, a sharded transaction mechanism similar to the `Lightning Network' (see AME Blockchain Technology ``Lightning Transaction'' for details) is provided between the Worker Ring servers and the clients. This mechanism can batch the user spends following certain strategies (such as timed payment, one-time login, etc.), which greatly reduces the transaction pressure on the main chain.

\subsubsection{Unity Server Booting}

\paragraph*{Unity Server Booting in Unrestricted Network}\mbox{}
\vspace{\itemsep}

\noindent When a Unity server on the Worker Ring DHT goes online, it first requests the Manager Ring for the information of joining the Unity. After obtaining this information, it joins the network and synchronizes data with other servers. As the Worker Ring has a stereoscopic node structure, there are usually multiple physical servers working for the same Unity. The new server can download all existing (key, value) pairs from them to populate its own DHT table.

In practice, each Unity server on the Worker Ring also maintains a ``adjacency list'', which contains information about other known Unities, such as their resources, abstract addresses, IP addresses and ports. Therefore, by issuing an initial query and obtaining adjacency lists from other servers, the new server is able to gradually expand its own list, and periodically remove obsolete entries as well.

However, when a Worker Ring Unity server has just launched, it may not know other existing Unity servers. This may happen, and hence the server cannot access any previously cached Unity servers or software hard-coded Unity servers. In this case, the Unity will send the datagram to a special ``channel'' from some relevant Unities. This approach does not need to know the recipient's public key in advance (but the sender's identity and signature should still be attached in the message), and thus the sent message will be transmitted without encryption (e.g., by sending an email). This approach is usually only used to request for the identity of the recipient (possibly creating a one-time identity for this purpose), and then conduct future communication in a more secure way.

Once at least one Unity is known, we can obtain more information by sending a special request to these known servers, and hence it is straightforward to populate more entries to the ``adjacency list'' and ``routing table''. Not all Unity needs to handle the datagrams sent to that channel, but those responsible for booting have to support this functionality.

\paragraph*{Unity Server Booting in Restricted Network}\mbox{}
\vspace{\itemsep}

\noindent The Worker Ring DHT network provides the solution for connections from restricted networks, which is useful for handling scenarios where servers in some particular areas are unable to connect to the network, due to the physical firewall blocking. When there exists a physical firewall barrier, the AME system provides dedicated encryption and traffic obfuscation plug-ins to enable communications between Unities. Please refer to the appendix for the details about the review evasion.

\subsubsection{Application Support with IM Example}

Here we take the IM application as an example to illustrate how the Worker Ring works during the lookup process. Suppose that Alice and Bob are two accounts that have logged-in on the Worker Ring network. If Alice wants to send a message to Bob, he needs to lookup Bob's login information on the Worker Ring, which ensures that the IM message can be correctly delivered. Alice then sends server A (where he is located) a lookup request with RB as the key and B as the value. This value can be the ID (i.e., a 256-bit hash value) of the target object that the IM message is sending to. The server A then checks whether B is in its own Group: if yes, the target resource server can be directly located in the intra-group table within the Group; otherwise, the server finds the index server responsible for the target server Group from its inter-group table. After that, the index server helps to provide the corresponding intra-group table that can locate the target resource server. This resource server then retrieves and returns the information (i.e., the value B) about Bob's address, such as IP and port. Finally, the Alice's login server starts to connect with Blob, and Alice's message can be delivered to Bob's login server.

\begin{figure}
\centering
\includegraphics*[width=4.98in, height=2.02in, keepaspectratio=true]{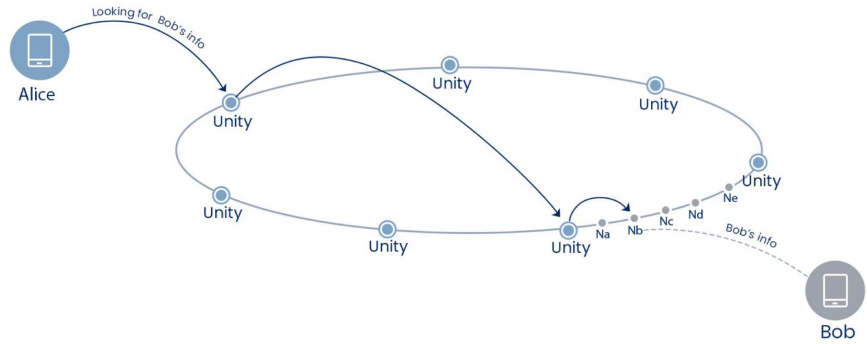}
\caption{Example: How Alice Finds Blob on the AME Network}
\end{figure}

\clearpage
\section{AME Consensus Mechanism}
\setcounter{figure}{0}
One of the fundamental building blocks in blockchain system is the consensus mechanism which creatively tackles the problem of replicated agreement on transaction recording in an open-access, weakly synchronized network with the tolerance to arbitrarily behaving nodes. The consensus mechanism is not simply the consensus for algorithms and data sharing among computers, but more importantly the consensus for collaborations among partners. The consensus mechanism enables participants in the blockchain to collaboratively manage a shared ledger with an agreement, which guarantees the correctness, consistency and continuity of accounting among all collaborators. We will discuss the existing consensus mechanism in \textbf{Section 2.1} and then propose our novel consensus protocol in \textbf{Section 2.2}.

\subsection{Consensus in Blockchain Network}

Networks in the context of distributed system, maintaining the canonical blockchain state across the P2P network can be seemed as a fault-tolerant deterministic replicated state machine that hosts all transaction. An agreement on a unique common view of the blockchain is expected to be achieved by the consensus nodes in the condition of Byzantine failures. In blockchain networks, Byzantine failures cause faulty nodes to exhibit arbitrary behaviors including malicious attacks/collusions (e.g., Sybil attacks and double-spending attacks), node mistakes and connection errors. In other words, the consensus is the process of determining the transaction sequence and filtering illegal transactions ensuring that transactions are objectively recorded over the entire network and cannot be tampered. A blockchain updating protocol is said to achieve the consensus in a Byzantine environment if the following properties are satisfied:

\begin{itemize}
\item\textbf{Validity}: If all the honest nodes activated on a common blockchain state propose to expand the blockchain by the same block, any honest node transiting to a new local replica state adopts the blockchain headed by that block.

\item\textbf{Agreement}: If an honest node confirms a new block header, then any honest node that updates its local blockchain view will update with that new block header.

\item\textbf{Termination}: All transactions originated from the honest nodes will be eventually confirmed.

\item\textbf{Total Order}: All honest nodes accept the same order of transactions as long as they are confirmed in their local blockchain views.
\end{itemize}

The consensus protocols vary with different blockchain networks. Since the permissioned blockchain networks admit tighter control on the synchronization among consensus nodes, they may adopt the conventional Byzantine Fault-Tolerant (BFT) protocols to provide the required consensus properties. In a network of $n$ consensus nodes, the BFT-based protocols are able to conditionally tolerate faulty nodes up to $\frac{n}{3}$. On the contrary, permissionless blockchain networks admit no identity authentication or explicit synchronization schemes. Therefore, the consensus protocol therein is expected to be well scalable and tolerant to pseudo identities and poor synchronization. Since any node is able to propose the state transition with its own candidate block for the blockchain header, the primary goal of the consensus protocol in permissionless networks is to ensure that every consensus node adheres to the ``longest chain rule''. Namely, when the blocks are organized in a linked list, at any instance, only the longest chain can be accepted as the canonical state of the blockchain. Due to the lack of identity authentication, the direct voting based BFT protocols no longer ensure the consensus properties in permissionless blockchain networks. Instead, the incentive based consensus schemes such as the Nakamoto consensus protocol are widely adopted.

\subsubsection{Existing Consensus Mechanism}

To ensure proper functioning of a permissionless blockchain network, Satoshi Nakamoto innovatively combines a consensus protocol based on a framework of cryptographic block-discovery racing game with economic incentives to probabilistically award the consensus participants based on an embedded mechanism of token supply and transaction tipping in the Bitcoin system, which is known as the Proof of Work Scheme. The Proof of Work requires miners to perform a moderate amount of computational work to solve the hash puzzle, which raises the cost of being a malicious node (e.g., publishing a fake block) and ensures that any consensus node will suffer from finical loss whenever it deviates from truthfully following the protocol. Inspired by Satoshi Nakamoto, many blockchain consensus mechanism have been proposed. In this section, we discuss the consensus protocols used by mainstream blockchain projects.

\begin{enumerate}
\item  \textbf{Proof of Work (PoW)}

To obtain the eligibility of a block commitment, each node has to solve a computation puzzle involving the hash of the previous block, the hash of the transactions in the current block, i.e., finding a random nonce that satisfies the constraints. That node can then add a valid block and broadcast it to all other nodes in the network after being verified. Other miners adopt and add block to the longest chain, which has the greatest Proof-of-Work effort invested in it. The advantage of PoW is that it is completely decentralized, and nodes can freely join and leave the network. The disadvantages and limitations are also obvious: Bitcoin has already attracted most of the computation resources in the world, which makes other PoW-based blockchain applications difficult to obtain comparable resources and achieve similar security levels; the block mining wastes huge amounts of electrical energy and other relevant resources; it takes a long period to reach the global consensus, which is not suitable for commercial applications.

\item  \textbf{Proof of Stake (PoS)}

In the Proof of Stake system miners do not compete, instead a validator set is maintained. Anyone, who owns blockchain's coins, can join this set by locking all his coins, called the stake, into a deposit. The validators participate then in the block creation process, where two major types of consensus algorithms are used. In Chain-based PoS the validator, who has the right to create the block, is periodically pseudo-randomly selected. In Byzantine-fault-tolerant-style PoS the validators can propose blocks, the right to do so is randomly assigned to them, further the validators then agree or disagree on the proposed blocks by voting. The block creator gets transactions fees instead of block rewards. Therefore, all coins are created in the beginning, and their number never changes. Advantages of PoS are that less energy is needed for consensus and the increased protection against attacks.

\item  \textbf{Delegated Proof of Stake (DPoS)}

DPoS is similar to the board voting, where the stakeholders select a certain number of nodes as their delegation for verification and accounting. In EOS, for example, there is a new block produced every 3 seconds, and only one delegated node can produce the block at any point of time. A block will be skipped if it is not produced within a specified time period. There are 21 delegated nodes taking turns to produce the blocks. At the beginning of each round, 21 unique nodes are selected by the system as block producers. The selected producers then start to produce blocks following a pseudo-random sequence. In general, there will not be any forking in the DPoS-based blockchain, since block producers work in a collaborative way rather than competitively. Therefore, this might be a better solution for our system. Advantages: significantly reduce the number of participating nodes for block verification and accounting, and achieve second-level consensus verification. Disadvantages: the consensus protocol entirely relies on tokens, while many commercial applications do not require tokens.

\item  \textbf{Proof of Elapsed Time (PoET)}

The Proof of Elapsed Time consensus protocol is proposed by Intel, and it randomly elects a leader node from a number of validators to produce the new block. The election method relies on a secure timer (from the Intel SGX Secure Guard Extensions framework) running on each node. The first node that has its timer expired is elected as the leader for submitting the next block. In the PoET protocol, the secure timer is just a simple counter, which means that it requires only a small amount of computation power to achieve consensus between thousands of nodes.The Intel Sawtooth platform applies the PoET consensus protocol and achieves significant advantages in terms of performance and scalability. However, as PoET relies on the primitives in Intel chips and lacks incentives for non-business participants, it may not be widely adopted on public networks. For private and federal networks, PoET might be a feasible alternative to PoW.

\item  \textbf{Practical Byzantine Fault Tolerance (PBFT)}

PBFT\cite{2_2} is a message-based consensus protocol. In normal case, it runs a three-phase protocol: pre-prepare, prepare, and commit. A client sends a request to one of the peers, who in turn broadcasts pre-prepare messages to the other peers. In the prepare stage, a prepare message is multicasted to all other nodes. When a replica receives $2f$ prepare messages, it matches with the pre-prepare message and multicasts a commit message. Once the replica receives commit messages, that match the pre-prepare message, it changes the state to committed and executes the message operation. Once the message is executed, a reply is sent to the client. The main purpose of a Byzantine fault tolerant consensus algorithm is to allow the system to be able to survive and continue work despite some of the machines exhibiting arbitrary faults. Although, PBFT is a consensus algorithm with proven security and liveness properties, the network overhead during consensus round does not allow scale the consensus protocol, limiting the throughput of the whole system. It was shown that PBFT can be attacked by an adversary using a simple scheduling mechanism, halting the consensus either completely, or forcing to wait long timeout when leader is partitioned and unsynchronized.

The following figure shows the workflow of 4 nodes reaching the agreement, in which node 0 is the leader, and the node 3 is a faulty node that does not respond or send any message. When the last node's status becomes committed, it indicates that the current round of agreement has been successfully reached.

\begin{figure}[htbp]
\centering
\includegraphics*[width=5.62in, height=2.44in, keepaspectratio=true]{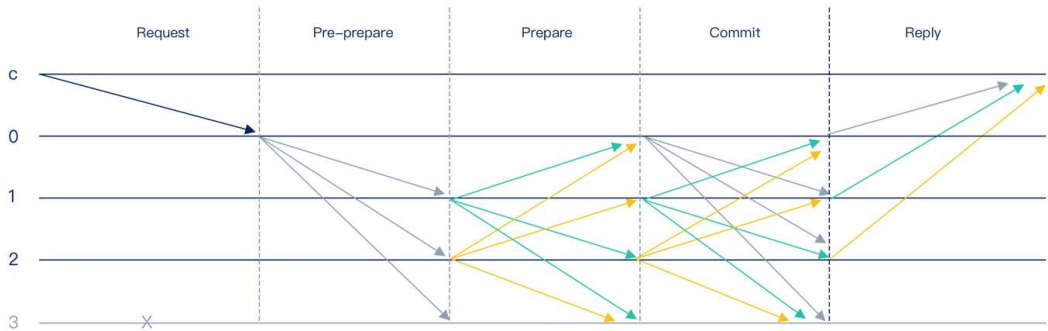}
\caption{PBFT Protocol}
\end{figure}
\end{enumerate}

In summary, the PoW (Proof of Work) consensus protocol used in Bitcoin is essentially a competition of computation resources. Since the total computation power in the Bitcoin network today is much higher than when the network was created, it is now more computationally difficult for a node to produce a block. As a result, miners have to consume a huge amount of computation resources for mining, which inevitably wastes lots of electrical energy. This is a severe waste of global resources, and is absolutely environmental-unfriendly. Moreover, to improve the mining efficiency in the Bitcoin network, some special and dedicated computers -- mining machines -- are assembled, which further form large mining pools jointly. Due to the existence of miners and mining machines, the community has a substantial concern about the centralization trend of the decentralized algorithm, i.e., there is a rumor that 60\% to 70\% of the computation power is located in China. It is difficult for normal clients to compete with these mining pools. This phenomenon results in the consequent that the Bitcoin network is becoming more and more centralized, and the network security is therefore declining. Besides, the existence of PoS is mainly supported by considerations and innovations from an economic point of view, such as the concepts of equity and interest. Although the PoS (Proof of Stake) consensus protocol solves the resource wasting problem in PoW, it has its own limitation: if a client only holds a small amount of coins, the probability for him to mine a block is also very small, leading to a Matthew effect. Thereafter, the DPoS appears, which does not require massive computation power to distribute equity. However, as an alternative protocol, it is unlikely that the DPoS is capable of replacing PoW, PoS or PoW + PoS. After all, what is actual is rational: each protocol has its own technical and business considerations (or implications) during a specific historical time period.

\subsection{AME Consensus}

In this section we describe ACP, an efficient blockchain consensus protocol for partial synchronous setting tolerating $f<{n}/{3}$ fault. ACP targets both high-performance in fault-free executions and correctness if Byzantine nodes exhibit arbitrary behavior. There has been many existing distributed consensus protocols that tolerate faulty nodes and later the ones that tolerate Byzantine nodes. However, their application was limited to a small scale. And a stable leader is required to facilitate the agreement in classic distributed consensus protocols from Multi-Paxos\cite{2_4}, Raft\cite{2_5} to ZAB\cite{2_43} protocol. In the permissionless blockchain setting, any node in the system may be Byzantine to exhibit arbitrary malicious behaviors. Therefore, the assumption that there will be a stable leader is vulnerable. Comparing with others blockchain consensus protocols, the key aspects of ACP can be summarized as follows:

\begin{enumerate}
\item  \textbf{Early Stopping Consensus}

An attacker powerful enough to control up to ${1}/{3}$ of the nodes are commonly assumed in the Byzantine threat model. However, the assumption is rather pessimistic. In the design of ACP,  leveraging the governance of the AI module of AME blockchain, we can assume a relative steady composition of AME network without frequent join and leave of temporary nodes. Therefore, ACP aims to achieving the agreement as soon as possible in the case of Byzantine nodes far less than ${1}/{3}$, while guarantees the safety and liveness when up to ${1}/{3}$ nodes are Byzantine. That is, the consensus algorithm should have the ``early-stopping'' feature.

\item  \textbf{Parallel Byzantine Agreement Instances}

Classic PBFT protocol relies on a stable leader to begin each instance. In case the leader node is Byzantine, it can slow down or stall the system by causing frequently view changes. To address this challenge, ACP runs multiple Byzantine agreement instances in parallel for a single block proposal. The overall efficiency is improved since a small scale is selected to run the PBFT protocol.

\item  \textbf{Prior or Posterior Strategies}

In the blockchain consensus, some posterior techniques, such as VRF\cite{2_33}, are usually employed to improve security; on the contrary, in order to improve efficiency and reduce complexity, some prior techniques such as well-known static grouping and round-robin scheme are needed. ACP combines the characteristics of other components in AME to make reasonable trade-offs, with the prior and posterior strategies, to ensure the safety, liveness and efficiency.

\end{enumerate}

\subsubsection{System Model and Problem Definition}

We first specify our system, threat model and the notion of consensus and finally state the goals of consensus for the ACP Protocol within our formalization framework.

\paragraph*{System Model}\mbox{}
\subparagraph{Node}
Each node represents a full node in AME Blockchain and can communicate with other nodes in a peer-to-peer fashion. We use the term non-faulty to refer to nodes in the network that follow the protocol's instructions without error and obey the AME governance strategy, and are perfectly capable of sending and receiving messages. Conversely, a node is Byzantine if he can deviate from the protocol in a completely arbitrary way, from simple crashes, to malicious behavior aimed at disturbing the consensus, fully coordinated between all Byzantine nodes.

The system consists of $n$ nodes, out of which up to $t<{n}/{3}$ may be Byzantine, i.e., behave arbitrarily and collude together. Denote by $f\leq t$ the actual number of Byzantine nodes in a given run. A few types of synchronous environments we refer to throughout the paper are given hereafter.

\subparagraph{Synchronous Network}

A network is said to be strongly synchronous if there exists a known fixed bound $\Delta$ such that every message delays at most $\Delta$ time when sent from one point in the network to another.

\subparagraph{Partial Synchronous Network}

A network is said to be partially synchronous if there exists a fixed upper bound $\Delta$ on a message's traversal delay and a fixed upper bound $\Phi$ on relative processor speeds and one of the following holds:
\begin{enumerate}
\item
$\Delta$ always holds, but is unknown.
\item
$\Delta$ is known, but only holds starting at some unknown time.
\end{enumerate}

We assume that the communication is partially synchronous in this work and state the goals of consensus for the ACP Protocol below.

\paragraph*{Byzantine Consensus}\mbox{}
\vspace{\itemsep}

\noindent The Byzantine consensus problem consists of each node $i$ having an initial value $v_i$ from a finite set $V$ (i.e., $v_i\in V$). Each node $i$ also has an output value $o_i\in O$. Two properties should hold:

\begin{enumerate}
\item
\textbf{Agreement}: $o_i=o_j$ for any two non-faulty nodes $i,j$ (thus we can talk about the output value of the algorithm);
\item
\textbf{Validity}: if all non-faulty nodes start with the same initial value $v$, then the output value of the algorithm is $v$.
\end{enumerate}
It is well-known that consensus can not be solved in asynchronous systems\cite{2_1}. Distributed consensus is implemented in partially synchronous systems.

\paragraph*{Formal Consensus Goals}\mbox{}
\vspace{\itemsep}

\noindent Assume a system of $n$ nodes, where each node $n_i$ has a private value $v_i$, and the following must be achieved:

\begin{enumerate}
\item  \textbf{Agreement}: All non-faulty nodes must agree on the same value.

\item  \textbf{Validity}: If all non-faulty nodes have the same initial value $v$, then the agreed upon value by all non-faulty nodes is $v$.

\item  \textbf{Termination}: All non-faulty nodes must eventually decide on a value.
\end{enumerate}

\subsubsection{Notations and Parameters}

\begin{itemize}[label={}]
\item
$r$: the current round number.

\item
$N_{all}$: the total number of nodes in the system at the beginning of round $r$.
\item

$N_{pc}$: the participating numbers of Potential Committee.
\item
$N_{fc}$: the participating numbers of Final Committee.
\item
$N_{fc-valid-leader}$: the expected numbers of nodes issue PBFT instance with valid block as initial value.
\item
$N_{fc-empty-leader}$: the expected numbers of nodes issue PBFT instance with empty block as initial value.
\item
${RS}_r$: the random seed of round $r$.
\item
$PK_i$: the public key of node $i$, which are known to all nodes in AME Blockchain.
\item
${SK}_i$: the secret key of node $i$, which are stored locally.
\item
${SE}^r_i$: the secret string of node $i$ in round $r$. $SE$ is a fixed-length bit string updated periodically by each node, and used for generating the random seed $RS$.
\item
${rep}^r_i$: the reputation of node $i$ in round $r$.
\item
$pw^r_i$: the weight of node $i$ in round $r$.
\item
$RI$: the secret string refresh interval.
\item
${\tau }_{pc}$: the expected numbers of Potential Committee.
\item
${\tau }_{fc}$: the expected numbers of Final Committee.
\item
${\sigma }^r_i$: the credential of node $i$ in round $r$.
\item
${\lambda }_{pc}$: the upper-bounds to the time needed to broadcast a message to the whole Potential Committee.
\item
${\lambda }_{fc}$: the upper-bounds to the time needed to broadcast a message to the whole Final Committee.
\item
${\lambda }_{all}$: timeout to broadcast to the whole network.
\item
$B^r_i$: the block proposed by node $i$ in round $r$.
\item
$B^r_{cb}$: the candidate block in round $r$.
\item
$B^r_{\epsilon }$: the empty block in round $r$.
\item
$B^r_{pbft-input}$: the block passed to PBFT in round $r$.
\item
$H$: a cryptographic hash function.

$SBR$: a synchronization barrier.
\end{itemize}
\subsubsection{ACP Overview}

In this section, we provide an brief overview of ACP protocol. We begin by presenting an figure showing the basic structure of ACP Protocol and then describe the individual building blocks. Each round of ACP consists of 4 stages as illustrated below:

\begin{figure}[htbp]
\centering
\includegraphics*[width=5.23in, height=2.42in, keepaspectratio=true]{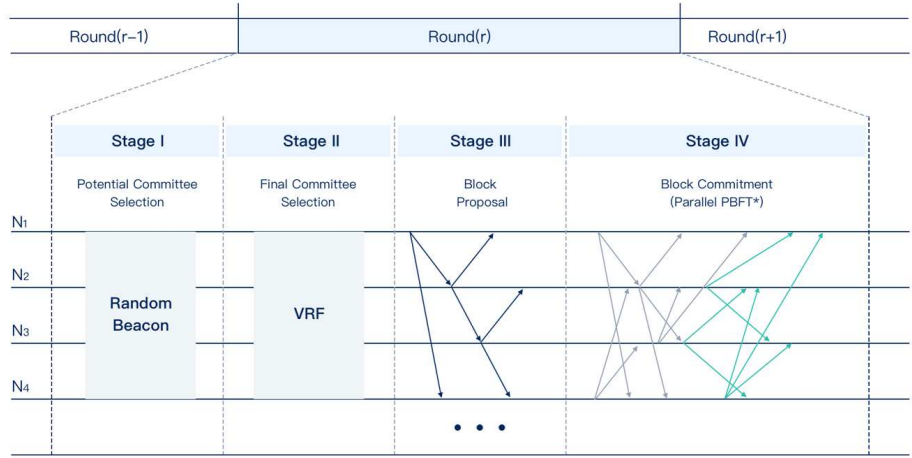}

\caption{An Overview of  ACP Protocol}

\end{figure}

The ACP protocol proceeds in round. At the end of each round, either a valid non-empty block containing a set of transactions or an empty block is agreed among all participating nodes and appended to the blockchain. Each node identifies current round number from local newest block.  Each round of ACP consists of the following 4 stages as illustrated in \textbf{Figure 2.2}.

\paragraph*{Stage \RomanNumeralCaps{1}} This stage is to select Potential Committee from AME Blockchain, in order to provide a balance between efficiency and resource usage, avoiding broadcasting messages to the whole network. This is one of the key techniques we exploit to overcome scalability challenge. The selection is based on a randomness beacon. Each node computes hash values (the outputs of a pre-specified hash function $H_{rb}$ of all nodes and selects a subset of nodes to proceed to next stage as the Potential Committee according to a given threshold $N_{pc}$.

\paragraph*{Stage \RomanNumeralCaps{2}} This stage is to select a much smaller subset Final Committee of the set Potential Committee, aiming at increasing resilient against adaptive attacks. Instead of adopting a hash function to determine eligibility in the Potential Committee Selection, we rely on a Verifiable Random Function (VRF)\cite{2_33} instead. The VRF ensures that the adversary cannot predict in advance which nodes are the block proposers. After the selection, each node selected in Final Committee propose their candidate block proposals from local pending transaction pool and then broadcast a signed message including their respective candidate block proposals, signatures, selection hashs and the hash proofs ($B^r_i$, $sig(B^r_i)$, ${\sigma }^r_i$) to all members in Potential Committee.

\paragraph*{Stage \RomanNumeralCaps{3}} After waiting an amount of time $\lambda_{pc}$ , each node $i$ in Final Committee chooses a candidate block from his received block proposals $B^r_{i\in \{0,\ldots,fc-1\}}$, denoted by $B^r_i$. And then Final Committee members start a two-step Reduction procedure\cite{2_35}. At the end of Reduction Procedure, each Final Committee member outputs a valid candidate block $B^r_{cb}$ that received at least $2N_{fc}/{3}+1$ votes in the second step of Reduction procedure or an empty block $B^r_{\epsilon}$ if no hash received enough votes. The Reduction procedure converts the problem of reaching consensus on an arbitrary value (the hash of a block) to reaching consensus on one of two values: either a specific proposed block hash, or the hash of an empty block.

\paragraph*{Stage \RomanNumeralCaps{4}} Each Final Committee member $i$ acts according to his output value at the end of stage \RomanNumeralCaps{3} as follows.

\begin{itemize}
  \item  If $i$ outputs $B^r_{cb}$ and $H({\sigma }^r_i)$ is among the $N_{fc-valid-leader}$ least value in $H({\sigma }^r_{i\in \{\mathrm{1,...},fc-1\}})$, then $i$ runs the optimized $\text{PBFT}^{*}$ instance as leader node with $B^r_{cb}$ as initial value. At the same time $i$ runs the others optimized $\text{PBFT}^{*}$ instances with the least $N_{fc-valid-leader}$ hash values as initial value in parallel.

  \item  If $i$ outputs $B^r_{\epsilon }$ and $H({\sigma }^r_i)$ is among the $N_{fc-empty-leader}$ largest value in $H({\sigma }^r_{i\in \{\mathrm{1,...},fc-1\}})$, then $i$ runs the optimized $\text{PBFT}^{*}$ instances as leader node with $B^r_{\epsilon }$ as initial value. At the same time $i$ runs the others optimized $\text{PBFT}^{*}$ instances with the largest $N_{fc-empty-leader}$ hash values as initial value in parallel.
\end{itemize}

Final Committee members propagate the agreed-upon block once they received the agreed-upon result from any optimized $\text{PBFT}^{*}$ instance. In this case, the node reaches Final Consensus. On the other hand, tentative Consensus means that it has not yet received the agreed-upon hash and broadcast an empty block $B^r_{\epsilon }$ after waiting a large amount of time $SBR$.

\subsubsection{ACP Details}

We now describe each of the components of the ACP protocol in more detail.

\paragraph*{Initialization of The Protocol}\mbox{}
\vspace{\itemsep}

\noindent The protocol starts with $r=0$. The initial random seed ${RS}_0$ is generated by a (Public)-VSS Coin Tossing scheme\cite{2_44} and hard-coded into the genesis block.
\begin{figure}[htbp]
\centering
\includegraphics*[width=5.23in, height=2.82in, keepaspectratio=true]{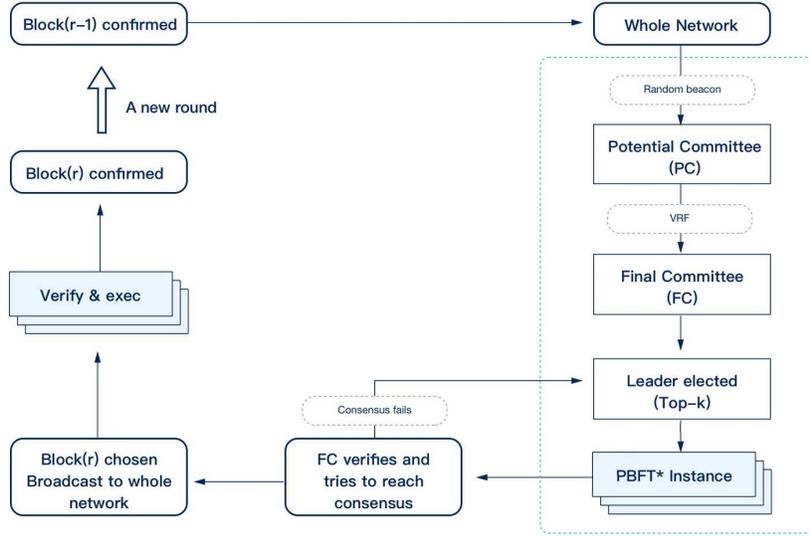}
\caption{The Flowchart of ACP}
\end{figure}

\paragraph*{Potential Committee(PC) Selection}\mbox{}
\vspace{\itemsep}

\noindent Each Node in AME Blockchain calculates for all nodes the Potential Weight(PW)\cite{2_34}:
\begin{equation}
{pw}^r_j=\dfrac{H({RS}^{r-1},r,{PK}_j)}{{rep}^r_j}
\end{equation}
Where $j=\{0,1,...,n-1\}$ is the indicator of nodes, ${RS}^{r-1}$ is the random seed of previous round, $r$ is the current round number, ${PK}_j$ is the corresponding public key of node $j$, ${rep}^r_j$ is the reputation of node $j$ in round $r$, $H$ is the pre-specified hash function. According to the weights and the expected Potential Committee size $m$, we select the $m$ nodes with lowest potential weight into Potential Committee.

\paragraph{Final Committee(FC) Selection}\mbox{}
\vspace{\itemsep}

\noindent Final Committee Selection is implemented using VRF\cite{2_33} to randomly select Final Committee Members in a private and non-interactive way. A VRF is a triple of algorithms Keygen, Evaluate, and Verify.
\begin{itemize}[label={}]
\item
\textbf{VRFGen}: On a random input, the key generation algorithm produces a public key PK and a private key SK pair.

\item
\textbf{VRFEvaluate}: The evaluation algorithm takes the private key SK, a message X as input and produces a pseudorandom output string Y and a proof $\rho$.

\item
\textbf{VRFVerify}: The verification algorithm takes the public key PK, the message X, the output Y and the proof $\rho$ as input. It outputs 1 if and only if it verifies that Y is the output produced by the evaluation algorithm on inputs SK and X.
\end{itemize}
In the Final Committee Selection Stage, each node in Potential Committee checks whether he is selected in Final Committee. If this is the case, he collects a block of transactions from pending transaction pool as his candidate block proposal BP and then broadcast a signed message to Potential Committee which includes the candidate block proposal BP, the selection output hash and its proof of selection.

After waiting an amount of time ${\lambda }_{pc}$, each Final Committee member chooses a candidate block from his received block proposals $B^r_{i\in \{0,...,fc-1\}}$.

The chosen mechanism is described below:

\begin{itemize}
\item  Chooses $B^r_i$ with the largest transaction size among $B^r_{i\in \{0,...,fc-1\}}$.

\item  Chooses $B^r_i$ with the least hash value among $H({\sigma }^r_{i\in \{0,...,fc-1\}})$ in case of multiple nodes with the same transaction size.
\end{itemize}

And then each node in Final Committee starts a two-step Reduction procedure\cite{2_35}. The Reduction procedure satisfies two properties:

\begin{itemize}
\item  If agreement is alert = true, there are non-faulty processes with different initial values from $V$. In this case, all non-faulty processes use a predefined default value from $V$ as the result of the following steps.

\item  If agreement is alert = false, then all non-faulty processes have the same initial value from $V$. This value is the result of the following steps.
\end{itemize}

In the first step of Reduction procedure, each Final Committee member votes for the hash of the candidate block chosen by the above mechanism . In the second step, Final Committee members vote for the hash that received at least ${2N_{fc}}/{3}+1$ votes in the first step, or the hash of the default empty block if no hash received enough votes. After the second step, each Final Committee member outputs a valid candidate block $B^r_{cb}$ that received at least ${2N_{fc}}/{3}+1$ votes in the second step or an empty block $B^r_{\epsilon }$ if no hash received enough votes. Note that a valid block $B^r_{cb}$ output and an empty block $B^r_{\epsilon }$ output corresponds to the above properties with true alert and false alert, respectively. Let us denote the output of Stage \RomanNumeralCaps{3} by $B^r_{pbft-input}$.

\paragraph*{Reach Consensus}\mbox{}
\vspace{\itemsep}

\noindent After Final Committee members having outputting their candidate block respectively, reaching agreement on the final block among all non-faulty Final Committee members remains the main problem. Classic Byzantine fault tolerance consensus protocols requires a stable leader node to facilitate the agreement. A Byzantine leader can cause frequent view changes which would prevent forward progress\cite{2_45,2_46}. In ACP, each Final Committee runs multiple $\text{PBFT}^{*}$ instances in parallel to circumvent the Byzantine leader problem.

\begin{figure}[htbp]
\centering
\includegraphics*[width=5.82in, height=1.95in, keepaspectratio=true]{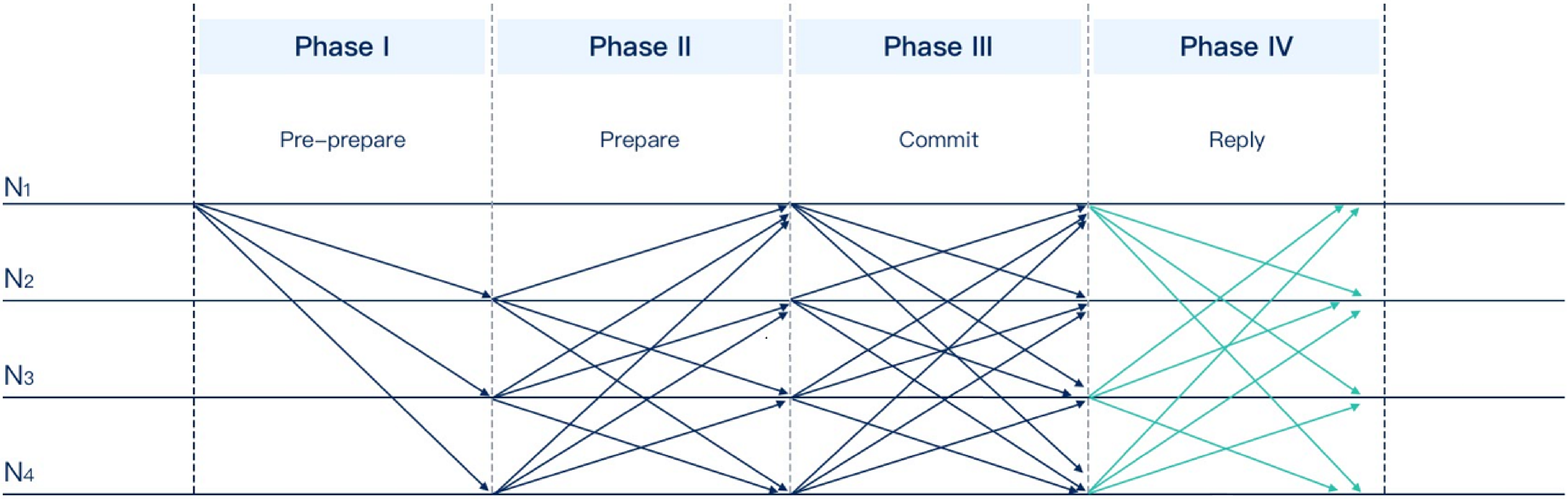}
\caption{$\text{PBFT}^*$ Protocol}
\end{figure}

\textbf{$\text{PBFT}^{*}$}: Each round of classic PBFT\cite{2_2} consists of 4 phases: Pre-prepare, Prepare, Commit and Reply. In the Reply phase, nodes send the committed result to the client and the client is aware of the agreement with replied result. In AME blockchain network, each node is an equal amongst others and hence, whether the agreement is achieved should be known to every node as soon as possible. We modify the Reply phase to achieve this goal by broadcasting committed result to the whole Final Committee instead. As a result,  each node in Final Committee can verify whether the agreement is achieved and enter next round as early as possible.

\textbf{Parallel Instances}: Most of the existing PBFT-based consensus protocols perform single PBFT instance to achieve consensus in each round. If the leader node itself is a Byzantine node, the communication complexity will boost to a staggering $O(n^3)$\cite{2_47}. In ACP protocol, multiple nodes are selected from Final Committee as leaders to execute multiple $\text{PBFT}^{*}$ instances in parallel. Although the leaders chosen by each node may be different, it will not hamper the safety of the protocol. The process is described below.

\begin{enumerate}[label=P\arabic*:]
\item
Each node in Final Committee outputs either $B^r_{\epsilon}$ or $B^r_{cb}$.

\item
Each node in Final Committee runs multiple $\text{PBFT}^{*}$ instances in parallel. If a node's local alert is equal to false, it rejects to run the $\text{PBFT}^{*}$ instance with $B^r_{\epsilon}$ as initial value.

\item
Denote the set of instances with $B^r_{cb}$ and $B^r_{\epsilon}$ as initial value by $\alpha $ and $\beta $, respectively. Use 1 to represent the state of reaching agreement, 0 otherwise. We enumerate the following potential agreement scenarios:

\begin{enumerate}[label=(\arabic*)]
\item $\alpha=1$, $\beta=0$;
\item $\alpha=0$, $\beta=1$;
\item $\alpha=0$, $\beta=0$; 
\end{enumerate}
\begin{enumerate}[label=P3.\arabic*]
\item
In case (1) and (2), all non-faulty nodes with successful completion of any optimized $\text{PBFT}^{*}$ instance will broadcast the agreed-upon result to the whole network. This scenario is defined as final consensus.

\item
In case (3), each node can not distinguish this scenario from the others two scenarios. So each node has to wait a relative large amount of time $SBR$ and broadcasts an empty block $B^r_{\epsilon}$ if it has not yet received the agreed-upon result till the completion of wait. This scenario is defined as tentative consensus.
\end{enumerate}

\item
The agreed-upon block will be appended to blockchain and transactions contained in $B^r_{cb}$ will be confirmed instantly in the case of achieving final consensus. Transactions from a tentative block will be confirmed only if and when a successor block reaches final consensus.

\item
ACP produces two kinds of consensus: final consensus and tentative consensus. The introduction of these two notion is to guarantee the liveness. With a negligible probability, the presence of network partition can create a fork in the network and forks will severely impact the liveness in turn. To mitigate this problem, a recovery protocol is proposed below.

\begin{enumerate}[label=P5.\arabic*]
\item
Each node becomes aware of potential forking by checking whether the previous block hash in current block proposal is the same with the last local block hash.

\item
A node aware of potential forking proposes an empty block whose predecessor hash is the last final consensus block $B_{last-final-block}$ observed by him so far.

\item
Finally, the node invokes ACP to reach consensus on forked blocks, choosing the block with highest $B_{last-final-block}$ as his candidate block in the Stage \RomanNumeralCaps{3} instead.
\end{enumerate}
\end{enumerate}
In the process of executing $\text{PBFT}^{*}$ protocol, each node in Final Committee signed the message broadcasted by itself. Once reaching agreement, Final Committee members will broadcast the agreed-upon block with these signatures, allowing any nodes to validate the correctness of a block.

\clearpage
\paragraph*{The ACP Protocol}\mbox{}
\par\noindent\rule{\textwidth}{0.5pt}

\begin{center}
\textbf{Stage 1: Potential Committee (PC) Selection}
\end{center}

\par\noindent
Instructions for every node in the network: Node $i$ calculates for all nodes the Potential Weight(PW):
\begin{equation}
{pw}^r_j=\frac{H({RS}^{r-1},r,{PK}_j)}{{rep}^r_j}
\end{equation}
where $j=\{\mathrm{0,1,...},n-1\}$ is the indicator of nodes, ${RS}^{r-1}$ is the random seed of previous round, $r$ is the current round number, ${PK}_j$ is the corresponding public key of node $j$, ${rep}^r_j$ is the reputation of node $j$ in term $r$.  $H$ is the pre-specified hash function.

According to the weights and the expected Potential Committee size $m$, node $i$ checks where $i\notin {PC}^r$ or not.

\begin{itemize}
\item  If $i\notin {PC}^r$, then $i$ stops his own execution of ACP right away.

\item  If $i\in {PC}^r$, then $i$ moves to Stage 2.
\end{itemize}
\par\noindent\rule{\textwidth}{0.5pt}
\clearpage
\par\noindent\rule{\textwidth}{0.5pt}
\begin{center}
{\textbf{Stage 2: Final Committee (FC) Selection}}
\end{center}

\par\noindent
Instructions for every node in PC: Node $i$ computes its hash output ${\sigma }^r_i$ with round number $r$ as input string and checks whether $i\in {FC}^r$ or not according to the expected Final Committee size.

\begin{itemize}
\item  If $i\notin {FC}^r$, then $i$ stops his own execution of Stage 2 right away.

\item  If $i\in {FC}^r$, then $i$ broadcasts $(1,B^r_i,sig(B^r_i))$ to all Potential Committee members and moves to Stage 3.
\end{itemize}
\par\noindent\rule{\textwidth}{0.5pt}

\clearpage
\par\noindent\rule{\textwidth}{0.5pt}
\begin{center}
\textbf{Stage 3: Candidate Block proposal}
\end{center}
\par\noindent
Instructions for every node in FC:

\begin{enumerate}[label=3.\arabic*]
\item
After waiting an amount of time ${\lambda }_{pc}$, node $i$ votes the hash of the candidate block chosen from his received block proposals $B^r_{i\in \{0,...,fc-1\}}$by the following mechanism:

\begin{itemize}
\item  Chooses $B^r_i$ with the largest transaction size among $B^r_{i\in \{0,...,fc-1\}}$.

\item  Chooses $B^r_i$ with the least hash value among $H({\sigma }^r_{i\in \{0,...,fc-1\}})$ in case of multiple nodes with the same transaction size.
\end{itemize}

\item
After waiting an amount of time ${\lambda }_{fc}$, $i$ votes for the hash that received at least $\frac{2N_{fc}}{3}+1$ votes in the Step 3.1.

\item
After waiting an amount of time ${\lambda }_{fc}$, $i$ outputs a valid candidate block $B^r_{cb}$ that received at least $\frac{2N_{fc}}{3}+1$ votes in Step 3.2 or an empty block $B^r_{\epsilon }$ if no hash received enough votes.
\end{enumerate}
\par\noindent\rule{\textwidth}{0.5pt}

\clearpage
\par\noindent\rule{\textwidth}{0.5pt}
\begin{center}
\textbf{Stage 4: Reach Consensus}
\end{center}
\par\noindent
Instructions for every node in FC:

\begin{enumerate}[label=4.\arabic*]

\item
Node $i$ acts according to his output value at the end of Stage 3 as follows.

\begin{itemize}
\item  If $i$ outputs $B^r_{cb}$ and $H({\sigma }^r_i)$ is among the $N_{fc-valid-leader}$ least value in $H({\sigma }^r_{i\in \{\mathrm{1,...},fc-1\}})$, then $i$ runs the $\text{PBFT}^*$ instance as leader node with $B^r_{cb}$ as initial value. At the same time $i$ runs the others $\text{PBFT}^*$ instances with the least $N_{fc-valid-leader}$ hash values as initial value in in parallel.

\item  If $i$ outputs $B^r_{\epsilon }$ and $H({\sigma }^r_i)$ is among the $N_{fc-empty-leader}$ largest value in $H({\sigma }^r_{i\in \{\mathrm{1,...},fc-1\}})$, then $i$ runs the $\text{PBFT}^*$ instances as leader node with $B^r_{\epsilon }$ as initial value. At the same time $i$ runs the others $\text{PBFT}^*$ instances with the largest $N_{fc-empty-leader}$ hash values as initial value in parallel.
\end{itemize}

\item
After waiting a relative large amount of time $SBR$, $i$ checks whether it has received an agreed-upon block from any PBFT instance or not.

\begin{itemize}
\item  If $i$ has received an agreed-upon block, then $i$ propagates the received final consensus block $B^r_{cb}$ or $B^r_{\epsilon }$.

\item  If $i$ has not received an agreed-upon block, then $i$ broadcasts an empty tentative consensus block $B^r_{\epsilon }$.
\end{itemize}
\end{enumerate}
\par\noindent\rule{\textwidth}{0.5pt}
\clearpage
\par\noindent\rule{\textwidth}{0.5pt}
\begin{center}
\textbf{Stage P: Propagate Consensus}
\end{center}

\par\noindent
Instructions for every node in the network:

\begin{itemize}
\item  If node $i$ has received an final consensus block, then $i$ appends the agreed-upon block to the blockchain.

\item  If node $i$ has received an tentative consensus block, then $i$ keeps  the received tentative consensus block as pending state until it received a successor final consensus block. After the reception of a successor final consensus block, it appends the tentative consensus block and the successor final consensus block to the blockchain.
\end{itemize}
\par\noindent\rule{\textwidth}{0.5pt}

\clearpage
\par\noindent\rule{\textwidth}{0.5pt}
\begin{center}
\textbf{Stage R: Recovery Protocol}
\end{center}
\par\noindent
With a negligible probability, the presence of network partition can create a fork in the network and forks will severely impact the liveness in turn. To mitigate this problem, a recovery protocol is proposed below.

\begin{itemize}
\item  Each node monitors potential forking by checking whether the previous block hash in current block proposal is the same with the last local block hash.

\item  A node aware of potential forking proposes an empty block whose predecessor hash is the last final consensus block $B_{last-final-block}$ observed by him so far.

\item  The node aware of potential forking invokes ACP to reach consensus on forked blocks, choosing the block with the highest $B_{last-final-block}$ as his candidate block in the Stage 3 instead.
\end{itemize}
\par\noindent\rule{\textwidth}{0.5pt}
\clearpage
\subsubsection{Proof of Safety Property}

Assume that the numbers of Final Committee is $n$. After Reduction stage, there are two possible states of the output of each Final Committee member.

\begin{itemize}
\item  alert = false, which corresponds to a non-empty block proposal, denoted by $B^r_{cb}$.

\item  alert = true, which corresponds to a empty block proposal, denoted by $B^r_{\epsilon}$.
\end{itemize}

We now enumerate the possible value of the numbers of Final Committee members with non-empty block initial value, denoted by $p$, and the numbers of Final Committee members with empty block initial value, denoted by $q$.

\paragraph{Case1}
If $p\geq \frac{2n+1}{3}$, then only the $\text{PBFT}^*$ instance with $B^r_{cb}$ initial value will complete successfully, due to the fact that the majority nodes with alert=false will reject the $\text{PBFT}^*$ instance with a $B^r_{\epsilon }$ initial value .

\paragraph{Case2}
If $\frac{2n+1}{3}>p>0$, then all of the instances may complete with a failure result for the presence of both instances with $B^r_{cb}$ initial value and with $B^r_{\epsilon }$ initial value.

Without loss of generality, we can assume that instances with $B^r_{cb}$ initial value are $\text{PBFT}^*_{x1}$, $\text{PBFT}^*_{x2}$, $\text{PBFT}^*_{x3}$ and instances with $B^r_{\epsilon}$ initial value are $\text{PBFT}^*_{y1}$, $\text{PBFT}^*_{y2}$, $\text{PBFT}^*_{y3}$.

PBFT instances with $B^r_{\epsilon}$ initial value, that is $\text{PBFT}^*_{y1}$, $\text{PBFT}^*_{y2}$, $\text{PBFT}^*_{y3}$, will abandon their running instances in any cases once they are aware of the existence of instances with $B^r_{cb}$ initial value, that is $\text{PBFT}^*_{x1}$, $\text{PBFT}^*_{x1}$, $\text{PBFT}^*_{x3}$.

The only case that $\text{PBFT}^*_{y*}$  has achieved agreement while $\text{PBFT}^*_{x*}$ have not sent the first message yet is that the number of PBFT instances with $B^r_{cb}$ initial value is very few. And furthermore $\text{PBFT}^*_{x*}$ and $\text{PBFT}^*_{y*}$ are located in different network partition. In this case, we let each node wait an amount of time $\lambdaup$ after sending a broadcast and tag a logic index for each received message $r-s:4$-phase:$i$, which means the $r$-th round, $4$-stage in $i$-th phase. Assume that the start time of a phase is $T_{phases-i}$, then in [$T_{phases-i}$, $T_{phases-i}$ + $\lambdaup$],  a node will receive most message sent from honest node and discard the message after $T_{phases-i}$ + $\lambdaup$. With this method, We can ensure that only one instance can achieve agreement.

\paragraph{Case3}
p=0, then all PBFT instances are with $B^r_{\epsilon}$ initial value, and thus all instances will complete successfully. We lists the instructions for each node $i$ in Final Committee:

\begin{itemize}
\item  If alert=false, then $i$ rejects PBFT instances with $B^r_{\epsilon}$ initial value.

\item  If alert=true and there exists running PBFT instances with $B^r_{cb}$ initial value, then $i$ rejects PBFT instances with $B^r_{\epsilon}$ initial value if no agreement has achieved, otherwise, abandon the current running instance.

\item  Node $i$ has to wait an amount of time to receive most of the message sent from honest node after broadcasting a message in $T_{phases-i}$, and discard all overtime messages.
\end{itemize}

\subsubsection{Security Analysis}

To achieve scalability and keep resilient against large scale DDOS attacks, two committees---the Potential Committee and the Final Committee are randomly selected from the total set of nodes. The parallel $\text{PBFT}^*$ protocol is then run within Final Committee. The agreement is achieved in the Final Commit and broadcast to the entire network. Therefore, when the Potential Committee and Final Committee can be normally selected and the Byzantine nodes is less than a third of the selected committee members, the nature of safety and liveness can be guaranteed by $\text{PBFT}^*$. The selection of Potential Committee is driven by the decentralized random beacon which ensures that the generation of random seed is provably immune from manipulations and unpredictable. After revealing the random seed, the membership of Potential Committee are known to all nodes. As to the membership of Final Committee, only after each node in Final Committee constructs his block proposal and broadcasts his identity, the membership of Final Committee are known to all nodes in Potential Committee. There is therefore potential attacks towards Potential Committee and Final Committee members. Below we will analyze the various types of attack and the economic incentives of these attacks to demonstrate that large-scale attacks on members of the PC cannot be performed successfully in the AME network.

\paragraph*{Nothing at Stake}

The Nothing at Stake attack is when the validator casts different blocks in the voting stage without penalty, which may causes the ACP consensus protocol to generate an empty block. In response to this attack, the ACP consensus solution is to provide economic incentives to validators in Potential Committee and Final Committee only in the case that valid block are generated without forking.

\paragraph*{Selfish Mining}

Selfish mining refers to the behavior of the selected block proposer prioritizing or only packaging the transaction that is beneficial to itself, so as to complete the confirmation of the self-interest transaction. Since all transactions are valid, the verification nodes cannot distinguish this behavior. In addition, because of its small size, the broadcast speed is faster and easier to be confirmed by other nodes. From this perspective, Final Committee members tend to selfish mining in order to prioritize their own interest-related transactions, reduce waiting time, and have a greater chance of becoming a final miner. To address this issue, we associate the volume of transactions in the blocks proposed by Final Committee members with the gains they can earn as validator.

\paragraph*{Sybil Attack}

Sybil attack in blockchain network is an attack where a single adversary is controlling multiple nodes in the network. Each node in AME network has unique identity, and the AI governance module in AME network also requires each node to report the corresponding identity, ip and other network attributes, which will make Sybil Attack difficult to appear in the AME network.

\paragraph*{Eclipse Attack}

Eclipse attack, which targets a specific node and sends them blocks of a private fork, while attempting to eclipse them from the rest of the network so that they don't see the main blockchain. According to the design of ABNN, each node has a large number of neighbor nodes and counterpart nodes, which mitigates the eclipse attack in AME network. In addition, to defend against Eclipse Attack, the AI governance module in AME network detects whether the interconnection graph of each node is abnormal.

\paragraph*{Adaptive Adversary}

The AI governance module in AME will adjust the reputation of each node according to his long-term performance. Once an malicious behavior is detected by the governance module, the Byzantine node will be punished by reducing its reputation. Therefore, it is controlling a large number of nodes over the threshold in the Potential Committee and Final Committee to manipulate the generation of the block for a long time can be eliminate in AME system.

\paragraph*{Incentive as Countermeasure of Threat}\mbox{}

\begin{figure}[htbp]
\centering
\includegraphics*[width=4.17in, height=3.57in, keepaspectratio=true]{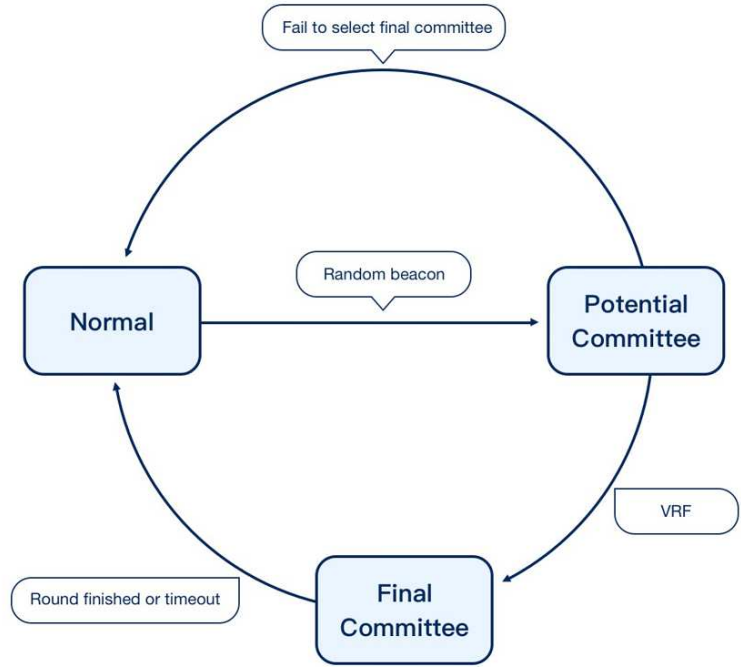}
\caption{Node State Transition}
\end{figure}

\subparagraph{PC Incentives}
The Potential Committee is selected in the first stage of ACP protocol. Potential Committee members further selected Final Committee through VRF calculations. Economic incentives should be granted for their devoted works. More importantly, Potential Committee members are motivated to ensure that they do not maliciously broadcast Final Committee membership to other nodes in the network, causing Final Committee members to be attacked by Byzantine nodes. When a node in Potential Committee knows that it has not selected in Final Committee in this round and has no chance to get the ``mining'' reward, there may be malicious attack toward Final Committee members. Therefore, we propose that Potential Committee members will be rewarded in next round given that the round they participated has successfully completed. Thus, ACP can avoid the Nothing at Stake attacks.

\subparagraph{FC Incentives}
Final Committee is selected in the second stage of ACP protocol. Final Committee members bear the responsibility of constructing block and establishing consensus through the $\text{PBFT}^*$ algorithm, which is also the ``mining'' in the traditional sense. In order to encourage the miners to work honestly and reduce malicious behavior, economic incentives should be granted to Final Committee members after reaching an agreement on a valid block in each round.

\subsubsection{Complexity Analysis}

In this section we analysis the communication complexity and latency of ACP protocol in each round.

In normal case operation, there are following message transmissions in each round of ACP protocol: the message of block proposals broadcasted by selected Final Committee members; the 2 message delay for running Reduction procedure; the 4 message delay for running PBFT; the message of agreed-upon result broadcasted to the whole system. That is, there are 8 message delay in each round of ACP protocol and the message volume is:
\begin{equation}
N^2_{pc}+2N^2_{fc}+{3N}^2_{fc}N_{fc-valid-leader}+3N^2_{pc}N_{fc-empty-leader}+N_{fc}N_{all}
   \label{Eq:equation1} 
\end{equation}
There are multiple $\text{PBFT}^*$ instances running in parallel. Each instance has a leader node respectively. Unless all of the leader nodes running $\text{PBFT}^*$ are Byzantine nodes, causing an extra waiting periods $SBR$, the protocol can completes at the end of any $\text{PBFT}^*$ instance process.

\subsubsection{Evaluation}

We provide our estimation of AME throughput and latency in this section. We start with some assumptions:

\begin{enumerate}
\item  1000 ms is needed to calculate the hash values of all nodes in the system.

\item  500 ms is needed to broadcasting a message to Potential Committee with 512 nodes.

\item  200 ms is needed to broadcasting a message to Final Committee with 16 nodes.

\item  3000 ms is needed to broadcast a message to the whole system with $100,000$ nodes.

\item  200 ms is needed to a message transmission between two nodes.
\end{enumerate}

The agreement time of ACP protocol can be estimated as below:

\[1000 + 500 + 6*200 + 3000 = 5700\text{ms} = 5.7\text{s.}\]

The throughput and latency of ACP can be estimated as below:
\begin{table}
\centering
\begin{tabular}{|m{0.9in}<{\centering}|m{1.0in}<{\centering}|m{1.6in}<{\centering}|m{1.6in}<{\centering}|}
\hline 
 &Block size & \parbox{1.6in}{\centering With a 3s \\ agreement time (tps)} &\parbox{1.6in}{\centering With a 5.7s \\ agreement time (tps)}\\
\hline 
\multirow{2}*{Bitcoin} & 4M & 1028 & 541 \\
\cline{2-4}
 ~& 8M & 2056 & 1082 \\
\hline 
\multirow{2}*{Ethereum} & 4M & 2608.4 & 1372.8 \\
\cline{2-4}
 ~& 8M & 5216.8 & 2745.6 \\
\hline 
\end{tabular}
\end{table}

In order to estimate the most suitable hop number for specific network setting, we estimate the block time under various setting by following equation:

\begin{equation}
(\dfrac{8\cdot BS_{block\_size}\cdot\sqrt[h]{N}}{B_{bandwidth}}+\dfrac{RTT}{2})\cdot h \leq T_{block}
\end{equation}

Where $N$ is the numbers of whole network; $B_{bandwidth}$ is the bandwidth; $RTT$ is the round-trip time; $h$ is the hop number; $BS_{block\_size}$ is the block size; $T_{block}$ is the block time. The equation is deduced following steps below:

\begin{enumerate}
\item
$\sqrt[h]{N}$ is the number of receiving end to which each node has to send at each hop.
\item
$BS_{block\_size}\cdot\sqrt[h]{N} $ is the amount of data to be sent at each hop. $\frac{8\cdot BS_{block\_size}\cdot\sqrt[h]{N}}{B_{bandwidth}}$ is the time required for sending all data at each hop.
\item
Per-hop time equals to the sum of network transmission time and the time required for sending all data locally.
\item
The block time equals to the multiplication of the number of hops and per-hop time.
\end{enumerate}
We list the estimated block time under various setting below: 

\begin{center}
\includegraphics*[width=6.2in, height=4.72in, keepaspectratio=true]{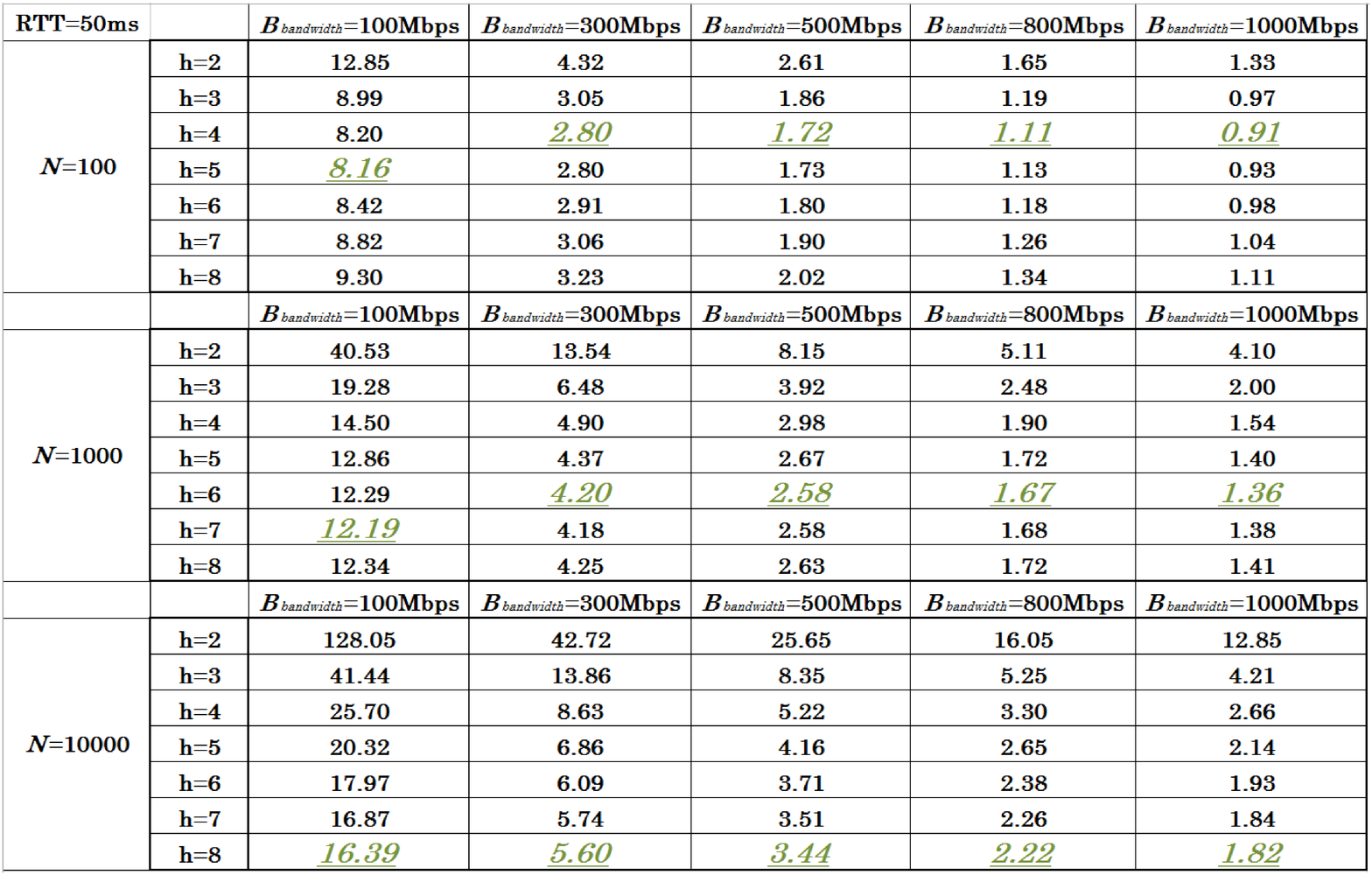}
\end{center}

\begin{center}
\includegraphics*[width=6.2in, height=4.72in, keepaspectratio=true]{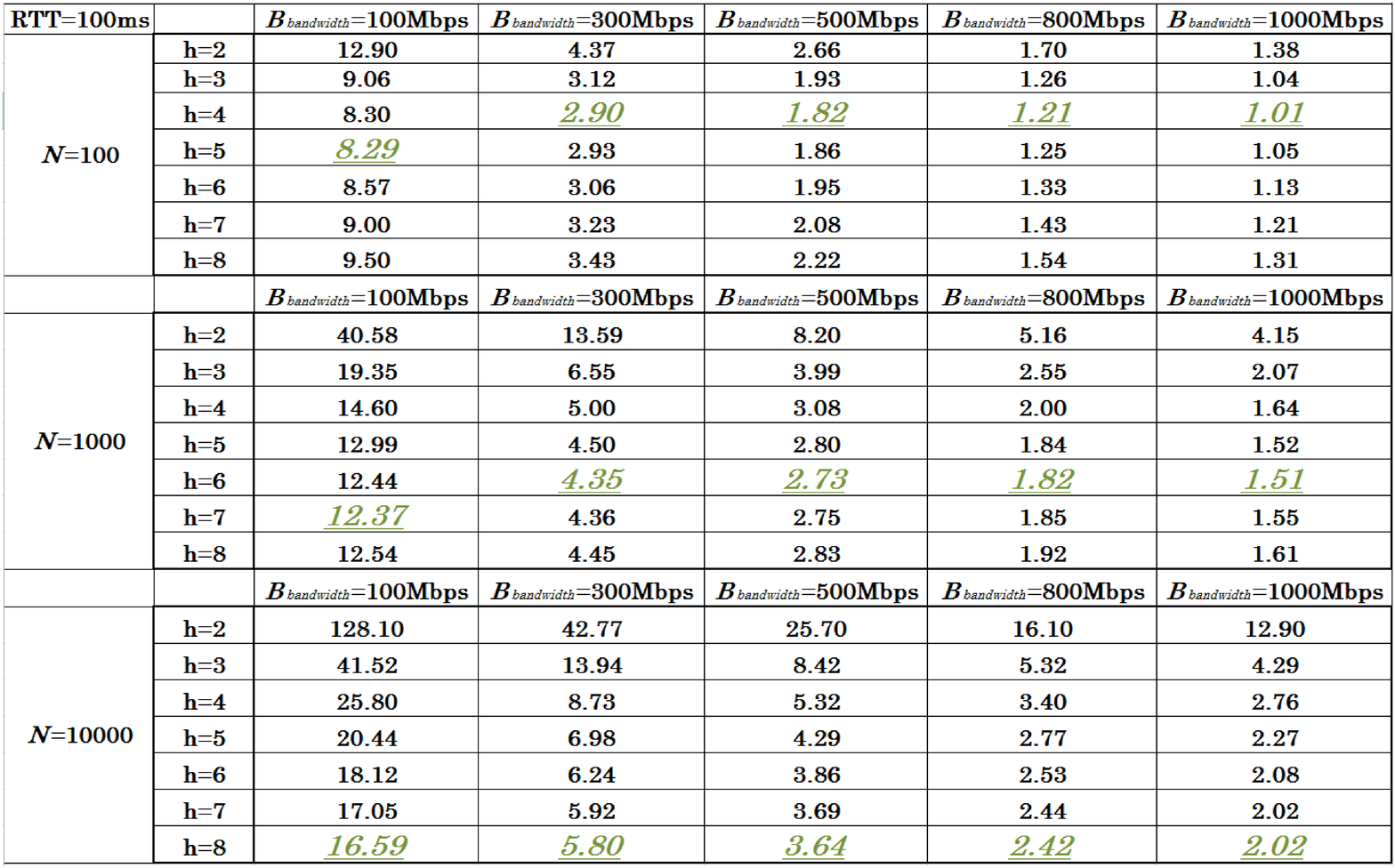}
\end{center}
\begin{center}
\includegraphics*[width=6.2in, height=4.72in, keepaspectratio=true]{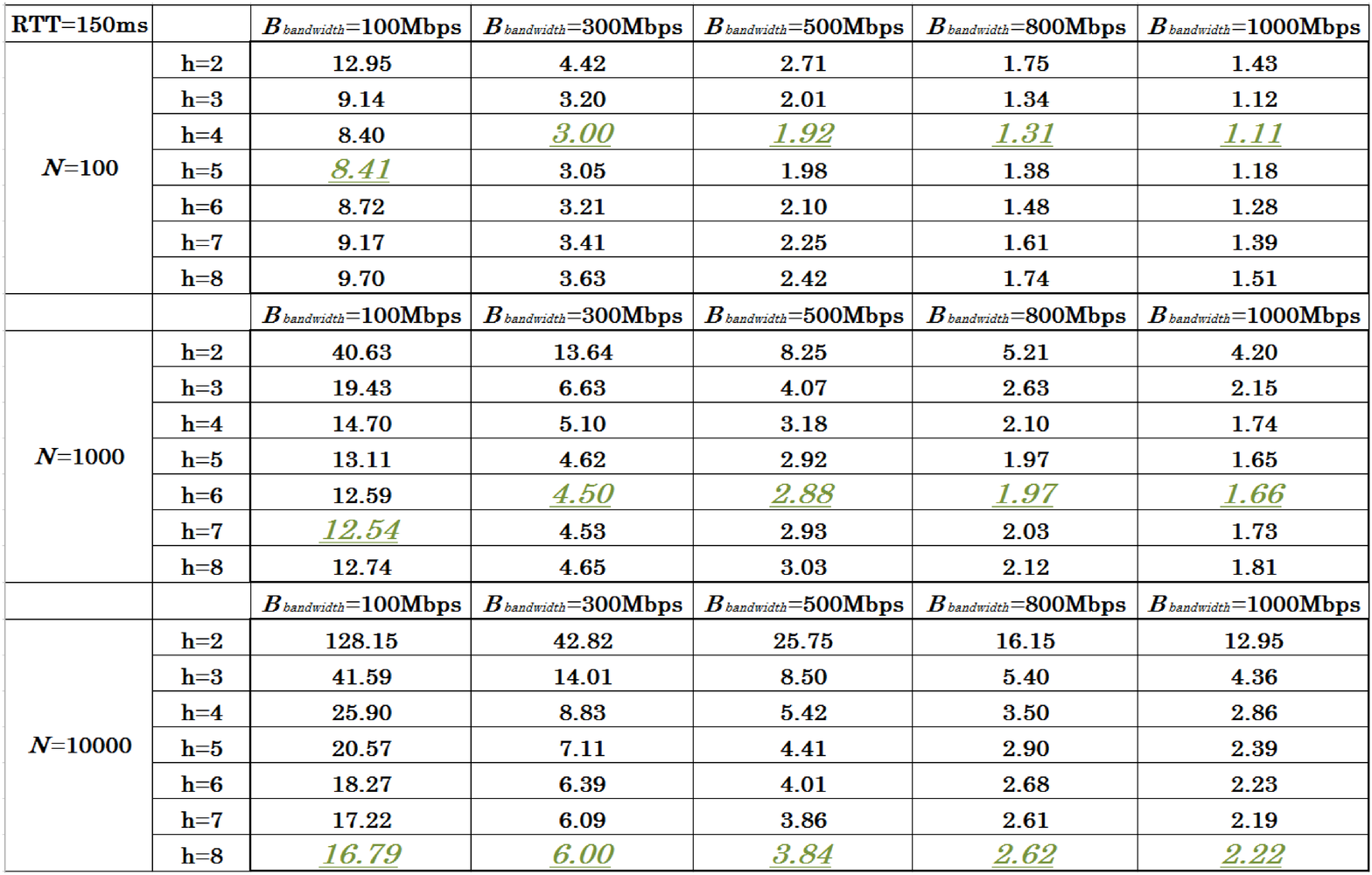}
\end{center}
\begin{center}
\includegraphics*[width=6.2in, height=4.72in, keepaspectratio=true]{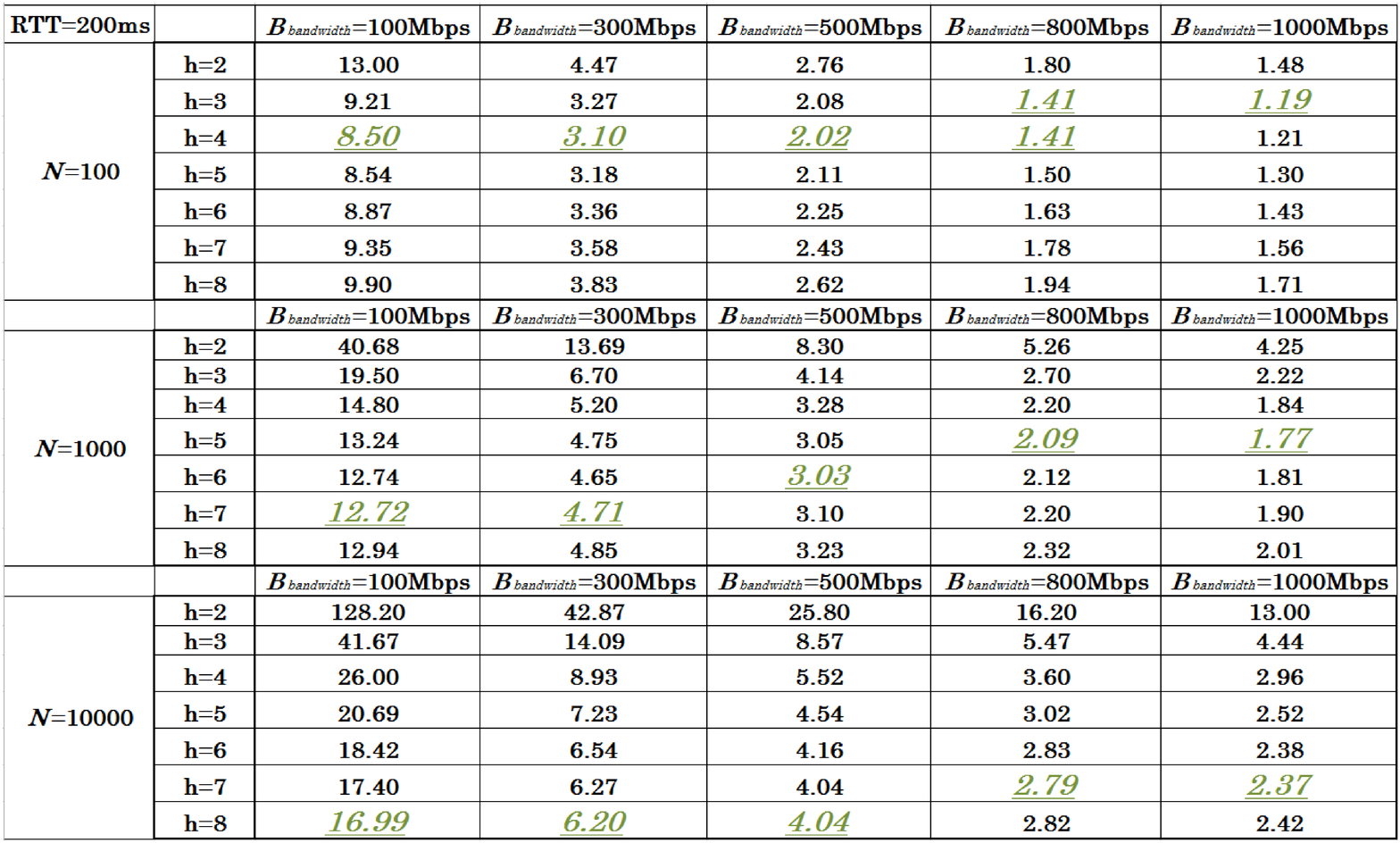}
\end{center}

\subsection{Conclusion and Future Work}

In this chapter, we have described the limitation of existing blockchain consensus and proposed a novel blockchain consensus protocol ACP with economic incentive compatible. The ACP consensus protocol offers the ability of instant transaction confirmation and the state-of-the-art throughput performance, while preserving high security and scalability and maintaining decentralization. In future, we plan to upgrade the protocol in following aspects:

\begin{enumerate}
\item  Design a novel transfer mechanism that message and block are transferred separately. Further, we will combine the push/pull and multi-level node model in the transmission process to improve network transmission performance.

\item  Employ threshold-signature technology in the last stage of ACP to accelerate the process of achieving agreement.

\item  Conduct study on dynamic replacement in Final Committee to improve the security of ACP.
\end{enumerate}

\clearpage
\section{AME Blockchain Native Storage -- ABNS}
\setcounter{figure}{0}
In this chapter, we will introduce a storage solution ABNS (AME Blockchain Native Storage) in the AME system optimized for blockchain applications. We present the architecture of our solution in \textbf{Section 3.1} and challenges of blockchain storage in \textbf{Section 3.2}. Then we will discuss in detail our implementation in \textbf{Section 3.3}.

\subsection{Architecture}

From the technical perspective, the blockchain is a large distributed storage system with Byzantine fault tolerance. However, the emerging blockchain applications expose unprecedented challenges to traditional distributed storage, due to its unique data characteristics and application models. To address the challenges and optimize the distributed storage, we look into two layers, i.e., architecture layer and storage engine layer. From the view of architecture, the optimization idea is relatively simple, similar to common approaches used in other system designs: extract and move storage related operations that cause high overhead out from the system critical path. This is also a common approach for handling data storage in mainstream blockchain application frameworks, and is called off-chain storage. From the view of storage engine, we propose the design of AME Blockchain Native Storage (ABNS), based on our deep understanding of blockchain storage that combines cutting-edge research achievements and practices.

\begin{figure}[htbp]
\centering
\includegraphics*[width=5.63in, height=2.98in, keepaspectratio=true]{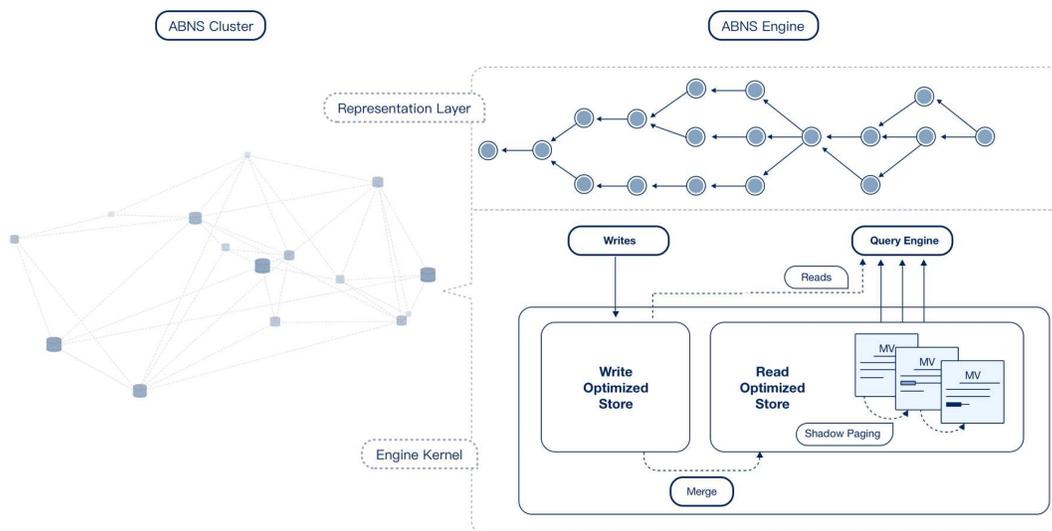}
\caption{ABNS Architecture}
\end{figure}
As shown in \textbf{Figure 3.1}, the ABNS cluster on the left represents the entire storage cluster. It is a decentralized storage system consisting of multiple nodes, each of which is an ABNS Engine. On the right, we illustrate the internal design of the ABNS Engine, composed of two layers: the bottom layer (i.e., Engine Kernel) is a high-performance, high-concurrency storage engine implemented with cutting-edge database and distributed system technologies; the upper layer (i.e., Representation Layer) achieves many blockchain-friendly features, such as multi-versioning, forking and immutability.

\subsection{Challenges and Requirements for Blockchain Storage}

\subsubsection{Challenges for Traditional Storage Systems}

First, the data stored in the blockchain system is very special compared to data in other systems. Currently, the mainstream blockchain systems store data as key-value pairs, where the key is usually the hash of the value part (also known as data fingerprint or digest). The keys stored in the key-value stores are therefore scattered in the entire key space. This trait poses a huge challenge for LSM-tree-based\cite{3_1} key-value storages, because the insertion of a new key-value pair may result in huge amounts of sorting workloads. Since ``real-time global ordering'' is one of the core features in the storage system, these sorting workloads cannot be avoided, though might be delayed to a certain extent. Moreover, as a single transaction usually involves reading and writing multiple key-value pairs, the looseness in the key distribution will cause requested keys to be scattered over different locations in the underlying physical storage, i.e., with poor data locality. This phenomenon brings about a huge number of random reads and writes, which leaves a heavy burden on the design of internal cache policy in the storage.

Second, the blockchain system usually needs to maintain different types of data, including blocks, transaction details and global states. The blockchain has special access patterns for these stored data. For example, the blockchain data access follows such a pattern that has high read ratio, low write ratio and writes cannot be blocked by reads. Therefore, the conventional optimizations in databases have no obvious improvement for data accesses in the blockchain.

Last but not least, due to the emergence of blockchain-based applications, the analytical queries over blockchain data become increasingly demanding. However, the widely used LSM-tree-based key-value stores have very limited functionalities and performance for such analytical queries, e.g., the lack of concurrent query or range query capabilities.

Besides the data storage challenges brought by the blockchain, there are also other limitations from the mainstream storage engine themselves. For example, LevelDB\cite{3_2} is one of the best stand-alone storage engines today, but it still has following limitations or disadvantages:

\begin{itemize}
\item  Almost all data files need be rewritten when conducting the full compaction, and the storage requires at least double space compared to the actual data size;

\item  The compaction is an extremely time-consuming operation and cannot be aborted, as re-execution of an aborted compaction needs to start from scratch;

\item  The read/write amplification effect is very severe, which directly degrades the performance of normal read and write operations.
\end{itemize}

According to relevant research findings\cite{3_3}: In the experimental setting that a LevelDB manages records with 16-byte keys and 1K-byte values, the write and read amplifications are about 14 and 327 times respectively when the data reaches 100GB.

\subsection{Storage Requirements for Data Models}

In the previous section, we discussed the challenges faced by storage systems for handling blockchain data characteristics. In this section, we discuss the blockchain storage requirements from the data model perspective.

\subsubsection{Multi-Versioning with Traceability}

The blockchain data model consists of two core concepts (i.e., blockchain = block + chain), which together form a weakly centralized ``global state'' as ``blockchain''. A ``Block'' records a specific change in the ``global state'', and the ``chain'' guarantees the traceability of the ``global state'' changes. From a macro perspective, the blockchain and the traditional storage can be conceptually mapped: a block contains multiple operations (i.e., transactions) that change the system status. In fact, each operation can also be packed as a single-record block. Packing multiple operations as a block is actually a specific optimization. This can be viewed as batch operations or group submissions in traditional storages. In addition, the blockchain generates a new version after a transaction is committed in the new block, which is similar to the transaction processing in traditional storages. However, one major difference is that most storages only maintain the latest version of data, while the blockchain maintains all historical transactions that are visible and verifiable from the user side.

In summary, when designing the storage engine for a blockchain system, we need to take the particularity of the data model into consideration. It is desirable to support such data models natively in the underlying storage engine, which might lead to an optimal technical solution. If blockchain data is maintained by a generic storage engine, it requires additional data transformation and adaptation at the application layer, apart from the schema design and storage engine.

\subsubsection{Multi-branched Data Versioning}

In the blockchain, there is a very important module called the consensus protocol, which is also called consistency protocol in the traditional distributed storages. Since the Byzantine problem is not considered in traditional storages, the mainstream strong consistency protocols are Paxos\cite{3_4}, Raft\cite{3_5}, Viewstamped\cite{3_6}, etc. In contrast, Byzantine fault tolerance is a necessity in blockchain systems, and their consistency protocols are mainly categorized into three classes as follows:

\begin{center}
\begin{tabular}{|m{0.9in}<{\centering}|m{3.1in}<{\centering}|m{1.52in}<{\centering}|} \hline 
\rowcolor{tbcolor}
Category & Basic Principle & Typical Implementation \\ \hline 
Algorithm Class & Achieve consensus by passing messages between nodes and verifying constraints & PBFT\cite{2_2}, Tangaroa\cite{3_8}, etc. \\ \hline 
Engineering Class & Achieve consensus from an engineering or sociological perspective, using economic game theory and complex computations & PoW, PoS, DPoS, etc. \\ \hline 
Synthesis Class & Achieve consensus using synthesis of algorithm, cryptography, hardware and engineering & Algorand\cite{3_9}, ByzCoin\cite{3_10}, Chainspace\cite{3_11}, etc. \\ \hline 
\end{tabular}
\end{center}

The ``Algorithm Class'' consistency protocols are rigorous and have been well proven with mathematical derivation. The ``Synthesis Class'' consistency protocols are similar to those from ``Algorithm Class''. They are mainly designed to address the scalability issue in ``Algorithm Class'' protocols, and all provide strong consistency. On the contrary, the ``Engineering Class'' protocols mainly provide eventual consistency, proven by existing theoretical analysis. For example, it is possible for two miners to simultaneously find the hash values meeting the requirements, hence the blockchain might be temporary forked in normal PoW protocols. Consequently, as a distributed storage, the blockchain has a unique property that its historical states might not be linear, i.e., forks may occur. Such scenarios seldom occur in traditional strong storages.

\subsubsection{Easy Detection of Historical Data Tampering}

Another characteristic of the blockchain is that the data cannot be tampered (or, more accurately, is extremely expensive to tamper). Therefore, for a blockchain-based application, the most critical problem is how to quickly identify whether the blockchain has been tampered with or not. From a technical perspective, nothing is impossible to tamper with. However, if tampering can be quickly detected and participants no longer trust the affected data, the blockchain becomes tamper-proof from the engineering point of view. Consequently, for a blockchain-friendly storage engine, it must have the ability to detect data tampering efficiently and timely.

\subsubsection{Append-Only Data Property}

Since the blockchain needs to maintain the entire evolution history of the global state, it is different from the conventional storages. A blockchain-friendly storage engine cannot conduct in-place-update on a historical version, instead it needs to transform each modification into a new version (i.e., new state). For example, a user executes following commands in Redis\cite{3_12}: \textit{set abc `x'; set abc `y'}. For this modification sequence, the response of querying \textit{abc} will be \textit{abc = `y'}, after \textit{set abc `y'} command has been successfully executed. The user does not know that the previous version is \textit{x}, and the provenance information that \textit{y} is derived from \textit{x}. However, in the blockchain, any transaction must be traceable, hence we need to maintain not only the latest state of the data, but also the entire evolution history of the state. The historical data is not only used to verify the validity of blocks and transactions, but also opens up the opportunity of supporting AI applications and analytical queries in the future.

In summary, we argue that the current mainstream storage engines do not have native Blockchain-oriented properties. In addition, due to the rapid development and industrialization of AI technology, we believe that in the future, the blockchain applications must be deeply integrated with AI techniques, which further derive advanced blockchain applications that are weakly centralized or fully de-centralized. Moreover, the AI applications require efficient retrieval of various data stored in the blockchain, which forces the underlying storage engines to have efficient retrieval capabilities. In other words, the storage engines require multi-dimensional query functionalities to reflect and fulfill the uniqueness of the blockchain. Therefore, blockchain backend technology is the base for large-scale popularization of blockchain-based applications. In particular, under the premise of maintaining the main chain with blockchain characteristics, the improvement of scalability, robustness and performance of the main chain will become the core competitive strength of blockchain backend technology.

\subsection{ABNS Technical Solution}

ABNS is a blockchain native storage system with the complete support of ACID\cite{3_25} properties. It is inspired by the design of \textit{delta-main update}\cite{3_13} concept, and provides excellent write performance while perfectly satisfying the read-intensive nature of the blockchain system. ABNS adopts the MVCC (Multi-Version Concurrency Control)\cite{3_14} mechanism based on the shadow paging\cite{3_15,3_24} technology, which implements latch-free write/read and latch-free read/read, and further improves the system performance via zero-copy read operations. In addition, ABNS fully integrates widely used blockchain data models, such as MPT (Merkle Patricia Tree)\cite{3_16}. For the indexing methods, ABNS adopts the unique CoW\cite{3_17} and speculative inheritance\cite{3_18} GC approaches, which ensure that the scattered hash keys in the blockchain system still preserve good data locality and cache locality\cite{3_26}. As far as we know, ABNS is one of the few storage engines in current stage that focus on the core ecology for blockchain technology and have deep exploration for native support.

\subsubsection{Design Principle}

In this section, we briefly introduce the design principles behind the ABNS. During the design of a general-purpose storage engine, we limit the possible ways of reading and writing data, once the physical data layout on storage devices (e.g., disk, flash, memory and cache) are designed and finalized. Therefore, in most cases, we have to consider the trade-off among three directions\cite{3_19,3_20}, i.e., read optimized, write optimized and space optimized:

\begin{figure}[htbp]
\centering
\includegraphics*[width=4.4in, height=3.in, keepaspectratio=true]{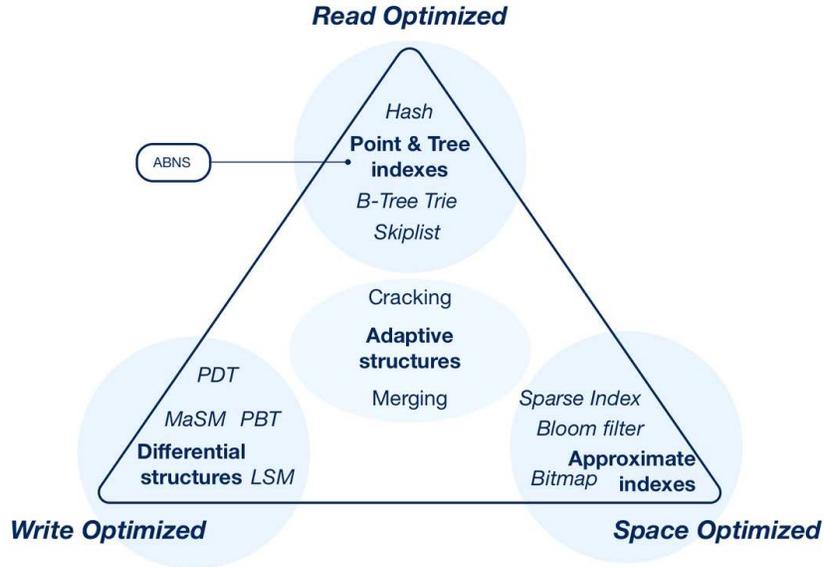}
\caption{ABNS Design Trade-off}
\end{figure}

As shown in \textbf{Figure 3.2}, optimizations towards any two directions will have negative impact on the third direction. Therefore, the design priority of ABNS is as follows: point lookup (read optimized), write, range lookup (read optimized), space optimized. Note that write is not the direction with highest priority, because we believe that the main bottleneck of the blockchain systems will still be in network communication and consensus mechanism for a long period of time from now. Besides, the trade-off between write throughput and latency can be tuned by flexibly configuring the write batch size; moreover, not all data requires real-time synchronous persistency, and we can exploit page cache and memory to alleviate the write pressure of the entire system. The point lookup has the highest priority in our design, because in order to improve transaction efficiency in the block system, it needs to conduct a large number of key-value queries during the execution of transaction processing and verification. The range lookup is of the third place, because ABNS is designed to support rich features and provide various possibilities for analyzing underlying blockchain data in different upper-layer applications. The primary goal is to ensure the high efficiency of blockchain transaction execution and to prevent the storage engine from becoming a bottleneck of the entire system.

After all, the design priority is a relative measurement, and we need to carefully consider the whole picture when designing a specific component. Any obvious limitation or bad design may affect the applicability of the ABNS. Therefore, the ranking of design priority does not mean that our design lacks the deep consideration or optimization on low priority directions.

\subsubsection{Design and Implementation}

The overall design of ABNS mainly consists of four layers, from bottom to top respectively: Engine Kernel (EK), Blockchain Feature Representation (BFR), Data Access Pattern (DAP) and Semantic Views (SV). EK is our self-developed key-value storage engine, whose design details will be covered in subsequent chapters. BFR is designed for common data structures and features of the blockchain system generalized from the EK design, such as: Merkle tree model abstraction, generic data validation abstraction, etc. All these abstractions can be further extended continuously. DAP is a high-level API, including point query, range query, point write, batch write, etc., which is actually the SDK provided in this layer. For the SV layer at the top, since different blockchain applications have own focuses depending on their specific domain knowledge and models, they need to use the API provided in this layer to customize differentiated data views.

\begin{figure}[htbp]
\centering
\includegraphics*[width=3.5in, height=2.5in, keepaspectratio=true]{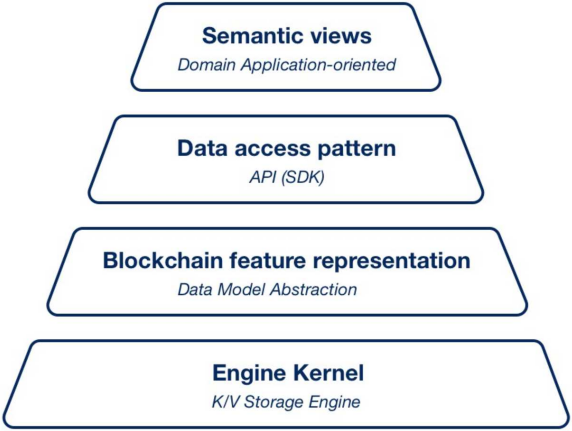}
\caption{Layered Design of ABNS}
\end{figure}

With the layered architecture like UStore\cite{3_22} and Forkbase\cite{3_23} shown in \textbf{Figure 3.3}, ABNS is able to deeply and independently optimize each layer, which offers high horizontal scalability without internal dependencies.

\begin{figure}[htbp]
\centering
\includegraphics*[width=4.1in, height=2.5in, keepaspectratio=true]{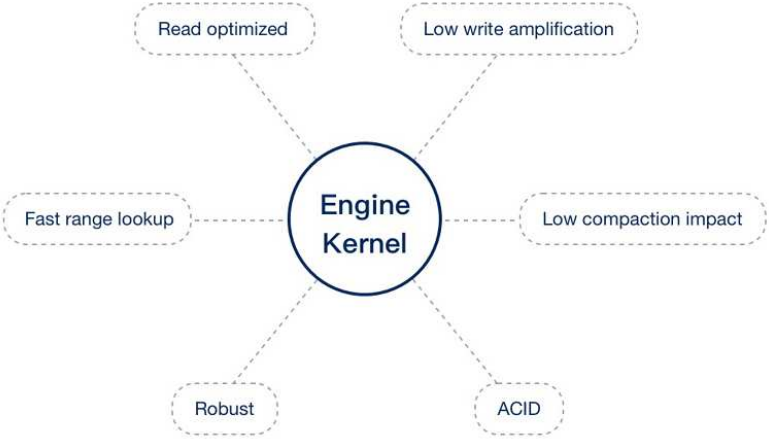}

\caption{Engine Kernel Features}

\end{figure}

\paragraph*{Engine Kernel Layer (EK)}\mbox{}
\vspace{\itemsep}

\noindent EK is the most critical part of the ABNS Engine and the base component of the entire ABNS Engine system. First, EK is a generic key-value storage engine, as shown in \textbf{Figure 3.4}, which provides rich features: read-optimized, low write amplification, low compaction impact, ACID, robustness, fast range lookup, etc.

The core data structure of EK is an implementation of B+-tree with append-only and shadow paging concepts\cite{3_24}. The index structure in B+-tree satisfies the requirement of intensive reads and fast range lookups. The conventional append-only approaches lead to very large storage overhead, and bring in large write amplification and compaction impacts. EK resolves the overhead and negative impacts by continuously recycling expired data pages via a unique speculative inheritance GC\cite{3_18} technology.

\begin{figure}[htbp]
\centering
\includegraphics*[width=4.9in, height=3.13in, keepaspectratio=true]{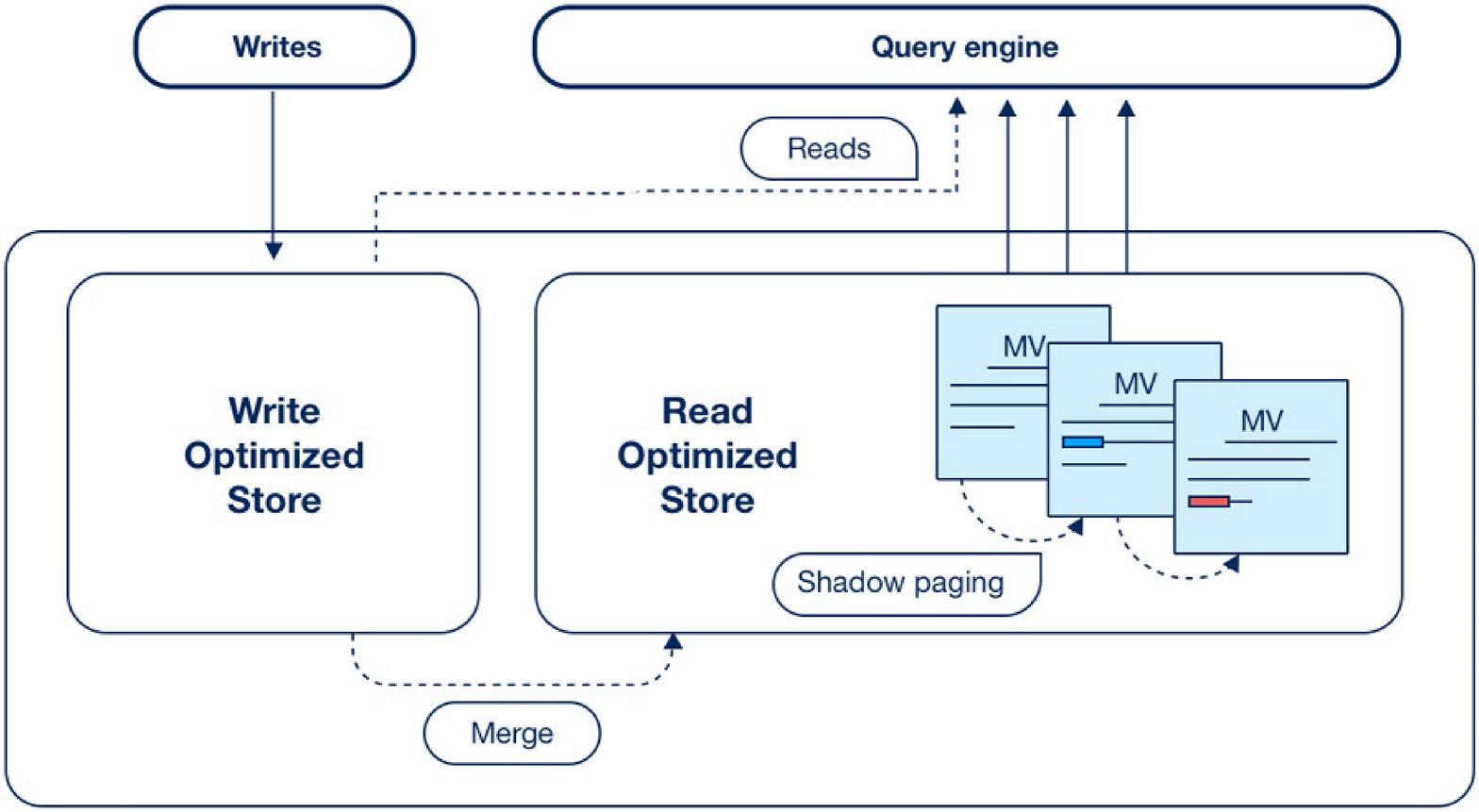}
\caption{Architectural Design of ABNS Engine Kernel}
\end{figure}

As shown in the figure above, EK offers latch-free write/read and latch-free read/read. Within each single partition, EK adopts a single-writer/multi-readers model, while for the entire system, it provides multi-writers/multi-readers model through partitioning. For data update, EK adopts the design of \textit{delta-main update} concept, which lowers the storage cost and improves the capability of concurrent processing. For physical data layout, EK adopts row-oriented layout, which provides superior point get/put operations and is more suitable for blockchain data access patterns. For data version management, by applying the shadow paging technology, EK organizes different versions of the same key as a chained sequence, which is friendly for conducting GC on old versions.

\paragraph*{Blockchain Feature Representation Layer (BFR)}\mbox{}
\vspace{\itemsep}

\noindent Since append-only and multi-version features are already included in the underlying EK design, these features can be directly utilized at the Blockchain Feature Representation (BFR) layer. In our design, two most important abstractions in this layer are: Merkle Tree Abstraction and Fast Validation Abstraction. Merkle Tree plays a very important role in the blockchain technology. Both Bitcoin and Ethereum adopt Merkle Tree implementations and optimizations, which are however heavily coupled with other modules. In addition, the underlying storage engines (e.g., LevelDB\cite{3_2} and RocksDB\cite{3_21}) do not natively support these features, i.e., append-only and multi-version. Therefore, in state-of-the-art blockchain systems, the support of these features has low efficiency and high complexity during the implementation.

In fact, For the Merkle Proof, its basic idea is similar to shadow paging, both of which aim to involve as less storage and computation related to the branch change as possible during the procedure. Hence, apart from the adoption of append-only and shadow paging in the B+-tree, we also consider the possible ultimate blockchain storage model in the long run\cite{3_27}. Possibly, the most ideal approach is not the key-value model, but a more blockchain native model. That is why Blockchain Feature Representation layer needs to provide functionalities such as existence checking of a specific key and verification of the associated value.

\paragraph*{Data Access Pattern Layer (DAP)}\mbox{}
\vspace{\itemsep}

\noindent Data Access Pattern layer provides SDK for upper-layer application developers. The provided APIs in the SDK can be flexibly extended according to different application scenarios and requirements.  The currently implemented APIs mainly include:

\begin{center}

 Put(key, base\_version, value) -$\mathrm{>}$ $\mathrm{\{}$version$\mathrm{\}}$

 Get(key, version) -$\mathrm{>}$ $\mathrm{\{}$value$\mathrm{\}}$

 GetPrevious(key, version) -$\mathrm{>}$ $\mathrm{\{}$values$\mathrm{\}}$

\end{center}

\paragraph*{Semantic Views Layer (SV)}\mbox{}
\vspace{\itemsep}

\noindent In Semantic Views layer, different blockchain applications focus on different aspects based on their domain knowledge and expert models. They can utilize the APIs provided in SV to achieve application-specific differentiated data views. The common data views include: fine-grained access control, data security, subscription/publishing of data updates, etc.

\subsection{Conclusion and Future Work}

In this chapter, we have outlined the ABNS, an advanced blockchain native distributed storage system. ABNS combines the cutting edge database technology with distributed system technology with focus on adapting blockchain feature in the design of ABNS. It perfectly addresses the performance, security, capacity, query processing and scalability issues faced by blockchain storage. In future, taking the intrinsic problem of blockchain, i.e., scalability and capacity into consideration, ABNS will focus on the following research directions:

\begin{enumerate}
\item  ABNS rebuilds genesis blocks by epoch, where the epoch interval is configurable

\item  ABNS stores complete blockchain data through technology of erasure coding, and any block can be reconstructed from other nodes to achieve low-consumption storage.

\item  By means of rebuilding and erasure coding technique, ABNS can purge cold-data safely, such that the scalability and capacity problem in blockchain is solved perfectly.
\end{enumerate}

\clearpage
\section{Service and Application}
\setcounter{figure}{0}
Here we mainly describe the service framework and application design based on AME platform. In the second half of this chapter, we will use AME-IM as a typical service case to introduce the specific content of services and applications.

\subsection{Service Architecture}

As a decentralized basic service platform, we summarize the service architecture of AME as follows:

\begin{enumerate}
\item  Decentralized application support platform based on Blockchain DNS, Open ID. 

\item  Above  the basic support platform, it is the abstract resource layer of the system, including resources such as storage, bandwidth and computing power etc.

\item  Applications are supported on top of the abstract resource layer.
\end{enumerate}
\begin{figure}[htbp]
\centering
\includegraphics*[width=4.7in, height=2.63in, keepaspectratio=true]{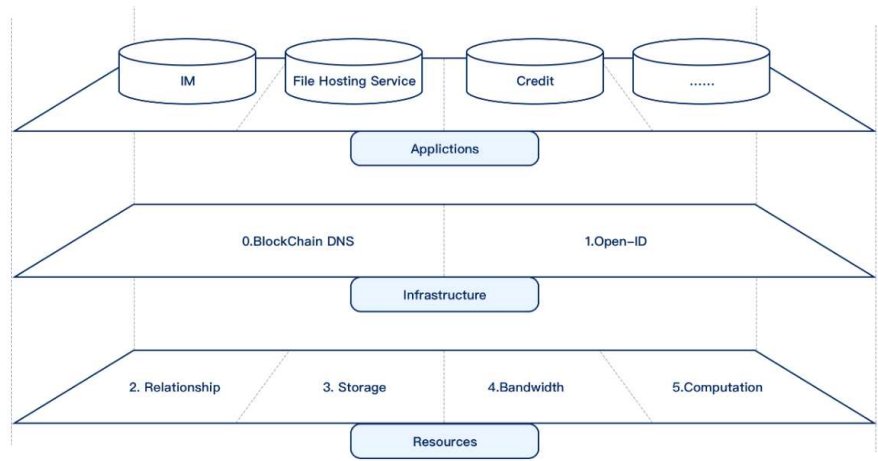}

\caption{AME Service Architecture}
\end{figure}

When a client requests a service from AME system, whether it is storing files or hosting services, or even relying on nodes on the network to complete computing tasks, it is essentially using three resources on the AME system including storage, bandwidth and computing. Therefore, we abstract these three resources and provide the OpenID service and decentralized DNS service to enable a uniform resource identifier and user-friendly domain name mapping. We describe these two function provided to decentralized application by AME system bellows.

\subsection{Decentralized Application Support Platform}

\subsubsection{Blockchain DNS}

It is an important aspect in a decentralized system about how to communicate. Earlier we mentioned that the Worker Ring can find the corresponding resource through the key. Meanwhile, the key is a 256-bit hash value which is unfriendly for users to use this public key (or its base58 or base64 encoding string) as an identifier to identify users. In order to address this problem, AME provides a fully decentralized DNS service at the support layer. BC-DNS (Blockchain-DNS) is a blockchain-based domain name service that comes with the system. The domain name in the BC-DNS domain name resolution service is not limited to the particular formatting rule ending in a fixed format such as .com or .org. It provides a user-friendly, readable, variable-length custom identifier (domain name) to the system account as well as the mapping between this identifier and its corresponding account.

\paragraph*{DNS Quick Query based on Manager Ring Servers}\mbox{}
\vspace{\itemsep}

\noindent It is relatively slow to query the BC-DNS domain name mapping by retrieving persistent storage through the get method of the BC-DNS smart contract. The system needs to provide a faster method to response BC-DNS domain name queries. In this case, we build a distributed in-memory BC-DNS database on the Manager Ring servers. Each Manager Ring server contains all the domain name mapping data of the fragment that it is currently responsible for. By providing a paid (or free) fast domain name retrieval service through the Manager Ring servers, the system nodes can quickly retrieve the BC-DNS domain name mapping through RPC or other means.

\paragraph*{DNS Query Process}\mbox{}
\vspace{\itemsep}

\noindent It is relatively slow to query the BC-DNS domain name mapping by retrieving persistent storage through the get method of the BC-DNS smart contract. The system needs to provide a faster method to response BC-DNS domain name queries. Therefore, we build a distributed in-memory BC-DNS database on the Manager Ring servers. Each Manager Ring server contains all the domain name mapping data of its corresponding shard. By providing a paid (or free) fast domain name retrieval service through the Manager Ring servers, the system nodes can quickly retrieve the BC-DNS domain name mapping through RPC or other means.

Assuming that the BC-DNS domain name query servers have completed the above service statement and are available, meanwhile, the client has obtained a list of servers that support the BC- DNS domain name query service. Below is a typical domain name retrieval process:
\begin{enumerate}
\item  The client retrieves the local BC-DNS cache. If there is an entry corresponding to the target domain name, exits and completes the query. Otherwise go to step 2.

\item  The BC-DNS dedicated server queries the cached local BC-DNS domain name mappings. If there is an entry corresponding to the target domain name, exits and completes the query. Otherwise, go to step 3.

\item  BC-DNS dedicated server initiates BC-DNS domain name retrieval to the Manager Ring server. If the corresponding entry is found, it will be cached into BC-DNS dedicated server. Then go to step 4.

\item  If the valid domain name mapping is obtained in step 3, it will be returned to the client. If there is no such domain name mapping, the ``domain name does not exist'' error is returned to the client.
\end{enumerate}

\paragraph*{Validity Period of BC-DNS Domain Name Mapping Cache}\mbox{}
\vspace{\itemsep}

\noindent Theoretically, the modification of domain name mapping data is very rare, but we still need a mechanism to handle BC-DNS domain name mapping updates in BC-DNS domain name query services.

Firstly, we set a timeout period for each domain name mapping on the BC-DNS domain name query server. For all queries, if the corresponding domain name mapping does not exceed the timeout period, it is considered valid and the mapping value is returned as the query result.

Secondly, the BC-DNS domain name query server must set the timeout period for returned results based on the current validity period of the domain name mapping. After receiving the result, the client considers that the domain name mapping is valid within the timeout period. Otherwise, delete the corresponding cache on the client and the client initiates a new BC-DNS domain name query request.

Based on the BC-DNS domain name resolution service, we can build a bunch of applications. We can treat the corresponding node’s IP address which appears in the URL https://yourdomain/index.html as the server IP by using BC-DNS domain name resolution service.

\subsubsection{Open ID}

The abstraction of accounts and resources in the AME system can ultimately be mapped to the same address space and obtain a unique identifier. This is the OpenID, which is essentially a 256-bit string. The address space of AME OpenID is about 2 to the power of 256, which is about 1.16 times 10 to the power of 77. This address space is large enough to ignore the problem of index conflict. The OpenID of each resource is unique in the AME system. It is also an index. With OpenID, users can easily and quickly find various resource in AME system including but not limited to an account, a file (digital fingerprint), as well as a service (address mapping), or even a node server (inter-node communication). For example, there is a service deployed in the AME system. According to the characteristics of the service, the system finally assigns it an index of 'hlkGDG34fjglk-2kdC0djfTlkaDsAdaDfdQs', which is a kind of OpenID. With this OpenID, users can find and get the corresponding service easily.

\subsection{Typical Case: BCM-IM as a Service}

AME-IM is an instant messaging application based on AME blockchain platform. With unique decentralized distributed technology, it can provide highly secure service to ensure free global chat with perfect forward secrecy (PFS). It also provides built-in wallet function which makes sending red envelopes and transferring money as easy and fast as chat.

The main features of AME-IM are:

\begin{itemize}
\item
Global decentralized server which does not require VPN to guarantee communication quality in various places.

\item
Support end-to-end encryption scheme to protect privacy.

\item
Support for sending and receiving most mainstream cryptocurrencies.

\item
Support Moments, Paid Group, Official Account and other functions.
\end{itemize}

\subsubsection{Open ID-based IM Account System}

In the AME-IM instant messaging application, the mapping between the basic support platform user account key value and user-readable, friendly “nickname” (domain name in BC-DNS) is established through the OpenID system (see \textbf{Section 4.2.1} BC-DNS). It enables users to build custom, recognizable personalized nicknames on the OpenID system. For the specific process of nickname registration and modification based on the open account system, please refer to \textbf{Section 4.2.1} BC-DNS.

\subsubsection{IM Backend as Decentralized Service}

Based on the decentralized application support platform (see \textbf{Chapter 4.2} Decentralized Application Support Platform), we treat the IM backend service as a special application on the support platform. The IM backend server joins the Worker Ring DHT and registers the IM service with the AME blockchain platform. In this way, in the Worker Ring DHT network, each IM backend server is functionally independent on the IM backend service, and can provide services independently. At the same time, through the DHT network interconnection, a decentralized AME-IM background service group is formed.

\subsubsection{IM Client Access Method}

By querying from the AME blockchain platform for the list of available servers supporting the IM background service, the IM client selects an available IM server to access the AME-IM network. Meanwhile, it ensures the long connection between the client and the current IM server to complete the accession of an IM client. It should be noted that when the IM client obtains the IM backend server list from the AME blockchain platform, the client caches it in the local storage. Therefore, when the client is off and restarts, it can get the last active server address and initiate a connection to it. Moreover, when the currently active IM backend server is unavailable, another IM backend server can be selected to ensure the reliability of the background service.

\clearpage
\section{AME's Security Technology Scheme}
\setcounter{figure}{0}
AME system is based on mature cryptography technology and components. It uses the best security practice technologies which are widely used in industry to protect users' security and privacy. AME adopts end-to-end encryption technology based on Signal protocol to protect users' communication content from being eavesdropped by any third party (including hackers, communication operators, national teams, etc.). AME also uses zero-knowledge proof and Tor-based anonymous network communication technology to prevent network traceability, thus protecting users' privacy. AME also employs quantum resistant encryption algorithm technology. With the coming of quantum computing, our encryption algorithm will still be able to withstand attacks from quantum computers. This section will describe three parts corresponding security techniques in details.

\subsection{End-to-End Encrypted Communication}

AME uses the Signal protocol to encrypt the transmission channel between the two clients. In this case, any third party, hackers, communication operators, national teams, including AME development team, cannot view the communication content. The security of the user's message content is not only guaranteed by the ethic of these third parties, but also by technical mechanisms. AME applies Signal protocol in both end-to-end communication and group communication, which ensures the security of transmission of message, picture, audio, video and other files. The AME system also provides forward security and backward security. Even if the key of a message is leaked, the hacker still cannot decrypt the previous and subsequent messages.

\subsubsection{Transmitting Media and Other Attachments}

Large attachments of any type (video, audio, images, or files) are also end-to-end encrypted:

\begin{enumerate}
\item  The sender (AME user who sent a message) generates an ephemeral 32 byte AES256 key, and an ephemeral 32 byte HMAC-SHA256 key.

\item  The sender encrypts the attachment with the AES256 key in CBC mode with a random IV, then appends a MAC of the ciphertext using HMAC-SHA256.

\item  The sender uploads the encrypted attachment to a blob store.

\item  The sender transmits a normal encrypted message to the recipient that contains the encryption key, the HMAC key, a SHA256 hash, and a pointer to the encrypted message in the blob store.

\item  The recipient decrypts the message, retrieves the encrypted blob from the blob store, verifies the SHA256 hash of it, verifies the MAC, and decrypts the plaintext.
\end{enumerate}

\subsubsection{Group Message Security Communication Technology}

Traditional unencrypted messenger apps typically employ ``server-side fan-out'' for group messages. When a user sends a message to a group, the server distributes the message to each group member. And ``client-side fan-out'' is the client sends a message to each group member. AME's group message is build on the pairwise encrypted sessions outlined above to achieve efficient server-side fan-out for most messages sent to groups. This is accomplished using the ``Sender Keys'' component of the Signal Messaging Protocol.

The first time a AME group member sends a message to a group:

\begin{enumerate}
\item  The sender generates a random 32-byte Chain Key

\item  The sender generates a random Curve25519 Signature Key key pair.

\item  The sender combines the 32-byte Chain Key and the public key from the Signature Key into a Sender Key message.

\item  The sender individually encrypts the Sender Key to each member of the group, using the pairwise messaging protocol explained previously.
\end{enumerate}

For all subsequent messages to the group:

\begin{enumerate}
\item  The sender derives a Message Key from the Chain Key, and updates the Chain Key.

\item  The sender encrypts the message using AES256 in CbC mode.

\item  The sender signs the ciphertext using the Signature Key.

\item  The sender transmits the single ciphertext message to the server, which does server-side fan-out to all group participants.
\end{enumerate}

The ``hash ratchet'' of the message sender's Chain Key provides forward security. Whenever a group member leaves, all group participants clear their Sender Key and start over.

\subsubsection{Verifying Keys}

AME users additionally have the option to verify the keys of the other users with whom they are communicating so that they are able to confirm that an unauthorized third party (or AME) has not initiated a man-in-the-middle attack. This can be done by scanning a QR code, or by comparing a 60-digit number. The QR code contains: a version, the user identifier for both parties and the full 32-byte public Identity Key for both parties. When either user scans the other's QR code, the keys are compared to ensure that what is in the QR code matches the Identity Key as retrieved from the server. The 60-digit number is computed by concatenating the two 30-digit numeric fingerprints for each user's Identity Key. To calculate a 30-digit numeric fingerprint: Iteratively SHA-512 hash the public Identity Key and user identifier 5200 times. Take the first 30 bytes of the final hash output. Split the 30-byte result into six 5-byte chunks. Convert each 5-byte chunk into 5 digits by interpreting each 5-byte chunk as a big-endian unsigned integer and reducing it modulo 100,000. Concatenate the 6 groups of 5 digits into 30 digits.

\subsection{Privacy Protection}

For AME, protecting the privacy of users is always at the top of the list. We will protect the privacy of our users and prevent them from being traced back to the network from both business logic and network aspects.

\subsubsection{Privacy Protection Technology at The Business Logic Level}

Bitcoin is designed as an anonymous currency. However, Bitcoin actually cannot achieve true anonymity (pseudonymity). Because the Blockchain is open and accessible to everyone. User identity information can be associated with certain addresses through techniques such as block browsers and data mining. If a merchant that supports Bitcoin transactions publishes its own Bitcoin address, by using the block browser, one can easily find the merchant's revenue, capital flow, and transaction details with the customer, etc. There are also many people who post their Bitcoin addresses in forums and blogs to accept donations, etc. This also directly links the address with personal identity, and further related transaction information about the identity is also leaked. Some countries, such as Australia, also require organizations that engage in Bitcoin-related businesses (such as exchanges) to provide KYC (Know Your Customer, understand user's real information) and AML (Anti- Money Laundering) for regulation. AME adopts Zero-knowledge proof at the business level to protect user privacy and prevent network traceability.

\paragraph*{Zero-Knowledge Proof}\mbox{}
\vspace{\itemsep}

\noindent Zero-knowledge proof is to prove something to others without reveal the information of the specific thing. For example, if Alice wants to prove to Bob that she has the key to the room, but does not want to give the key to Bob, then she can bring something in the room (such as Bob knows that there is a silver iPad in the room) to Bob. And this can indirectly prove that she has the key to the room.

Of course, this example has certain requirements for business scenarios, as well as excessive interaction. The zero knowledge proof we need to achieve is bound to be universal. Thus, we use a non-interactive zero-knowledge proof -- zkSNARKs which can hide the input and output address and transaction amount in the transaction details.

In fact, for miners, they don't care how much money a transaction spends, as well as the sender and the recipient. Miners only care about whether the system's money is conserved. Then they just need to prove the following three questions:

\begin{enumerate}
\item  The sender's money belongs to the person who initiated the transaction.

\item  The money sent by the sender is equal to the money received by the receiver.

\item  The sender's money is indeed destroyed after the transaction is over.
\end{enumerate}

zkSNARKs technology can mathematically transform each problem that needs to be proved into a polynomial. In this case, the proof of a problem can be converted to: If you know this polynomial, you can prove the problem with this polynomial. Then, how can we prove we know this polynomial? We need to use blind evaluation of polynomials:

\begin{enumerate}
\item  The system generates a parameter with a random number and publicize it.

\item  A uses polynomial parameters to compute polynomial results and gives to B.

\item  B verifies the correctness of polynomial results.
\end{enumerate}

If the result is proved to be correct by B, then A can prove that he knows a polynomial, and at the same time prove that the problem corresponding to this polynomial is proved.

The concrete implementation process of zero knowledge proof has the following steps:

\begin{enumerate}
\item  Homomorphic Hiding

\item  Blind Evaluation of Polynomials

\item  From Computations to Polynomials

\item  The Pinocchio Protocol
\end{enumerate}

The specific steps are more complicated and involve a lot of mathematical knowledge. We will not give a detailed description here. However, zero-knowledge proof technology is a very important privacy protection technology in our products, and it is also a privacy protection technology that is highly valued in the Blockchain technology.

\subsubsection{Network Level Privacy Protection Technology}

It is still not enough to protect privacy at the business logic level. Users' IP will often reveal such kind of important information, for example, who the user is, where they live, which will still lead to the user being traced back to the network. AME uses Shadowsocks to prevent network traceability to protect user privacy.

\paragraph*{Shadowsocks}\mbox{}
\vspace{\itemsep}

\noindent Shadowsocks is a light weight and high performance socks5 proxy. It is very easy to set up a Shadowsocks. Typically, it just needs to install a client and configure a server. Shadowsocks splits the Socks5 protocol created by the original ssh into server and client. The request from the client communicates with the ss-local terminal based on the Socks5 protocol. Since this ss-local is generally a local machine or a router or other machine on the local area network which the firewall does not review, the problem of block by the firewall through feature analysis will be solved. Meanwhile, both ss-local and ss-server communicate through a variety of optional encryption methods. When the firewall is reviewed, it is a regular TCP packet. There is no obvious feature and the firewall cannot decrypt the data as well. Then, ss-server decrypts the received encrypted data, restores the original request, and then sends it to the service that the user needs to access. Finally, it obtains the response and sends back to the client. For the commonly keyword filtering system, a Shadowsocks plugin should be enough to bypass the censorship and access the free Internet.

\subsection{Quantum-Resistant Encryption Algorithm}

\subsubsection{The Threat of Quantum Computers on Blockchain}

Cryptographic algorithms, especially asymmetric cryptography, are the cornerstone of blockchain security. However, with the rapid development of quantum computer technology, this cornerstone is not strong enough anymore.

Currently, popular public key cryptography algorithms are based on three types of problems:(1)large integer decomposition problem (RSA);(2)discrete logarithm problem (El-Gamal, DSA, DH key exchange);(3)discrete logarithm problem based on elliptic curves (ECDSA). The algorithms based on these three types of problems are safe because their computational complexity is very high. Under the current computer power level, there is no way to brute force in a short time (100 years). However, with the introduction and development of the concept of quantum computers, this situation will change.

In CES2018, Intel demonstrated a 49 qubit processor, which is considered a milestone. On March 5, 2018, Google Quantum A.I. Lab Team announced their latest achievement which is ``Bristlecone'', a 72 qubit processor.

In the near future, there may be quantum computers that can crack existing encryption algorithms. Therefore, we must also respond to this to protect the security of our Blockchain technology.

\subsubsection{Quantum-Resistant Encryption Algorithm Technology -- Falcon}

Falcon means Fast-Fourier Lattice-based Compact Signature over NTRU, which is a lattice-based digital signature scheme with the properties such as compactness, efficiency and provable security, etc. It has high practical value.

Falcon's construction idea is mainly to instantiate the GPV framework (Gentry, Peikert, Vaikuntanathan proposed in 2008)\cite{5_6,5_7,5_8,5_9} to construct a lattice-based Hash-and-sign digital signature.

Instantiating this framework requires two parameters:(1)A class of cryptographic lattices;(2)Select a trapdoor sampler.

For these two parameters, Falcon selects the NTRU lattices and Fast Fourier Sampling. So the design of the Falcon scheme can also be simply described by the following formula:
\begin{equation}
Falcon = GPV framework + NTRU lattices + Fast Fourier sampling
\label{Eq:equation1}
\end{equation}
 
\paragraph*{GPV Framework}\mbox{}
\vspace{\itemsep}

\noindent The GPV framework is used to construct a lattice-based digital signature, which can be described as below:

\begin{enumerate}
\item  The public key contains a full-rank matrix $A\in Z^{n\times m}_q$ (with $m > n$) generating a q-ary lattice $\Lambda$.The private key contains a matrix $B\in Z^{n\times m}_q$ generating ${\Lambda}^T_q$ , where ${\Lambda}^T_q$ denotes the lattice orthogonal to $\Lambda$ modulo q: for any $x \in \Lambda$ and $y \in {\Lambda}^T_q$ , we have $\langle x, y \rangle = 0\, \text{mod}\, q$. Equivalently, the rows of $A$ and $B$ are pairwise orthogonal: $B \times A^t = 0$.

\item  Given a message m, a signature of m is a short value $s \in Z^m_q$ such that $sA^t = H(m)$, where $H:\{0, 1\} \ast \to Z^n_q$ is a hash function. Given A, verifying that s is a valid signature is straightforward: it only requires to check that s is indeed short and verifies $sA^t = H(m)$.

\item  Computing a valid signature is more delicate. First, an arbitrary preimage $c_0 \in Z^m_q$ is computed, which verifies $c_0A^t = c$. As $c_0$ is not required to be short and $m \ge n$, this can simply be done through standard linear algebra. B is then used in order to compute a vector $v \in {\Lambda}^T_q$ close to $c_0$. The difference $s = c_0 - v$ is a valid signature: indeed, $sA^t = c_0A^t - vA^t = c$ $\mathrm{-}$ 0 = c, and if $c_0$ and v are close enough, then s is short.
\end{enumerate}

The above description of the framework was not originally proposed by GPV, but was first proposed in GGH and NTRUsign. However, GGH and NTRUSign suffer of total break attacks, whereas the GPV framework is proven to be secure in the classical and quantum random oracle models assuming the hardness of SIS for some parameters. The difference between these frameworks is the way of computing v in the signing procedure. GGH and NTRUsign use the Round-off algorithm. The problem with this algorithm is that each signature will reveal some information about the matrix B. On the other hand, the calculation of v in GPV relies on a randomized variant of the nearest plane algorithm. This algorithm proves that it will not reveal information about the B matrix, so it is provably secure.

In addition, in order to prevent the same message from generating different signatures, GPV also introduces a fixed length of random salt in the signed message.

\paragraph*{NTRU Lattices}\mbox{}
\vspace{\itemsep}

\noindent After determining the framework, the next step is to instantiate the lattices. The main consideration for Falcon is the compactness. Because the NTRU lattices are chosen, and the performance is excellent in public key size and efficiency.

Let $\phi \in Z[x]$ be a monic polynomial, and $q\in N$. A set of NTRU secrets consists of four polynomials $f,g,F,G\in {Z[x]}/{(\phi )}$ which verify the NTRU equation:
$fG-gF = q \bmod \phi$
     Provided that $f$ is invertible modulo $q$, we can define the polynomial $h\leftarrow g\cdot f^{-1} \bmod q$.

Typically, $h$ will be a public key, whereas $f,g,F,G$ will be secret keys. Indeed, one can check that the matrices $\begin{bmatrix}1&h&0&q\end{bmatrix}$ and $\begin{bmatrix}f&g&F&G\end{bmatrix}$ generate the same lattice, but the first matrix contains two large polynomials ($h$ and $q$), whereas the second matrix contains only small polynomials, which allows to solve problems mentioned before. The security of NTRU relies on solving two small polynomials $f',g'$ such that $h=g'\cdot{(f')}^{-1}$ is a hard problem.

Instantiate the GPV framework over NTRU lattices:

\begin{enumerate}
\item  The public basis is $A=[1h]$

\item  The secret basis is $B=[g-fG-F]$, the matrices $A$ and $B$ are orthogonal: $B\times A=0 \bmod q$.

\item  The signature of a message m consists of a salt r plus a pair of polynomials ($s_1,s_2$), such that $s_1+s_2h=H(r\vee m)$. Since $s_1$ is completely determined by $m$, $r$ and $s_2$, the signature can simply be ($r,s_2$).
\end{enumerate}

\paragraph*{Fast Fourier Sampling}\mbox{}
\vspace{\itemsep}

\noindent When instantiating a GPV framework, we also need to select the Trapdoor Sampler to solve for the vector v. When selecting trapdoor the main measurement is efficiency, and how short the final signature s is, that is, how close is v to $c_0$. There are four main choices:

\begin{enumerate}
\item  Klein algorithm which is nearest plane algorithm with randomization. It is superior in security but relatively low in time and space efficiency $O(m^2)$.

\item  Peikert's round-off algorithm with randomization optimizes the space and time efficiency, but the security is worse than the Klein algorithm.

\item  A highly efficient and simple trapdoor sampling proposed by Micciancio and Peikert, but the compatibility with NTRU is unclear.

\item  The nearest planar algorithm similar to Fast Fourier Transform proposed by Ducas and Prest can be added with a randomization method, which is both safe and efficient, and can be used on the NTRU lattices.
\end{enumerate}

After comprehensively comparing the above four schemes, Falcon chose the fourth trapdoor sampler with randomized Fast Fourier nearest plane.

In addition, it is necessary to pay attention to the setting of the parameter standard deviation of the trapdoor sampling. If it is too small, the key information (basis of the matrix B) will be leaked. But if it is too large, the generated signature will not be ``short'' enough. Both of these situations will make the constructed signature scheme unsafe.

\paragraph*{Parameters Selection of Falcon}\mbox{}
\vspace{\itemsep}

\noindent According to the security level requirements, the selected parameters of Falcon submitted to NIST are:
\begin{center}

\begin{tabular}{| m{0.8in}<{\centering}| m{0.94in}<{\centering}|m{1.1in}<{\centering}| m{0.85in}<{\centering}|m{1.6in}<{\centering}|}
\hline 
 Level &   Dimension $n$ &  Polynomial $\Phi$ &  Modulus $q$ &  Acceptance bound $\beta^2$\\
\hline 
1-AES128 & 512 & $x^n+1$ & 12289 & 43533782\\
\hline 
2-SHA256 3-AES192 & 768 & $x^n-x^{n/2}+1$& 18433 & 100464491\\
\hline
4-SHA384 5-AES256 & 1024 &$x^n+1$ &12289 & 87067565 \\
\hline
\end{tabular}

\end{center}

In the AME account system, the parameters of the second security level (SHA256, AES192) will be selected. In addition, the hash algorithm that generates the signature message digest uses SHAKE-256.

\subsubsection{Cryptographic Algorithm Scheme Considering Both Pre-Quantum and Post-Quantum}

The AME account system will support both ECDSA and Falcon's public key system. Before the quantum computer matures, normal users can use ECDSA as the public and private key of their wallet account. After the quantum computers become mature, we can use the public and private keys generated by the new post-quantum signature scheme to protect the asset security of the wallet account.

\paragraph*{Wallet Address Generation}\mbox{}
\vspace{\itemsep}

\noindent Both ECDSA and Falcon will use Bitcoin-like wallet address generation process.

\begin{enumerate}
\item  First is to generate a public key using a private key.

\item  The public key generates a public key hash of 20 bytes in length through two Hash operations (SHA256 and RIPEMD160).

\item  Performing the SHA256 operation twice on the public key hash, and takes the first 4 bytes of the operation result as the check code of the wallet address.

\item  Select a prefix of the wallet address, and append the public key hash and the check code. Encoding it in Base58 to generate the final wallet address.
\end{enumerate}

The specific process is as follows:

\begin{figure}[htbp]
\centering
\includegraphics*[width=5.5in, height=3.3in, keepaspectratio=true]{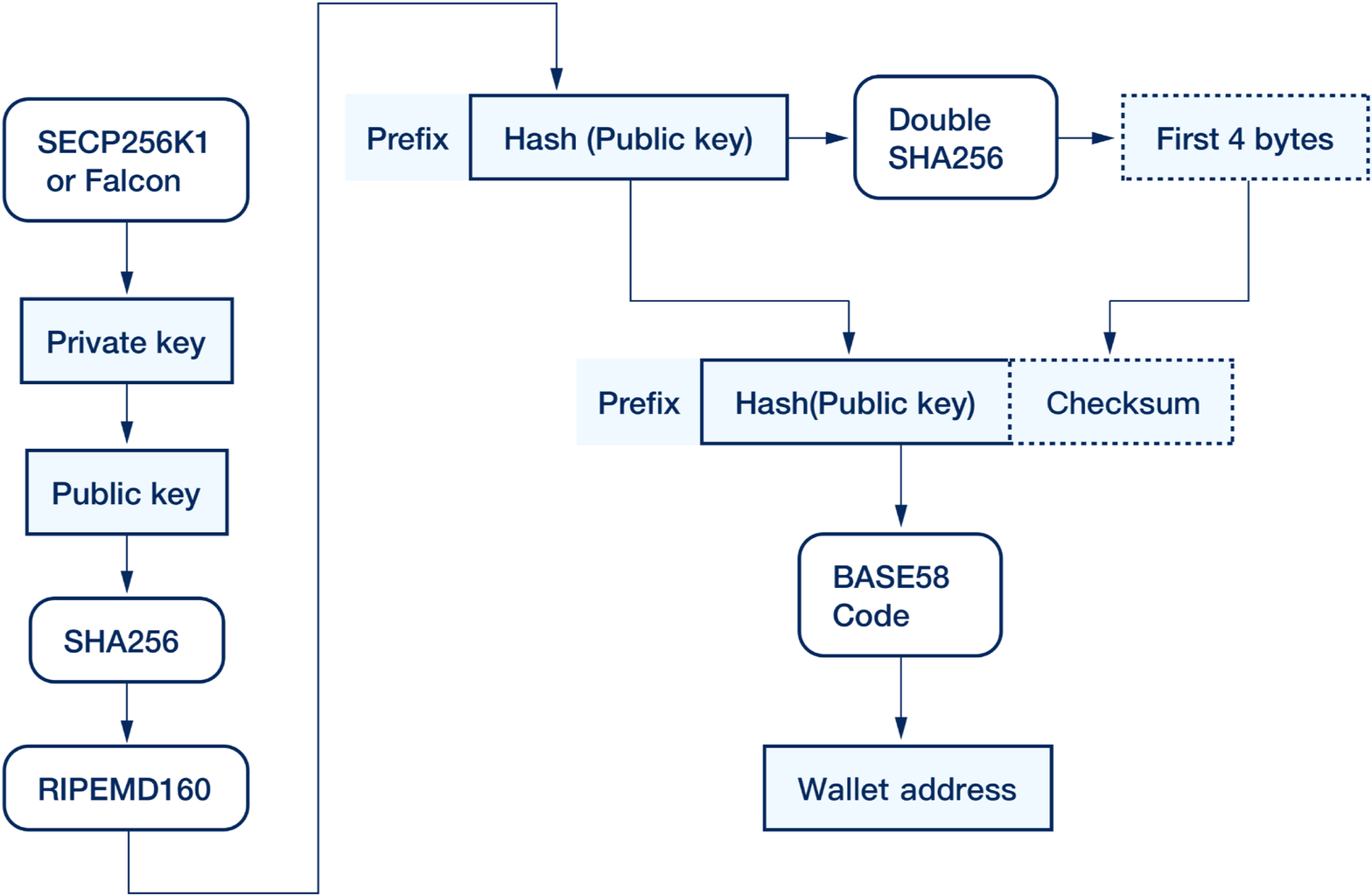}
\caption{Wallet Address Generation Process}
\end{figure}
In order to distinguish between ECDSA and Falcon's wallet addresses, these two addresses will use different prefixes.

\begin{enumerate}
\item  The wallet address generated using the ECDSA public key, will use prefix `A', and the wallet address example is as follows:
\begin{center}
AJA6FuwhMzkriA8mk2zkuKFFb1MvvoCifX
\end{center}
\item  The wallet address generated using the Falcon public key, will use prefix `F', and the wallet address example is as follows:
\end{enumerate}
\begin{center}
FJA6FuwhMzkriA8mk2zkuKFFb1MvvoCifX
\end{center}
Or we can use Bitcoin Cash's wallet format to distinguish two different addresses more clearly. We can use the word ``falcon:'' as a prefix. An example of a wallet address is:
\begin{center}
falcon:JA6FuwhMzkriA8mk2zkuKFFb1MvvoCifX
\end{center}
\paragraph*{Transfer Account Assets to Falcon Address}\mbox{}
\vspace{\itemsep}

\noindent After the emergence of a mature quantum computer, the assets of the user on ECDSA wallet address become unsafe. Hackers can forge legal signatures to spend the balance on an account. In order to cope with this situation, users need to transfer their assets to an anti-quantum account. There are two ways:

\begin{enumerate}
\item{\textbf{Use the RPC command (or curl, etc.)}}

Design an RPC command:
\begin{center}
sendalltofalconaddr [falcon wallet address],
\end{center}
     the AME client uses this command to iterate over all ECDSA accounts in the local wallet file. And it uses the private key to sign and transfer all account balances which are greater than zero to the falcon address in the command by one transaction. Example is shown below:
\begin{center}
./ame-cli sendalltofalconaddr falcon: JA6FuwhMzkriA8mk2zkuKFFb1MvvoCifX
\end{center}
You can also use a fine-grained command to make a split transfer:
\begin{center}
sendtoaddress [falcon wallet address] [amount]
\end{center}
Example is shown below:
\begin{center}
./ame-cli sendtoaddress falcon:JA6FuwhMzkriA8mk2zkuKFFb1MvvoCifX 1000
\end{center}

\item{\textbf{Use the wallet client}}

Users can use the ``Transfer to Falcon Wallet in One-click'' feature button on the full-node wallet or other third-party light-node wallet. Users enter the Falcon Wallet address and transfer all of the user's assets to a secure account. Similarly, Users can also use split transfers.
\end{enumerate}

\clearpage 
\section{AME AI Governance System}
\setcounter{figure}{0}
In this chapter we will introduce the novel artificial intelligent based built-in governance mechanism in AME, AME AI Governance system, that we design to mitigate the security issues, maintain the system responsive and resilient to the continuously changing state. We will present the objectives of AME AI Governance system in \textbf{Section 6.1} and its architecture in \textbf{Section 6.2}. Then we are going to introduce of modules contained in AME AI Governance system.

\subsection{Objectives}

The nature of openness of decentralized blockchains and the dynamical compositions of computing machinery resulting in design inherent weaknesses. Every user can submit transaction request to the network and every computer can join and leave the cluster. This results in inconsistency of the state of blockchain, including users dynamic online activities, the status of each node and the changing network composition, the potentially hostile machines residing within the corporate network. The objectives of AME AI Governance system is to adapt to the above characteristics of blockchain. We list the main objectives of AME AI Governance system below.

\begin{itemize}
\item  Establish a monitoring infrastructure to constantly monitor the behavior of the network and automatically detect anomalies or failures affecting the network to avoid wide spread damage;

\item  Assign users jobs to nodes according to their historical performance to maintain AME robust and resilient to changing workload;

\item  Control the admission, elimination and performance evaluation of nodes in AME network to guard system against adversary participants;
\end{itemize}

To realize all of the objectives is hard due to multiple challenges stemming from varying state of the composition of blockchain network. In particular, a key challenge is to design networks and distributed algorithms that guarantee reliable operation of the system in the face of faults or sophisticated adversarial attacks on certain participating node.

\subsubsection{Threats}

The participants in permissionless blockchain may be malicious (or compromised) and therefore should not be trusted with the confidentiality and integrity of data and relationship information. There are various attacking threats can happen in blockchain system, not excepting the AME blockchain. It includes the blockchain network security threats targeting the weakness of underlying peer-to-peer network, the malicious online activity and attacks against machine learning leveraged for intelligent governance in AME.

We list the threats that AME AI Governance aims to mitigate below\cite{6_1}.

\begin{enumerate}
\item{\textbf{Network Threats}}\mbox{}

The blockchain peer-to-peer nature of the network, which includes all the nodes who maintain and run the blockchain protocols and provides services come under the blockchain network. In case of the AME there are two types of nodes included in its p2p network ABNN: worker nodes in ring1, which are application servers offering instant messaging services, storage services and game services provided in AME light app platform CUBE ; and the management nodes in ring0, which run the blockchain protocol and manage the worker nodes including job allocation, result verification and reputation shifting. The adversary in AME may controls a number of nodes or peers in the network and can block or degrade the network itself and can feed malicious information into the network. The following are some attacks may happen at the AME network layer.

\textbf{Sybil Attack}, the attacker subverts the reputation system of P2P network by creating a large number of pseudonymous identities and then use them to gain a suspiciously large influence. A Sybil attack in blockchain network is an attack where a single adversary is controlling multiple nodes in the network. While economic incentives (rewards and punishments) can mitigate Sybil attacks against a blockchain system, it is hard to prevent a Sybil attack from occurring.

\textbf{Eclipse Attack}, which targets a specific node and sends them blocks of a private fork, while attempting to eclipse them from the rest of the network so that they don't see the main blockchain.

\textbf{Probing Attack}, scanning and probing behavior, as well as the attempted exploitation of high-profile vulnerabilities.

\item{\textbf{Attacks against machine learning}}\mbox{}

Several studies show that machine learning models may be vulnerable to well crafted malicious input in an adversarial environment\cite{6_4}. Researchers have investigated the vulnerabilities exposed by various types of attacks, e.g., adversarial attack, poisoning attack and membership against machine learning models. These types of attacks usually start with manipulating the input samples by adding certain noises or obfuscating features to baffle the model into misclassifying the malicious samples or misleading actions in case of reinforcement learning. Additionally, machine learning models can also leak various types of sensitive information contained in the training data.

\textbf{Adversarial Attack}, which is specially crafted inputs that have been developed with the aim of being reliably misclassified in order to evade detection. Adversarial inputs include malicious documents designed to evade antivirus, and emails attempting to evade spam filters.

\textbf{Data Poisoning Attack}, which involves feeding training adversarial data to the classifier. The most common attack type we observe is model skewing, where the attacker attempts to pollute training data in such a way that the boundary between what the classifier categorizes as good data, and what the classifier categorizes as bad, shifts in his favor. The second type of attack we observe in the wild is feedback weaponization, which attempts to abuse feedback mechanisms in an effort to manipulate the system toward misclassifying good content as abusive.

\textbf{Membership Inference Attack}, which is to determine the membership of a data record in the training data of the machine learning model, given just the data record and black-box access to the model. In some cases, it can directly lead to a privacy breach. More commonly, it is used to optimize opponent model against guardian models.

\end{enumerate}

Moreover, the malicious entities are considered to be Byzantine, and can launch both active and passive attacks. We assume that the adversary is unable to decrypt the content and addressing information of the packets from the user's traffic. We also assume that the hardware capabilities of the adversary limit him from controlling a majority of peers in the network.

\subsubsection{Solution Sketch}

To achieve the objectives of AME AI Governance system against these threats, multiple techniques for automatically monitoring the threat and maintaining the system robust and resilient with the presence of adversary are proposed. All of these techniques adapt to core AME blockchain function and follows the design philosophy:
\begin{itemize}
\item
Decentralized: does not rely on a functionality provided by a central server(s) to perform its tasks.
\item
Autonomous: can operate without user intervention or expert feedback.
\item
Privacy-Preserving: multi agent coordinate without reveal sensitive individual information.
\item
Adversary Resistant: can withstand dynamic adversarial attack.
\end{itemize}
We sketch our solutions as the following.
\paragraph*{Management Representatives Sampling}\mbox{}
\vspace{\itemsep}

\noindent Management representatives sampling are introduced by choosing a committee---a small set of management representatives randomly selected from the total set of management nodes in ring 0---to run each round of node status monitoring, result verification and management protocol. It allows nodes which are monitored to report their status only to their random chosen management nodes. Particularly, the management nodes can also gossip their received messages to others in the management committee. By repeating this process periodically, each management node is going to maintain a partial, yet continuously updating set of received other nodes profiles (i.e., a partial view of the network), that has participated in the system. After every task, the management nodes cast a vote on the proposed action according to the performance of the node and the state of network based on AI mechanism.

\paragraph*{Reputation}\mbox{}
\vspace{\itemsep}

\noindent Reputation system are proposed to guide the nodes to perform well in network. It can combine with the economic incentives to achieve the goal of encouraging participating node in AME to provide expected services and not to act maliciously. Nodes can gain reputation by working on the job owners' tasks. When the service is delivered, the workers that follow service level agreements, measured by quality, are rewarded with increased reputation. The others that violate the service level agreement are considered deceitful and thus lose reputation. This measurement of quality of service is done at the time of validation. Before validation is allowed to occur, enough replicas must have been returned and the sum of the respective workers' reputation must be high enough. The shifting of reputation and node role reduces the attack surface in AME blockchain system.

\paragraph*{Decentralized Machine Learning}\mbox{}
\vspace{\itemsep}

\noindent To achieve the goal of monitoring, diagnostic and taking response action to the state of AME system automatically without the intervene of experts, two machine learning algorithms are used:(1) Graph based anomaly detection are used to spot potential attacks early without knowing attack patterns and collecting labeled data; Specifically, node centric graph partition algorithm are leveraged to find the anomaly community in the hypergraph in the context of decentralized network.(2) Given the context of adaptive policy-driven scheduling, we opt for reinforcement learning algorithms to implement the proposed strategy learning model for AME governance. Reinforcement learning based strategy learning that can dynamically adapt to traffic variation, and to various task specific reward functions set by network operators, to optimally.

\subsection{Architecture of AME AI Governance Module}

The architecture of AME AI governance system is displayed on \textbf{Figure 6.1}. The system is designed to be entirely decentralized. It is composed of four main components: data extraction layer, graph-based feature extraction layer, task specific feature representation layer and reinforcement learning based strategy learning layer. As a result, it does not burden any single machine with excessive workload and at the same time does not require all the data to be centralized for execution.

\begin{figure}[htbp]
\centering
\includegraphics*[width=5.79in, height=3.14in, keepaspectratio=true]{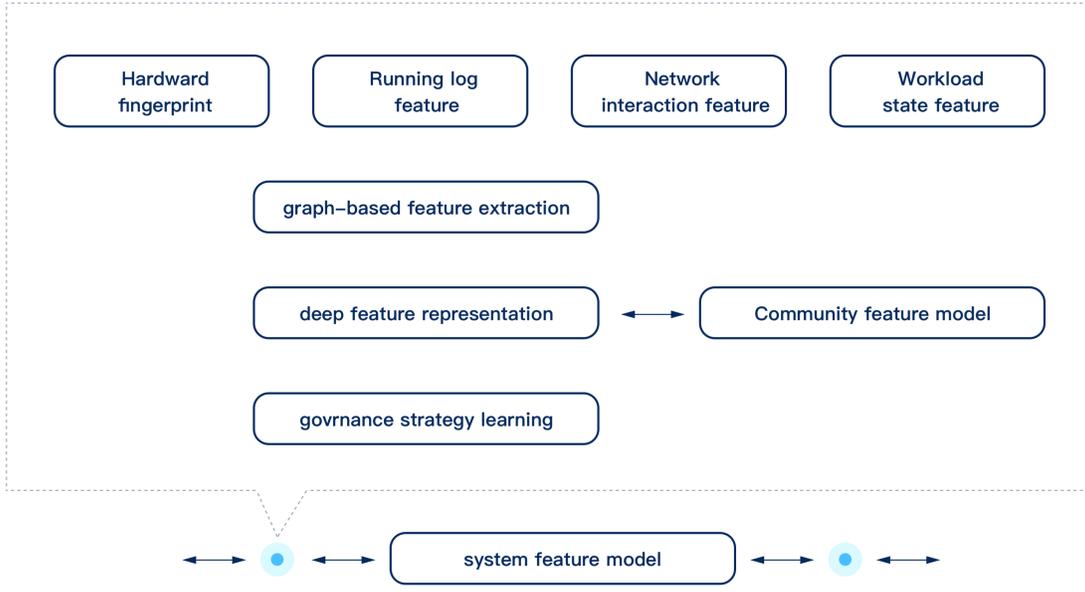}

\caption{AME AI Governance System Architecture}

\end{figure}
\subsubsection{Data Collection Module}

This module is responsible to collect information needed for monitoring, diagnostics and strategy learning. In the context of decentralized data collection, each node receives partial measurements of the state of nodes or users in AME blockchain, and seeks to asymptotically estimate the entire state by exchanging information with its neighbors in the network. Due to the possibility of having malicious peers disrupting the monitoring infrastructure by sending its neighboring peers with fake data, the data collection mechanism should make this malicious behavior harder for the adversary as a complementary method to economic punishment. We deal with collaboratively collecting and estimating the state of the system under the assumption that certain nodes are compromised by adversarial attacks. Specifically, we consider a Byzantine adversary model, where a compromised node possesses complete knowledge of the system dynamics and the network, and can deviate arbitrarily from the rules of any prescribed algorithm. We utilize a distributed filtering algorithm that enables each uncompromised node to asymptotically recover the entire state dynamics without explicitly detecting which nodes are under attack. We will discuss the algorithm and various type of data we collect in \textbf{Section 6.3}.
 
\subsubsection{Graph Based Anomaly Detection}

This module provides a big-data analysis framework to detect malicious and compromised nodes early without the need of relying on historical or labeled training data. The framework is based on large graph analysis and machine learning techniques. Graph Analysis and anomaly detection is performed locally on the peer side using the information collected by the nodes. Due to the dynamic nature of the observed network, it is difficult to determine a priori the expected values or behavior of the input data. Therefore, unsupervised machine learning techniques are required.

It first constructs a set of hyper-graphs to represent the activities of nodes and users. Each hypergraph node may correspond to a set of events or a set of users or server nodes in AME network, with edge attributes specifying their connectivity relationship. On top of these constructed graphs, the system applies community detection algorithms and performs large-scale graph analysis to determine a subset of anomaly users or server nodes and their activities with high confidence. Particularly, in the context of AME blockchain system, decentralized and scalable community detection mechanism are required. Random walks and diffusion-based techniques are adopted to extract disjoint communities\cite{6_15}, which can be implemented using node-centric programming model without requiring any global knowledge. The result set of detected high-confidence anomaly users or server nodes and activities are then used as self-generated training data to feed into the following machine learning components to derive a set of risk detection models or strategy learning models. Finally, these newly generated risk detection models can be used to detect the remaining set of undetected user accounts or account activities. In this framework, the graph analysis bootstraps the system to automatically generate training data on demand, without relying on historical training data obtained from manual labels or external detection components. As such, early detection of malicious users or server nodes and their activities in an unsupervised manner can be achieved. We will elaborates our graph based algorithms in \textbf{Section 6.4}.

\subsubsection{Reinforcement Learning Based Strategy Learning}

The AME Governance system requires adaptive governance strategies to handle the unpredictable events occurring in nodes where various tasks are executing. To this aim, we adopt deep reinforcement learning to deal with the complicated scheduling and reputation shifting problems of the nodes in AME network with large state space.

We can consider the management decision process as a Markov decision process (MDP) model\cite{6_23}. MDP model can help selecting the possible actions from the current state and observing the derived reward/cost from each transition in order to find a better scheduling/reputation shifting decision in AME network. More concretely, we consider the scheduling and reputation shifting decision as a life cycle in which the node progresses through this life cycle and goes from one state to another. For instance, a node can go through the following states in the MDP model: idle, scheduled, finished, and failed with its reputation score. We consider the mapping between these states and the entire network states over the possible actions to find an appropriate scheduling and reputation shifting policy as the process of the scheduling and reputation shifting decision selection in AME governance module. Following this approach, our proposed solution can estimate and compare all possible rewards that are earned when applying the actions from a given environment state. Furthermore, deep reinforcement learning based strategy learning framework allows to consider the dynamic events occurring in a system's environment and to adjust the decisions making procedures under uncertainty. We will discussed the reinforcement learning algorithm in \textbf{Section 6.5}.

\subsubsection{Privacy Preserving}

Given the decentralized nature of AME system, sharing and working on sensitive data in distributed settings is a challenge due to security and privacy concerns.\textbf{ }There are a number of different approaches, from homomorphic encryption to differential privacy. We choose to follow differential privacy scheme to preserve privacy in the process of distributed computation without losing our effectiveness and efficiency.

Differential privacy\cite{6_6} is one of the most popular definitions of privacy today. Intuitively, it requires that the mechanism outputting information about an underlying dataset is robust to any change of one sample, thus protecting privacy and assuring resistant to membership inference attack. In the AME Governance machine learning training implementation, we will follow a variation of the Laplacian mechanism to preserve privacy. Before the discussion of the mechanism we followed, we quote some definition of differential privacy.

A mechanism $f$ satisfies $(\epsilon ,\delta )$-differential privacy for two non-negative numbers $\epsilon $ and $\delta $ iff for all neighbors $d(D,D')$, and all subset $S$ of $f$'s range, as long as the following probabilities are well-defined, there holds

\begin{equation}
P(f(D)\in S)\le \delta +e^{\varepsilon }P(f(D')\in S)
\label{Eq:equation1}
\end{equation}

where $d(D,D')$ denotes the minimum number of sample changes that are required to change $D$ into $D'$. Intuitively speaking, the number $\delta $ represents the probability that a mechanism's output varies by more than a factor of $e^{\epsilon }$ when applied to a dataset and any one of its neighbors. A lower value of $\delta $ signifies greater confidence and a smaller value of $\epsilon $ tightens the standard for privacy protection. The smaller $\epsilon$ and $\deltaup$ are, the closer $P(f(D)\in S)$ and $P(f(D')\in S)$ are, and the stronger protection is.

Laplacian mechanism is a popular $\epsilon $-differentially private mechanism for queries $f$ with answers $f(D)\in \mathbb{R}$ , in which sensitivity plays an important role. The sensitivity is defined below:

Given a query $f$ and a norm function $\parallel\cdot\parallel$ over the range of $f$, the sensitivity $s(f,\parallel\cdot\parallel)$ is defined as
\begin{equation}
s(f,\parallel\cdot\parallel)=\operatornamewithlimits{max}_{d(D,D')=1}\parallel f(D)-f(D')\parallel
\label{Eq:equation1}
\end{equation}
Usually, the norm function $\parallel\cdot\parallel$ is either L1 or L2 norm. The Laplacian mechanism: given a query $f$ and a norm function over the range of $f$, the random function $\overline{f(D)}=f(D)+\eta$ satisfies $\epsilon$-differential privacy. Here $\eta $ is a random variable whose probability density function is $p(\eta )\propto e^{-\epsilon \parallel \eta \parallel/s(f,\parallel \cdot \parallel )}$. There is a variation of the Laplacian mechanism, which replaces Laplacian noise with Gaussian noise. On one side, this replacement greatly reduces the probability of very large noise; on the other side, it only preserves $(\epsilon ,\delta )$- differential privacy for some $\delta >0$, which is weaker than $\epsilon $-differential privacy. Variation of the Laplacian mechanism: given a query $f$ and a distance function over the range of $f$, the random function $\overline{f(D)}=f(D)+\eta$ satisfies $(\epsilon ,\delta )$-differential privacy. Here $\eta $ is a random variable from distribution $N$.
 
\subsubsection{Decentralized Optimization}

In our decentralized setting, instead of sending all of their data to a centralized server, there is a set of workers, each of which collects data from different data sources. Therefore, the training of machine learning model used in AME AI Governance system requires a decentralized implementation, where nodes pass updates to every other data shard in the cluster without having a shared parameter server. There are many distributed algorithms such as the dual averaging-based algorithm and the subgradient methods. However, most of existing works are built on the hypothesis that the network is deployed in benign surroundings without any intruder. Under the presence of adversary, the existing distributed optimization algorithms become vulnerable or even invalid, which may lead to system paralysis. We define the decentralized optimization with adversary nodes problem as follows:

Assume that each agent $i$ has its local cost function $f_i(x)\in \mathbb{R}$, where $x\in \mathbb{R}$ is the same variable owned by all agents. The local cost function $f_i(x)$, $\forall i\in V$ is a convex function. At the same time, it is supposed that the optimal point set arg min $f_i(x)$ is nonempty, bounded and closed. The derivative of the function $f_i(x)$, $\forall i\in V$ is represented by $f'_i(x)$. The optimization problem can be written as,
\begin{equation}
\operatornamewithlimits{min} \frac{1}{n}\sum^n_{i=1}{f_i(x)}
\label{Eq:equation1}
\end{equation}
When adversary agents exist, for all normal nodes, the problem becomes,
\begin{equation}
\operatornamewithlimits{min} \frac{1}{n_0}\sum_{i\in V \setminus V_a}{f_i(x)}
\label{Eq:equation1}
\end{equation}

Where $n_0$ is the normal agent. To address this problem of distributed optimization dynamics to failure and adversarial behavior, a resilient distributed filtering algorithm that guarantees that the non-adversarial nodes converge to the convex hull of the minimizers of their local functions are used in the implementation of AME AI Governance machine learning training, which we will discussed in \textbf{Section 6.3.1}.

\subsection{Data Collection}

In a fully-decentralized and highly dynamic P2P network, the absence of central or supernode like entities makes monitoring the network a challenge since any information required for understanding the behavior of the network is distributed over possibly hundreds or even thousands of participating peers. Gathering information from all peers becomes unscalable as the network grows and privacy concerns limits the nature of information that can be collected from peers. In addition to these issues, we assume that there are adversary participants in the network. Therefore, a node may receive fake data from its peers. In the process of decentralized machine learning training, each agent receives gradient from neighbors to update its local variable value and the arbitrary fake data sent by adversary peers may lead system paralysis. To tackle these challenges, the distributed filtering algorithm are used to filter malicious data.

\subsubsection{Distributed Filtering Algorithm}

Consider the linear dynamical system $x[k+1]=Ax[k]$, where $k\in \mathbb{Z}$ is the discrete-time index, $x[k]\in {\mathbb{R}}^n$ is the state vector and $A\in {\mathbb{R}}^{n\times n}$is the system matrix. The system is monitored by a network $G=(V,E)$ consisting of $N$ nodes. The $i$-th node has partial measurement of the state
\begin{equation}
x[k]:y_i[k]=C_ix[k],
\label{Eq:equation1}
\end{equation}
where $y_i[k]\in {\mathbb{R}}^{r_i}$ and $C_i\in {\mathbb{R}}^{r_i\times n}$. We denote
\begin{equation}
y[k]={[y^T_1[k],...,y^T_N[k]]}^T,
\label{Eq:equation1}
\end{equation}
and

\begin{equation}
C={[C^T_1,...,C^T_N]}^T .
\label{Eq:equation1}
\end{equation}

For any ${\lambda }_j\in sp(A)$, where $sp(A)=\{\lambda \in C\vee det(A-\lambda I)=0\}$denotes the set of all eigenvalues (modes) of a matrix A, let $z^{(jm)}[k]$ denote the $m$-th component of the vector $z^{(j)}[k]$, and let ${\hat{z}}^{(jm)}[k]$denote the estimate of that component maintained by node $i\in V$, where $V$ is the node set of the distributed network. For any node $i$, let the set of eigenvalues it can detect be denoted by $O_i$ , and let $UO_i$ = $sp(A)O_i$. Consider an unstable eigenvalue ${\lambda }_j\in UO_i$. For such an eigenvalue, node $i$ has to rely on the information received from its neighbors, some of whom might be adversarial, in order to estimate $z^{(j)}[k]$. To this end, a resilient consensus algorithm that requires each regular node $i\in V\setminus S_j$ to update its estimate of $z^{(j)}[k]$ using the following two stage filtering strategy:

\begin{enumerate}
\item  At each time-step $k$, each regular node $i$ collects the state estimates of $z^{(j)}[k]$ received from only those neighbors that belong to a certain subset $N^j_i\subseteq N_i$(to be defined later). For every component m of $z^{(j)}[k]$, the estimates of $z^{(jm)}[k]$ received from nodes in $N^j_i$ are sorted from largest to smallest.

\item  For each component m of $z^{(j)}[k]$, node $i$ removes the largest and smallest $f$  estimates (i.e., removes $2f$ estimates in all) of $z^{(jm)}[k]$ received from nodes in $N^j_i$, and computes the quantity:

\begin{equation}
\bar{z}^{(jm)}[k]=\sum_{l\in M^{(jm)}_i[k]}{w^{(jm)}_{il}}[k]{\hat{z}}^{(jm)}_l[k]
\end{equation}

where $M^{(jm)}_i[k]\subset N^j_i\subseteq N_i$ is the set of nodes from which node $i$ chooses to accept estimates of $z^{(jm)}[k]$ at time-step $k$, after removing the $f$ largest and $f$ smallest estimates of $z^{(jm)}[k]$ from $N^j_i$. Node $i$ assigns the weight $w^{(jm)}_{il}[k]$ to the $i$-th node at the $k$-th time-step for estimating the $m$-th component of $z^{(j)}[k]$. The weights are nonnegative and chosen to satisfy $\sum_{l\in M^{(jm)}_i[k]}{w^{(jm)}_{il}[k]=1}$, $\forall {\lambda }_j\in UO_i$ and for each component $m$ of $z^{(j)}[k]$. With the quantities $\bar{z}^{(jm)}[k]$ in hand node $i$ updates ${\hat{z}}^{(j)}_i$ as follows:

\begin{equation}
\hat{z}^{(j)}_i[k+1]=V({\lambda }_j)\bar{z}_i^{(j)}[k]\qquad\qquad\text{if } {\lambda }_j\in {UO}_i \text{ is real}
\end{equation}

\begin{equation}
\hat{z}^{(j)}_i[k+1]=W({\lambda }_j)\bar{z}_i^{(j)}[k]\qquad\qquad\text{if } {\lambda }_j\in {UO}_i \text{ is not real}
\end{equation}

where $\bar{z}^{(j)}_i[k]={\begin{bmatrix} \bar{z}^{(j1)}_i[k]&\ldots&\bar{z}^{(j{\sigma}_j)}_i[k]\end{bmatrix}}^T$, ${\sigma }_j=a_A({\lambda }_j)$  if ${\lambda}_j\in UO_i$ is real and ${\sigma}_j$ = $2a_A({\lambda}_j)$ if ${\lambda }_j\in UO_i$ is not real.
\end{enumerate}

\subsubsection{Types of Data}

With the filtering algorithm in hand, various data can be collected and computed to estimate the state of network and to train AI based model. In order to distinguish the malicious nodes and cooperative nodes and give a evaluation of the quality of the user, we are using multi dimensional data include on-chain data and off-chain data to monitor the state of AME network and to learn the historical malicious modes and cooperative nodes based on the historical logs.

Our goal is to provide detectors with network wide information collected at the same time that flow records are generated in order to minimize detection delay. We also wish to provide detectors with a combination of both volume and distribution information so that detectors can be generalized rather than specialized for specific attack signatures.

Information to be collected can be classified into:
\begin{itemize}
\item On-Chain Data: history transaction related data of nodes and users.
\item Current State of Network: related to the underlying network and the interaction of each node. For example, number of neighboring peers, peer uptime, etc.
\item Profile of Nodes: the reputation score, the workload and history performance of the node and the information related to the host. For example, memory and processor consumption.
\item Running Log: history running log of tasks that nodes received.
\end{itemize}
These data will encode the information such as the individual state of each node and the partial view of the state of entire network. All these related features are fused into the anomaly detection model and various task specific strategy learning model to measure the reputation of a node in a multiple-dimensional way and make decision about task allocation, reputation shifting and privilege management.

\subsection{Graph based Feature Representation}

The governance system is responsible for the monitoring, diagnosis and dynamic scaling the network in order to robust and resilient to changing state. The governance system needs to ensure the user/nodes has not been compromised before assigning task. Hence, monitoring the state of the network is the first step towards secure system governance. In addition to monitoring the state of the network, anomaly detection is needed to spot unusual system behaviors such as failures, different attacks and anomalous communication patterns. To automatically and reliably detect anomalies, it is required to characterize and construct a model of normal network behavior and identify abnormal behavior as it occurs. The normal behavior of a node is expected to be constantly evolving and a present notion of normal behavior might not be valid in the future.

However, investigating individual resource behavior may not be efficient in detecting abnormal behavior in large and complex data centers. By leveraging Graph-Mining techniques\cite{6_19}, unusual behaviors in data centers could be detected not only based on a per-resource behavior, but using a holistic view of inter-dependency and inter-communication pattern between different resources.

The constructs and analyzes several types of activity graphs, referred to as hypergraphs, to detect malicious (or compromised) accounts and malicious events without using training data. A global view of the connectivity structures among users and events allows the system to perform early detection of stealthy attack patterns that are difficult to identify when each user or event is examined in isolation. The hypergraph based detection can identify groups of malicious accounts without requiring labeled data provided by the customers. The labeled data are often hard to obtain, especially with new unseen attacks. With hypergraph analysis, the system can self-bootstrap the system with an initial list of malicious accounts or events. This step also has the ability to capture new attack campaigns automatically.
 
\subsubsection{Graph Construction}

To build the hypergraphs, the system first processes input data and derives a set of features and statistics for each node (or each node event). The combination of all features or statistics is referred to as a profile. For each user or node, the system can compute a corresponding profile. In addition, for each of one or more groups of users or nodes, the system can compute a corresponding group profile. Collectively across an entire user population available to the system, a global profile can be computed.

The set of computed feature profiles will be used to construct hypergraphs for graph analysis. Each node on a graph corresponds to a feature profile. Each feature profile can be constructed from a set of correlated events or a set of correlated users or nodes. The set of correlated events or correlated user or node is identified by taking the set of events with similar behaviors. The edges of the graphs may be computed in multiple ways. The edges can be computed by adding an edge between node $A$ and node $B$, if $A$ and $B$ share a similar feature. To determine if two features are similar, the system can perform the following procedures. If the feature corresponds to a numerical value, then the system can compare their respective values. In some implementations, the system checks whether the difference between two corresponding feature values is smaller than a pre-set threshold. Alternatively, in some other implementations, the system checks whether the ratio of two features value is smaller than a pre-set threshold.

The output graph component information can be combined with individual user or node or event information to generate an initial list of malicious users or nodes with a high confidence, as they have exhibited stronger global correlations in conducting malicious activities.

\subsubsection{Suspicious Graph Node Detection}

Once the system obtains a list of suspicious graph nodes, it proceeds to identify suspicious graph communities. Graph communities can be identified using several different graph algorithms. As mentioned in \textbf{Section} \textbf{6.2.2}, random walk and diffusion-based techniques which can be implemented in node-centric model are adopted to adapt to the decentralized feature of AME blockchain system.

\paragraph*{Graph Diffusion based Community Detection}\mbox{}
\vspace{\itemsep}

\noindent Classical community detection is formulated as a clustering problem. That is, given the full graph $G=(V,E)$, partition the vertex set into $K$ subsets $S_1,\ldots,S_k$, (a partitioning), such that $\bigcap_1^{k}S_i=\emptyset $ and $\bigcup_1^{k}S_i=V$. A quality metric $Q(\{S_1,...,S_k\})$ is defined over the partitions and a community detection algorithm will try to find a partitioning that maximize or minimize $Q$ depending on its nature. This is for non-overlapping community detection and one can simply remove the constraint $\bigcap_1^{k}S_i=\emptyset $ to get the overlapping version. Note that $Q$ is only an artificial surrogate to the axiomatic notion of community. The maximum $Q$ does not necessarily corresponds to the best community. However, the community detection problem becomes tractable via well-studied optimization frameworks by assuming a form of $Q$ e.g. Modularity, Conductance. Now consider the decentralized scenario. One node (observer) is limited to its local view of the whole graph. It is unreasonable to ask for a global partitioning in terms of sets of nodes.

\subsection{Fully Decentralized Reinforcement Learning}

Reinforcement learning is a machine learning method used to tackle sequential decision-making problems through a trial-and-error technique to search for effective actions\cite{6_23}. In the broadest sense, a machine learning system, using the reinforcement learning paradigm, interacts with a single environment ; it observers the state of that environment, selects an action, and receives a scalar reward or feedback for the action. The process is depicted in \textbf{Figure 6.2}.
\begin{figure}[htbp]
\centering
\includegraphics*[width=5.73in, height=2.44in, keepaspectratio=true]{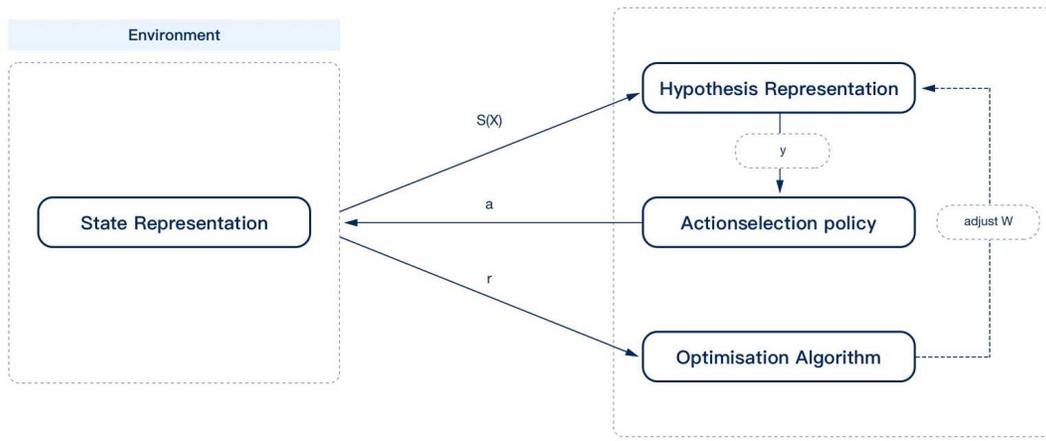}
\caption{Reinforcement Learning Paradigm}
\end{figure}

The environment, in this paradigm, is characterized by a set of states, $S$, in which every state is constructed from a vector of features (called state features). The machine learning system consists of a set of actions, $A$, that are applicable to perform on the environment (\textbf{Figure 6.2}). A machine learning system interacts with its environment at each time of a sequence of discrete or continuous time steps, $t=\mathrm{0,1,2,3,\cdots}$  The interaction takes place through a repeated cycle of three steps:

\begin{enumerate}
\item  sensing the state of the environment at $t$, $s_t\in S$;

\item  performing an action $a_t\in A(s_t)$, where $A(s_t)$ is a set of actions that are admissible for the state $s_t$; that is, $A(s_t)\subset A$ ;

\item  receiving a scalar reward, which in general cases is defined as $R\mathrm{:}S\times A\times S\to \mathbb{R}$, which specifies the reward obtained for a transition from one particular state to another, after performing an action.
\end{enumerate}

At each time step $t$, the machine learning system interacts with its environment with the goal of building an action selection policy (denoted ${\pi }_t$) , which maps the states to the actions $\pi \mathrm{:}S\to A$, where ${\pi }_t(a|s)$ is the probability of $A_t=a$ if $S_t=s$. The goal of this policy is to maximize the reward signal that represents a long-term objective. Thus, the policy is a fundamental step in understanding the characteristics of the reinforcement learning components before building any reinforcement learning application; the components are the environment state space, the machine learning system action space, and the reward space. These components of reinforcement learning can be formalised using a Markov decision process (MDPs) framework, especially if the state and action space are discrete. They are tuples ($S,A,P,R$) where:

\begin{itemize}
\item  $S$ is the set of environment states, which can take a broad range of forms. For instance, state spaces can be defined by continuous variables such as velocity, price, performance etc., called continuous state-spaces ($\lvert S\rvert \in \mathbb{N}$); Alternatively, they can be defined by a discrete state-space if the number of states is discrete.

\item  $A$ is the set of possible actions available to the machine learning system.

\item  $P$ is the state transition function. It is defined as $P(s_t,a_t,s_{t+1})\to [\mathrm{0,1}]$$\mathrm{\to}$ [0, 1], where $P$ represents the probability of reaching state $s_{t+1}\in S$ by applying action $a_t\in A(s_t)$ in state $s_t\in S$. A characteristics of this function is that it is deterministic. This refers to the probability of the learning system being in some state $s_{t+1}$ after taking action at from state $s_t$ , or $p^{a_t}_{s_ts_{t+1}}$ . State transition determinism occurs when  $p^{a_t}_{s_ts_{t+1}}$= 1. By contrast, if $p^{a_t}_{s_ts_{t+1}}<1$ the transition is non-deterministic or stochastic.

\item  R is the reward function: $R(s_t,a_t,s_{t+1})\to \mathbb{R}$. It provides an immediate indication when an action $a_t\in A(s_t)$ is taken in state st and moves the machine learning system into a subsequent state $s_{t+1}\in S$.
\end{itemize}

\subsubsection{State Space}

The state space S consisting of all possible state vectors $s^i$. Where each $s^i$ is the concatenation of a pair of sub-states such that $s^i=({c}^i,{o}^i)$. The control sub-state helps inform the controller of which metric or metrics are in most need of correction. The operating sub-state gives the controller information about the current network operating environment. The control sub state is a three element vector representing the three performance metrics: precision, recall, and forwarding. Each element of the vector is set to 1 or -1 indicating necessity for improvement of a particular performance metric where 1 indicates improvement required and -1 indicates none. The calculation of ${c}^i_t$ is accomplished by use of a artificial neural network (ANN) to map system performance and operator priorities to the possible sub-states. The ANN accepts as input, the value at time t of the three performance functions $F_q(t)$ and three operator defined performance goals $G(q)$. A control string indicating the priority order of the performance functions $G(q)$ is used to calculate weights. Where q = (1, 2, 3) = (Precision, Recall, Forwarding).

The operation sub state ${o}^i$ consists of a vector of state variables produced by the Network Preprocessor. This combination of information allows the system to respond to changes in network conditions while also providing the system information both specific to and independent from the underlying algorithm. The resulting state space is potentially high dimensional and continuous. In reinforcement learning continuous state spaces can be managed using function approximation methods.

\subsubsection{Action Space}

We define an action space A over a parameter space $P$ as a set of action vectors where each element of an action vector represents a possible governance action on the available governance parameters in the underlying detection system $(p^1,...,p^n)$. Each governance action consists of a direction and magnitude of change for that governance parameter. We discretize the action space into a set of $n$ dimensional action vectors $a=(a_1,...,a_n)$ where $n$ is the number of tunable parameters in the objective model and each element of $a_i=\pm 1$ where the sign represents a decision to increase or decrease $p^k$ . We also maintain a single $n$ dimensional vector $M$ to track the magnitude of the tuning changes. Each $m^k$ in $M$ is a dynamic range variable that increases or decreases with the series of sign changes in $a^i_k$ . Each $m^k$ is a range limit in the interval $[0,\delta k]$, where ${\delta }^k=\operatornamewithlimits{max}(p^k) - \operatornamewithlimits{min}(p^k)$ between minimum and maximum values according to the underlying algorithm. Each time the sign of $a^i_k$ stays the same, $m^k$ is incremented toward ${\delta}^k$ . Each time the sign of ${\delta}^k$ changes, $m^k$ is reset to 0. Ceiling and floor limits of $\operatornamewithlimits{max}(p^k)$, $\operatornamewithlimits{min}(p^k)$ are applied to ensure parameters stay within algorithmic limitations. The result of this approach is that when parameters are tuned on consecutive intervals in the same direction, the magnitude of the adjustments continues to increase. When parameters are tuned on consecutive intervals in the different directions, magnitude of the change is initially reset to 0 and gradually increased as appropriate. Additionally, while the direction of the adjustment to $p^k$ is determined by the discrete action output by the RL tunner, the magnitude of the change is independent of the other parameters in the action vector. It is determined by the aggregate history of sign changes of $p^k$ only.

\subsubsection{Reward Definition}

One of the primary objectives of this approach is to ensure that high priority metrics remain within established goals while lower priority metrics bear the tradeoff. The reward function is calculated to place emphasis on priority metric functions remaining within established goals. We first determine if the system is within operating parameters. If the weighted factor of H is used when the system is not within parameters to ensure the priority metrics are accounted for first. We calculate a system score at time $t$ :

\begin{equation}
F_q(t)-G(q).
\label{Eq:equation1}
\end{equation}
Where G(q), and $F_q(t)$ are as previously described. If over the interval $t\ldots t+1$ the score has increased, then the reward is 1 and 0 otherwise. If the system has crossed the threshold of operating standards from in standards to not in standards, the reward is 0. If the system has crossed from not in standards to in standards, the reward is 1. Otherwise the weighted r non weighted score is used to determine reward.

Once the training phase is done, various decisions can be made by the algorithm. The decisions include whether the observed state is anomalous and the reputation shifting decision.

\clearpage
\section{Economy System}
\setcounter{figure}{0}
\subsection{Introduction}

This chapter proposes an economic development plan for the AME blockchain ecosphere. The purpose is to enable participants of the entire ecology to perform their duties under reasonable economic incentives and strong fiscal governance to ensure the healthy operation of the AME blockchain's ecosystem. First of all, we will introduce the unique three-layer ecological structure, including the underlying AME blockchain technology platform, the application layer-blockchain messenger, and the top-level BCM-based game platform CUBE. Secondly, in order to link the entire three-layer ecological structure, we have established a set of economic operation systems, including the token issuance mechanism, consensus economics, and the economic cycle mechanism. We also illustrate how the economy within the entire ecology is running through practical examples. Thirdly, in order to ensure the healthy operation of the ecology and stabilize the value of the tokens, we have added several fiscal policies, including inflation, bond mechanism, and foundation system, forming a distinctive economic system with a combination of free markets and macro controls. Next, we will discuss some of the mechanisms used in the economic system, and make comparisons with other blockchain projects to elaborate in detail the trade-offs in the design of these mechanisms. The final section will summarize the ecosystems covered in this chapter and propose our vision for the future.

\begin{figure}[htbp]
\centering
\includegraphics*[width=3.2in, height=2.5in, keepaspectratio=true]{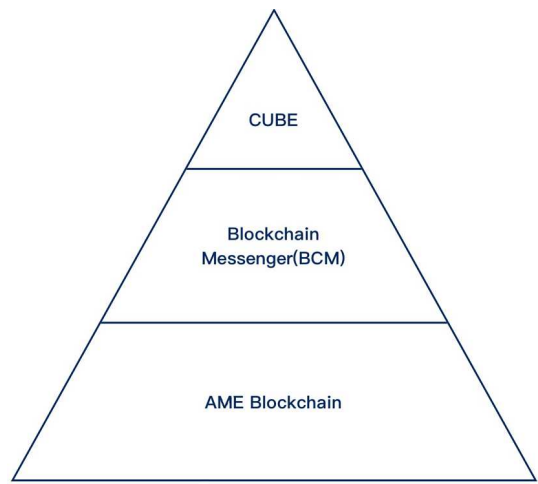}
\caption{Schematic Diagram of The Three-layer Ecological Structure}
\end{figure}
\subsection{Three-Layer Ecological Structure}

The three-layer ecological structure splits the entire ecosystem into three parts, namely the underlying AME technology platform, the middle level blockchain messenger (hereinafter referred to as BCM) and the top-level CUBE. Among them, the AME technology platform is the foundation of the entire ecology, as well as the provider of resources and services. BCM is an instant messenger based on the AME technology platform that provides users with secure and encrypted communication services. CUBE is a blockchain-oriented open application platform built on top of BCM, especially focusing on gambling games. The platform provides easy-to-use development tools and blockchain resources for developers. Developers only need to pay attention to the development of business logic. Meanwhile, the platform also provides comprehensive APIs, and welcomes as well as supports the third part games to access in CUBE. Obviously, the three parts of the ecology progressively form a pyramid structure. We will introduce the three parts of the pyramid in details, focusing on the participants in each part and their code of conducts, rather than specific technical details.

\subsubsection{AME Technology Platform}

As mentioned earlier, the AME technology platform is a secure, self-managed, and decentralized platform based on blockchain technology. Hence, the blockchain technology is the basic component of the AME technology platform. It adopts a double-ring network topology design, manages the node servers participating in the blockchain system, and separates the management logic from the business logic. Thus, a flexible and scalable-decentralized P2P system solution is realized. For different kinds of server nodes, roles and benefits of them are discussed as below.

\paragraph*{Manager Ring Node}\mbox{}

\subparagraph*{\textit{``Mining'' Rewards}}\mbox{}

In the blockchain, the server nodes in the Manager Ring, which assume the role of the manager, and is not responsible for the specific businesses, can respond to the client's transaction request, and assign the corresponding worker server to the client The transaction is collected and stored in the AME blockchain by the Manager Ring nodes in the form of ``mining''.

Generally speaking, in other blockchain projects, the successful ``mining'' node receives the mining rewards which are distributed in the form of tokens. For example, a winner in Ethereum's blockchain ``mining'' wins 5 new Ethereum rewards\cite{7_1}. However, it needs to be clarified that the ``mining'' rewards is not an endogenous rule of the blockchain, but a sociological means of economic incentives. For example, at the time of this white paper successful bitcoin miners grant 12.5 bitcoin rewards, but the total amount of bitcoins will reach 21 million when 2140. After that, miners will not be able to get ``mining'' rewards\cite{1_1}. The source of revenue for miners will only be the transaction fee. In addition, in the Ripple protocol, the server node does not need nor can't ``mine'', and naturally cannot get the ``mining'' rewards\cite{7_3}. In other words, when deciding whether to reward a server node for successful ``mining'', one must find the sociological basis.  

The AME blockchain is intended to issue a certain amount of token-rewards to server nodes that successfully ``mine''. This setting is mainly based on the following considerations. First of all, the Manager Ring nodes competed in ``mining'', packaged the transaction information, paid the computing power, and contributed to the blockchain system. Secondly, the combined effect of ``mining'' cost and ``mining'' rewards will increase the cost of malicious behaviors; using the behavior of nodes pursuing profits to restrict nodes from behaving maliciously. Finally, the ``mining'' behavior of the AME blockchain system exists only in the Manager Ring, and the nodes in the Worker Ring and the temporary Worker Ring are not eligible to participate in ``mining''. The AME blockchain system provides a swap channel for nodes of the Worker Ring and the Manager Ring. If a Worker Ring node provides better service, then he has the opportunity to compete into the Manager Ring. If a Manager Ring node fails in the competition, then he may be downgraded into the Worker Ring. In order to benefit, server nodes tend to provide better resources and services to enter or remain in the Manager Ring and obtain ``mining'' rewards.

\subparagraph*{\textit{Transaction Fee}}\mbox{}

In addition to the ``mining'' rewards, the transaction fee is another source of revenue for the Manager Ring nodes. When the client initiates a transaction, it is required to pay a transaction fee, which is allocated to the service providers- the Manager Ring nodes and the Worker Ring nodes. The purpose of setting the transaction fee is to prevent the client from abusing the resources of the underlying blockchain, and also to quantify the work of the blockchain nodes.

\subparagraph*{\textit{Service Fee}}\mbox{}

In addition to the transaction behavior, the Manager Ring nodes can also provide non-transaction behavior for the client, that is, a service that does not need to be broadcasted, in order to obtain a ``service fee''. This is an affirmation of the labor value of the nodes. Even if the Manager Ring nodes are not qualified to ``mine'' in the current packaging process, they can still obtain certain benefits through the work, ensuring the enthusiasm of all nodes in the Manager Ring, preventing none of the mine-qualified nodes from being passively absent or cheating.

\paragraph*{Transaction Fee Allocation}\mbox{}
\vspace{\itemsep}

\noindent As mentioned earlier, the main sources of revenue for the Manager Ring nodes are ``mining'' revenue and transaction fee. And, the ``mining'' income belongs to the nodes that successfully propose blocks. There are two forms of transaction-fee allocation which are discussed separately below.

The first form is to grant all of the transaction fee to the successful ``miners''. All transaction fee generated within a block time, along with rewards for ``mining'', are issued to the node that successfully proposes the block. In other words, in addition to the ``mining'' rewards, the nodes with successful ``mining'' also received all the transaction fee, while the remaining Manager Ring nodes were unable to obtain transaction fee rewards. For the nodes involved in the ``mining'' competition, it is within their psychological expectation to not get the block and not receive the transaction fee rewards. But what is frustrating is that this kind of rewarding method makes the nodes without ``mining'' qualifications have no hope of obtaining transaction fee rewards in this block time. Then, there will be a sociological problem. Since I don't have the possibility of getting a transaction fee, why should I participate in the managing task of the transaction?

The second form solves this problem by allocating transaction fee to all nodes on the Manager Ring. Fairly, in a certain period of time, all nodes on the Manager Ring are involved in the processing and managing of transactions, and it is reasonable to obtain transaction fee based on their contribution. The first form of transaction fee is technically implemented without any threshold, but it is easy to cause the ``non-miner'' Manager Ring nodes to cheat, and it is not effective to stimulate all nodes on the Manager Ring to work hard. The second way is to actually provide a reward-stimulation for all Manager Ring nodes, regardless of whether the node participates in ``mining'', which can effectively improve the enthusiasm of the Manager Ring nodes. But the second solution poses a tricky technical question, namely, how to confirm that the nodes involved in the ``mining'' or not, and whether their workload is consistent with what they claimed to be? Even if this problem is solved, the system will face challenges from many nodes, resulting in a big amount of energy being spent on anti-cheating tasks.

Therefore, we propose a third solution and try to solve this problem by using sociological methods instead of technical methodologies. In this scenario, the ``mining'' rewards and the current round of transaction fee are still packaged in the form of ``ABC'' and rewarded to the successful ``mining''-nodes, but nodes without ``mining'' qualification can still earn ``service fee'' in the form of ``ABIT'' by providing the client with non-transaction type of services. ``ABC'' and ``ABIT'', explained in \textbf{Chapter 3}, are both the tokens circulated in the blockchain ecosystem. It is worth noting that ``ABC'' can exchange into legal tender in the exchanges while ``ABIT'' can not, and hence, ``ABC'' is more valuable than ``ABIT''. This method ensures that the Manager Ring nodes have opportunities to profit at any moment, and prevents the nodes that are not qualified for ``mining'' from being negatively absent or cheating. Besides, the token issuance with two layers structure can make the influence on the whole economy brought by the profit allocation as little as possible.

\subparagraph*{\textit{Unworthy Law}}\mbox{}

``Unworthy law'' is one of the ten major laws of economics. The most intuitive expression is that things that are not worth doing are not worthy to be done well. The law reflects people's psychology: if a person is engaged in something that he or she thinks is not worthwhile, they will often hold perfunctory attitudes towards it, resulting in not only the small success rate, but also lack of sense-of-achievement even if they succeed.

The original intention of the service-providing blockchain nodes is to earn revenue. If the nodes do not have the expectation of earning revenue in a round of service, according to the ``unworthy law'', the blockchain nodes will think that this round of service is not worth being done well, maybe not even worth doing at all. In addition, some nodes tend to behave maliciously for those that they think are ``worthy to do'' in order to profit. Therefore, as mentioned above, we have proposed the concept of ``service fee'' to solve this problem.

\paragraph*{Worker Ring Node}\mbox{}
\vspace{\itemsep}

\noindent The Worker Ring nodes are responsible for the specific business, perform the work assigned by the Manager Ring, and obtain revenue by providing services and resources. (Including computing power, network bandwidth and storage, etc.) As mentioned above, the Worker Ring nodes do not participate in ``mining'', so their income includes only transaction fee and service fee.

According to the ``Double-Ring'' topology of service nodes, one manager ring node manages N worker ring nodes. Therefore, the manager node should share part of incomes with its workers. Besides, the worker, which is subject to many managers simultaneously, can complain to all the managers if some of his managers do not share with him. The whole process will be completed with the help of AI self-manage module. Next, we will introduce the specific process of the profit allocation.

\paragraph*{Interest-Distribution among Nodes}\mbox{}
\vspace{\itemsep}

\noindent The interest-distribution among nodes refers to how the transaction fee between the Manager Ring nodes and the Worker Ring nodes are distributed in one block period. The transaction fee and ``mining'' rewards we discussed are issued in the form of tokens (ABC, AME Blockchain Coin). Assume that successful ``miner'' is granted with M ABCs; All transaction fee in a block time are T ABCs; Manager Ring has $N_{1}$ nodes, and Worker Ring has $N_{2}$ nodes. Then,

\begin{equation}
E_m = (M + T \div 2) \div N_{1}                                      
\end{equation}
\begin{equation}
E_w = T \div 2 N_{2}                                              
\end{equation}
where $E_m$ is the revenue expectation of Manager Ring nodes, $E_w$ is the revenue expectation of Worker Ring nodes

Let T = 2M, that is, the ``mining'' rewards are 1/2 of the transactions rewards in one specific block period, then the formula above can be transformed into:
\begin{equation}
E_m = 2M \div N_1                                                
\end{equation}
\begin{equation}
E_w = M \div N_2                                               
\end{equation}
Among them, the upper limit of $N_1$ is 100, and $N_1 \ll N_2$, so $E_m \gg E_w$, that is, the revenue expectation of the Manager Ring nodes is far greater than the revenue expectation of the Worker Ring nodes. Therefore, to pursue revenue, the Worker Ring nodes hope to enter the Manager Ring in order to obtain higher revenue.

In order to form a healthy competition-ecology and make blockchain nodes striving to provide better services, AME blockchain provides a bidirectional channel between the Manager Ring and the Worker Ring. If the Worker Ring nodes provide better resources and services, they can be allowed to enter the Manager Ring and become Manager Ring nodes, leading to higher revenue. However, if the quality of service of the Manager Ring nodes are degraded or cannot meet the requirements of the Manager Ring nodes, they will be degraded to the Worker Ring to undertake the work of the business layer.

Specifically, the access mechanism of the Manager Ring and the Worker Ring includes the following three aspects: the workload, the stability of the service, and whether the node has a bad record. Regarding one of the three aspects, if the node has ever committed a bad behavior, it can never enter the Manager Ring to provide services. The dynamical adjustment mechanism is driven by the AI governance module.

\paragraph*{Disbursement and Collection of Fees}\mbox{}
\vspace{\itemsep}

\noindent When the client uses the service of the node, it needs to pay the real-time transaction fee or the service fee. The node cannot receive the fee immediately since the ``lightning network'' performs the timed settlement. The method reduces the pressure of the server.

For nodes, the amount of service fee they can receive is related to the amount of work they provided in that service. For Manager Ring nodes, in a transaction or service, they often provide a one-time ``introduction service'', which is to help the client find a suitable Worker Ring node providing resource services. After the ``introduction'' is completed, Manager Ring nodes' work is often over, and will not continue with the client and Worker nodes to the end of the service period. Therefore, it is reasonable to use the number of transactions or the number of services to evaluate the workload of the Manager Ring nodes. The Worker Ring nodes are the real executor of the service. Ethereum uses the size of the transaction, i.e. the number of bytes, to calculate the transaction fee. This is a simplified solution that considers storage and ADSL services only, and eliminates the computing power service. Its advantage is easy to carry out. If the computing power is also added to the evaluation criteria, then the computational complexity of the transaction needs to be evaluated. Therefore, the workload evaluation of the Worker Ring nodes includes two aspects: the number of bytes of the transaction and the computational complexity of the transaction.

\subsubsection{BCM IM}

\paragraph*{Function Description}\mbox{}
\vspace{\itemsep}

\noindent BCM is an instant messenger based on the AME blockchain platform that can run on iOS and Android operating systems. BCM is committed to making itself a blockchain game platform, providing fast access methods and dedicated development tools for developers to support third-party developments. More importantly, the vision of BCM is to realize decentralized operation. That means BCM is an application run on the AME blockchain independently and can not be controlled by anyone or any organization.

\paragraph*{The Tragedy of The Commons}\mbox{}
\vspace{\itemsep}

\noindent The Tragedy of the commons is a theoretical model proposed by Professor Garrett Harding in the United Kingdom in 1968\cite{7_4}. The core idea of the model is that the commons, as resource or property, has many owners. Each of them has the right to use, but has no right to prevent others from using it, resulting in excessive usage amount and exhaustion of resources. Excessively felled forests, overfished fishery resources, and heavily polluted rivers and air are typical examples of the ``tragedy of the commons''. The reason why it is called tragedy is that every party knows that resources are exhausted due to excessive use, but everyone feels powerless to prevent the situation from continuing to deteriorate. Moreover, because of greedy human nature, there is no easy way to recover the system without any external force. Therefore, the public property is, because of determination of property rights or the cost of defining is too high, inevitably used excessively or encroached competitively.

For BCM users, the AME blockchain platform is actually a ``commons''. If the BCM user use the AME normally, the underlying blockchain resources are totally satisfactory for all users' demands. However, if a malicious attack is encountered, or a large number of fake accounts send messages at the same time, it is easy to cause blockchain resources to be abused, resulting in the ``tragedy of the commons''.

\paragraph*{Problem Solving}\mbox{}
\vspace{\itemsep}

\noindent From a certain perspective, the cause of the ``tragedy of the commons'' can be attributed to the fact that people do not need to endure any costs when using resources, and do not need to pay any fees to the resource providers. As an IM application, BCM provides users with functions of sending messages and using services. Considering the user experience, it is naturally unnecessary to ask users for payments. So, how to prevent malicious users from sending messages madly or abusing blockchain resources while ensuring the free experience of normal users?

In order to solve this problem, we require that users' chat for a certain period of time to be a transaction, and the transaction requires users to spend ``ABITs'' (AME Blockchain Initial Token). ``ABITs'' is the chat fund that the system regularly issues to the user, and the amount of each payment is sufficient to ensure the normal usage of the user during a certain period of time. If the user uses an excessive amount, the system will use the CAPTCHA method\cite{7_5} to quickly verify the authenticity of the user to ensure the normal usage of real users. The ``ABITs'' spent by the user on the BCM service is transferred to the blockchain nodes providing the service and assigned as the way that's described in \textbf{Section 2.1}. For each block period, the system will measure the workload according to the ``ABITs'' received by each node, and then issue the corresponding amount of token rewards. These rewards are the transaction fee income part of the node.

\subsubsection{CUBE}

CUBE is a BCM-based platform for blockchain games. The platform provides easy-to-use development tools and blockchain resources for third-party developers who only need to pay attention to the development of application logic.

At the same time, the platform provides a convenient interface for third-party games, welcomes and supports games which have already existed on other markets to access into the platform.

\paragraph{Decentralized Game Platform}\mbox{}
\vspace{\itemsep}

\noindent Traditional game distribution channels, such as Facebook, AppStore, Wechat, etc., enjoy absolute authority as centralized platforms in the whole ecology of games. Game developers need to pay different percentages of their profits to the platform. As a participant in the value of the platform and as a value input, the game players' labor is not recognized, and the value generated on the platform is mostly acquired by the platform. As a game platform based on BCM, CUBE is committed to creating a free service platform that connects game developers and players with the idea of decentralization. Due to the investment properties and anonymity of cryptocurrency, CUBE aims to embrace the gambling games and capital related games.

\paragraph*{Game Developers}\mbox{}
\vspace{\itemsep}

\noindent From game developers' perspective, the platform does not charge any fees, and all game revenue is owned by the developers themselves. CUBE provides game developers with easy access. For third-party games, access to the platform is equivalent to an extra income. It is a very attractive policy for developers. At the same time, CUBE also provides developers with convenient game development tools. Based on the development tools and the underlying server of the blockchain, the game can be quickly deployed on the CUBE. Reasonably, if the developers use the blockchain service, they will have to pay a fee to the server.

\paragraph*{Players}\mbox{}
\vspace{\itemsep}

\noindent From the perspective of the gamers, while playing games through CUBE, that a certain amount of digital currency can be harvested, and the value of labor is recognized by the platform. Using tokens rewarded by the platform, game players can continue to buy platform games or in-game purchases to meet their daily gaming needs. For heavy rollers, the platform's rewards cannot meet their gaming needs. So the heavy rollers will buy tokens from exchanges for game purchases or in-game purchases.

\subsection{Economic System}

This part mainly focuses on the economic system of the ABC ecosystem.

\subsubsection{Token Issuance}
 
\paragraph*{ABC(AME Blockchain Coin)}\mbox{}
\vspace{\itemsep}

\noindent ABC, which is issued by AME blockchain, can be obtained in the exchanges. Blockchain nodes mining successfully and CUBE developer can earn ABCs while the users of AME blockchain and CUBE need to pay with ABCs.

The minimum unit of ABC is a, and 1a is equal to 0.000001 ABC.

ABC is also a warrant and it can be used to divide the percentage of platform revenue according to the number of ABCs held by users.

\paragraph*{ABIT(AME Blockchain Initial Token)}\mbox{}
\vspace{\itemsep}

\noindent ABIT is a token in the AME ecosystem that exists at the BCM layer. ABIT is required to use BCM's chat service, value-added services, and unlock various functions in CUBE.

The minimum unit of ABIT is bt, and 1bt is equal to 0.000001ABIT.

ABIT does not support transfers and also does not support redemption at exchanges.

\paragraph*{ABC and ABIT}\mbox{}
\vspace{\itemsep}

\noindent ABC can be converted to ABIT in one direction in BCM, while ABIT cannot be converted to ABC.

The exchange rate is tentatively set at 1 ABC = 1000 ABIT.

\paragraph*{Distribution Plan}\mbox{}
\vspace{\itemsep}

\noindent AME plans to issue billions of tokens in the initial period. Among them:

\begin{itemize}
\item  40\% - ecosystem construction fee;

\item  20\% - issued/offered to VC/PE;

\item  20\% - reserved for backup and establishing a foundation to keep currency prices stable;

\item  10\% - reserved for research and development expenses;

\item  10\% - rewarded to early developers;
\end{itemize}

\begin{figure}[htbp]
\centering
\includegraphics*[width=4.5in, height=3.30in, keepaspectratio=true]{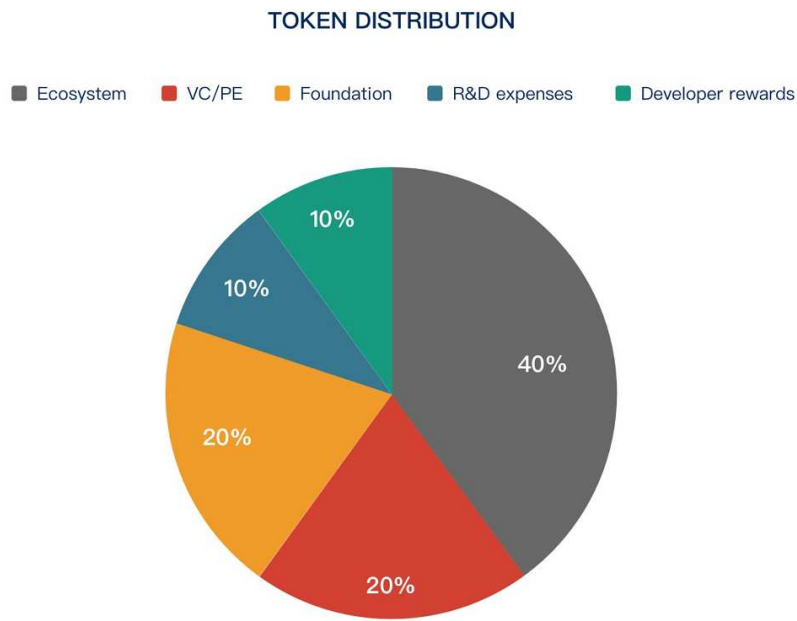}
\caption{Token Distribution Plan}
\end{figure}

\paragraph*{Token Unlock Plan}\mbox{}
\vspace{\itemsep}

\noindent The ecological construction cost refers to the service fee used to reward the node of the block, the so-called ``mining cost''. In order to meet the demand pool and inflation demand, the initial token is unlocked in one year, and 20\% is issued on the basis of the first period from the second year. The formula for issuing additional shares is:

\begin{equation}
{N}_{r-eco-abc} = N_{eco-abc} \times (1 + 0.2 \times (r - 1))
\end{equation}

In which $N_{r-eco-abc}$ is the initial ecological construction expenses, r is the serial number of the token issue period.

ABIT is unlocked at a scale of 1:1000 with ABC. The formula for issuing additional is:

\begin{equation}
N_{r-eco-abit} = 1000 \times N_{r-eco-abc}
\end{equation}
\paragraph*{Option Plan}\mbox{}
\vspace{\itemsep}

\noindent The option system of the ecosystem includes two mechanisms: the deposit requirements for registration and the temporary revenue freeze. At first approach, a specific minimum amount of frozen currency in its account is required for registering as a node in AME blockchain. The amount of required option for the Manager Ring node is more than the Worker Ring node account. With this mechanism, we can bound the number of identities a malicious user can create. At the same time, it further increases the cost of malicious behavior. At second approach, a small part of blockchain node's revenue will be temporarily frozen. This frozen ``ABC'' can be seen as a Vesting Schedule. The exercise period of the option is divided into many periods and the option to be exercised is divided into many parts. The same amount of ``ABC'' is returned to the node every day, and the return is completed in several days.

It should be noted that whether both of these two mechanisms are applied depends on the ecological context of ``ABC.''

\subparagraph*{\textit{Option Theory}}\mbox{}

Option is also translated as ``trading rights.'' It is a standardized derivative contract that gives the holder a certain amount of specific subject matter to the other party at a certain price for a certain period of time in the future, or a specific date in the future, but does not bear obligation to buy or sell\cite{7_6}.

The option plan for the ``node'' reward is based on the ``incentive'' theory in economics, and the establishment of the ``incentive'' principle first recognizes the property value of human capital and solves the entrusting-agent problem

The establishment of an option plan in the blockchain ecosystem has the following advantages:

\begin{itemize}
\item  To a large extent, it avoids the problem of insufficient mid/long-term incentives in the form of cash reward distribution.

\item  Avoid node-loss and restrict the behavior of nodes.

\item  Reduce the daily cash-offering and paying burden of the ecosystem and slow down the rate of inflation.
\end{itemize}

\paragraph*{Inflation}\mbox{}
\vspace{\itemsep}

\noindent In the AME blockchain ecosystem, the amount of money will increase with the development of the system. According to the Fisher equation, P will increase, that is, inflation will occur. According to classical economic theory, moderate inflation favors the functioning of the labor market (ie, the workers in the ABC ecosystem) and promotes the vitality of the economy. One reason is that it is difficult to renegotiate for price cuttings in certain cases, especially for wages and contracts. Therefore, if the price rises slowly, it would be easier for the relevant prices to adjust. There are a variety of commodity prices that will ``resist to price cuts'' and tend to keep rising. So trying to achieve zero inflation (prices are maintained) will lead to lower prices, profits, and numbers of employees in other industries. Therefore, the executive departments of several companies regard mild inflation as ``lubricating the commercial ship''. The pursuit of absolute price stability will lead to devastating deflation (continuing price dropping), which will lead to bankruptcy and economic recession (even the economic depression). As the ABC ecosystem will continue to have blockchain nodes, users and other currency value injections, it is necessary to issue a certain amount of currency to ensure ecological operation. But it needs to be ensured that the additional currency will stimulate the entire ecology mildly, so that the system will maintain a moderate 2\% inflation-policy every year.

In principle, we expect the amount of money deposited in the option pool to account for 20\% of the total currency in the same period to ensure a stable demand for currency within the system. In addition, annual inflation is expected to account for 20\% of the option pool, then the inflation of the entire system will be controlled at around 4\%. It is worth noting that the amount of tokens that is forgotten, discarded or lost in circulation each year is expected to account for about 2\% of the total amount of tokens. Therefore, the annual inflation rate should be controlled at 8\% to meet the demand of these three aspects, including option pool deposit, token destroyed, and additional issuance.

\paragraph*{Token Demands}\mbox{}
\vspace{\itemsep}

\noindent In the entire ecosystem, there are three demand parties for currency, namely, the option deposit of nodes, the game developers of CUBE, and the users of CUBE. The currency demand of CUBE game developers lies in the use of the underlying resources of the blockchain, which requires a certain fee to be paid to the blockchain nodes. The recharging/circulation currency on the CUBE game platform is ABC, and players need to spend a lot of ABCs to play games, so players will create a big amount of currency demand. In addition to the 20\% monetary option deposit, we expect the CUBE gaming platform to generate approximately another 20\% of the currency demand.
 
\subsubsection{Consensus Economy}

\textbf{Chapter 2} has elaborated and analyzed the ACP consensus mechanism. This section will analyze the distribution of economic benefits and the reasons for each role in the consensus process based on ACP mechanism.
\\
\paragraph*{ABC Incentive}\mbox{}
\vspace{\itemsep}

\noindent The FC is the final committee for the second round of elections, which bears the responsibility of packaging and generating consensus through the $\text{PBFT}^*$ algorithm, which is also the ``mining'' in the traditional sense. In order to encourage the miners to work honestly and reduce bad behavior, FC members should be granted with economic incentives after they have made a block and be confirmed successfully. Since the miner's ``mining'' motivation lies in the ABC that can be exchanged, the economic incentives at this time are issued using ABC.

As mentioned earlier, FC members can be divided into three roles:

\begin{itemize}
\item  Final Miner -- A successful packager.

\item  Final Leader -- The proponent of the block in $\text{PBFT}^*$.

\item  Final Verifier -- The verifier of the block in $\text{PBFT}^*$. All FC members are verifiers.
\end{itemize}

According to the contributions of the three parties, the overall income ratio of the three parties is defined as:

\begin{equation}
N_{abc-miner} : N_{abc-leader} : N_{abc-verifier} = N_{1} : N_{2} : N_{3}
\end{equation}

Therefore, the revenue model for different role in FC is:

\begin{itemize}
\item  $N_{abc-miner} = N_{r-eco-abc} \div ( 365 \times 24 \times 60 \times 6 ) \times N1 \div 10$

\item  $N_{abc-leader} = N_{r-eco-abc} \div ( 365 \times 24 \times 60 \times 6 ) \times N_{2} \div 10 \div N_{fc-leader}$

\item  $N_{abc-verifier} = N_{r-eco-abc} \div ( 365 \times 24 \times 60 \times 6 ) \times N_{3} \div 10 \div N_{fc}$
\end{itemize}

The income expectation model of the FC node is:

\begin{equation}
N_{fc-node} = (N_{abc-miner} + N_{abc-leader} \times N_{fc-leader} + N_{abc-verifier} \times N_{fc}) \div N_{fc}
\end{equation}

Where $N_{fc}$ is the number of nodes of the FC.

In addition, assuming that the upper limit of the capacity of one block in the AME blockchain network is $C_{limit}$ MB, and the size of the package proposed by Final Miner is $C_{final}$ MB, then the ABC reward that can be obtained is determined by the following formula:
\begin{equation}
N_{abc-miner-non-selfish} = N_{abc-miner} \times ( C_{limit} - 1 \div (C_{final} + k)) \div C_{limit}
\end{equation}
Among them, k is a constant, which is used to control the convergence speed of the benefit; N${}_{abc-miner-non-selfish}$ is the actual benefit that can be obtained according to the size of the package every time the node is successfully mined.

In order to encourage all members of the FC to package local transactions into blocks as much as possible, it is necessary to associate the interests of the remaining two roles in the FC, Final Leader and Final Verifier, with the size of the package, namely:
\begin{equation}
N_{abc-leader-non-selfish} = N_{abc-leader} \times ( C_{limit} - 1 \div (C_{final} + k)) \div C_{limit}
\end{equation}
\begin{equation}
N_{abc-verifier-non-selfish} = N_{abc-verifier} \times ( C_{limit} - 1 \div (C_{final} + k)) \div C_{limit}
\end{equation}
\paragraph*{ABIT Incentive}\mbox{}
\vspace{\itemsep}

\noindent The PC was elected to the committee in the first round of elections. Its members further elected FC through VRF calculations. They paid for their power and labor and should receive corresponding economic incentives. More importantly, the incentive for the PC to ensure that its members do not maliciously broadcast FC to the entire network, causing FC members to be attacked by malicious nodes. Obviously, before the FC wants to carry out the PBFT consensus, it is necessary to broadcast its identity to the PC to find all FC members, and the PC knows the identity of the FC. When a non-FC PC knows that it can't enter the FC in this round, leading to no chance of ``mining'' reward, and then they may act maliciously because the success of this round of packaging has nothing to do with itself. Therefore, we propose that if the last round is completed successfully, PC will be rewarded at the beginning of current round, thus ensuring that each round of PCs hope the success of the last round, so as to avoid the economic interests of evil behavior.

Since the number of PC nodes is relatively large, and in the long run, each node is equally likely to assume PC work. Therefore, the incentive for PC is a wide range of airdrop behavior, which requires the use of ABIT tokens with relatively low value and limited circulation.

Assuming that the blockchain has 100,000 participating nodes, according to the ACP consensus protocol, the theoretical period of one block is about 10 seconds. According to the ABIT unlock plan in \textbf{Section 3.1}, the number of ABITs issued in this round should be T times the ABC issued in the previous round. Therefore, FC members can receive ABIT rewards for each successful block cycle, which can be calculated by the following formula:

\begin{equation}
N_{abit-pc} = N_{r-abc} \times T \div (365 \times 24 \times 60 \times 6 \times N_{pc})
\end{equation}

Where $N_{r-abc}$ is the total amount of ABC rewarded to the nodes in the last round.

As a common blockchain node in the AME blockchain network, the daily ABIT revenue expectation is:
\begin{equation}
N_{abit-node/day} = N_{abit-pc} \times N_{pc} \div N_{all} \times 6 \times 60 \times 24
\end{equation}

\subsubsection{Economic Cycle}

\paragraph*{Background}\mbox{}
\vspace{\itemsep}

\noindent In this section, we provide a comparative analysis on the token systems of Bitcoin, Ethereum and EOS with a focus on their economic models in order to design a better token system for the ABC ecosystem.

\textbf{Figure 7.3} shows a simplified bitcoin flow model. When the blockchain node mines, bitcoin flows from the miners to the exchange, and then flows to the investors. The bitcoin that flows out does not return to the bitcoin network, re-creating value for the system. Therefore, the economy of Bitcoin is a non-closed, open flow model.
\clearpage
\begin{figure}[htbp]
\centering
\includegraphics*[width=2.1in, height=2.1in, keepaspectratio=true]{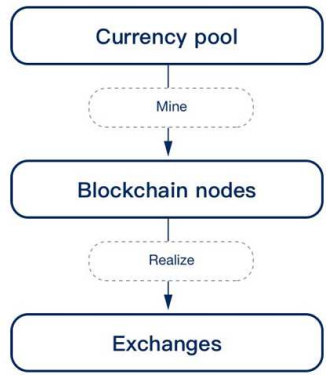}
\caption{Bitcoin Economic Operation Model}
\end{figure}
\textbf{Figure 7.4} shows a simplified Ethereum economic operation model. Compared with Bitcoin, Ethereum's economic model has made significant progress which mainly reflected in its recycling of fuel Gas, making the economic flow forming a simple closed loop.

\begin{figure}[htbp]
\centering
\includegraphics*[width=4.2in, height=3.2in, keepaspectratio=true]{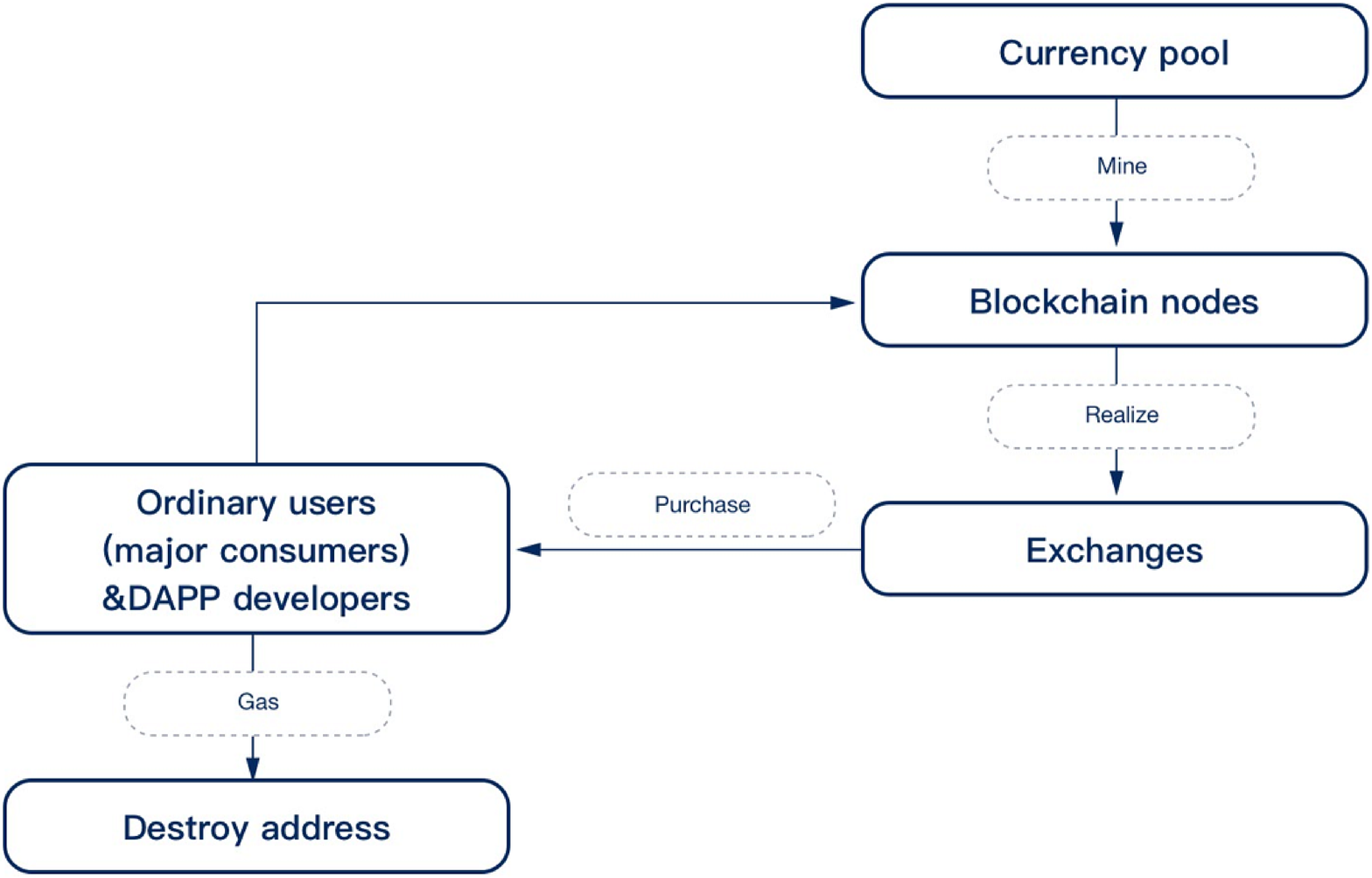}
\caption{Ethereum Economic Circulation Model}
\end{figure}

\textbf{Figure 7.5} shows the expected economic transfer model for EOS. On the supply side of the currency, it is still miners producing the money. On the demand side of the currency, the fuel consumption of Ethereum was replaced with the deposit of the blockchain accounts. Compared with fuel, deposit depends on the activity level of the entire blockchain community, the number and size of developers that are unpredictable at the beginning of the project. Once the project performs poorly, the demand for currency drops sharply. In other words, the recovery method of deposit is not direct or effective enough.
\begin{figure}[htbp]
\centering
\includegraphics*[width=3in, height=3.in, keepaspectratio=true]{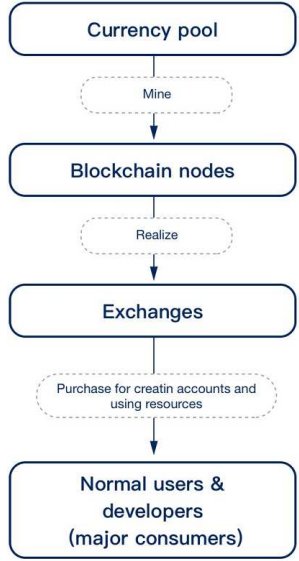}
\caption{EOS Economic Circulation Model}
\end{figure}
Despite the success of these three projects, there are some notable flaws in their economic models. From the comparative analysis described above, it can be seen that their ecologies do not form a closed loop of tokens, which is vital to any economic model. Therefore, we propose the CFE token system for the ABC ecosystem to address this common pain point of existing blockchain projects.

\paragraph*{CFE Token System}\mbox{}
\vspace{\itemsep}

\noindent For ease of understanding, the CFE is first disassembled into two parts, the BCM layer and the CUBE layer.
\begin{figure}[htbp]
\centering
\includegraphics*[width=4.45in, height=2.5in, keepaspectratio=true]{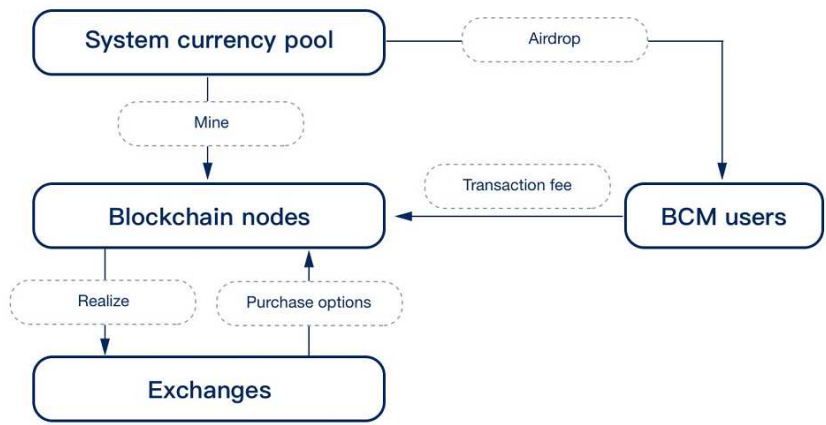}
\caption{BCM Layer Economic Cycle Model}
\end{figure}
\subparagraph*{\textit{BCM Layer}}\mbox{}

As shown in \textbf{Figure 7.6}, the BCM forms a closed loop of currency circulation around the blockchain nodes. The blockchain nodes obtain the currency rewards by ``mining'' and collecting the transaction fee of the BCM users. The nodes can sell the acquired currency to the exchange to realize the cash, or can save and convert it into options to obtain more rewards. If it is needed by the nodes, it is also feasible to buy the currency directly from the exchange and deposit it as options.

\subparagraph*{\textit{CUBE Layer}}\mbox{}

\textbf{Figure 7.7} shows the economic cycle model of the CUBE layer, which forms a closed loop of currency around gamers and game developers.

\begin{figure}[htbp]
\centering
\includegraphics*[width=4.6in, height=2.9in, keepaspectratio=true]{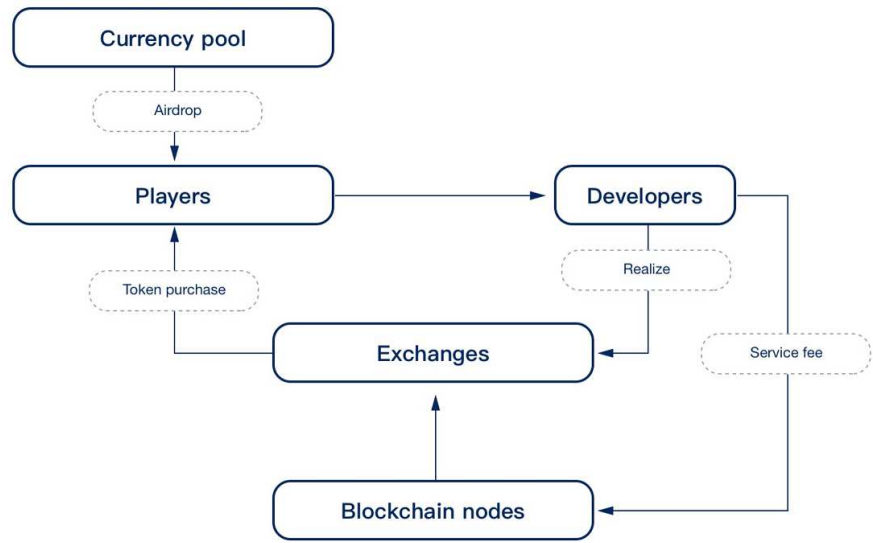}

\caption{CUBE Layer Economic Cycle Model}
\end{figure}

For the players, they can get a certain amount of ``ABIT'' airdrops by playing games, but this part of the currency cannot meet the demand of heavy rollers. The heavy rollers will choose to buy ``ABC'' from the exchanges and change them into ``ABIT'' to pay the game developers. The ``ABC'' that the developers received can be realized as cash in the exchanges.

For game developers who use services and resources of blockchain, they also need to pay a portion of the currency to blockchain nodes.

\subparagraph*{\textit{CFE Token System}}\mbox{}

\textbf{Figure 7.8} combines the BCM and CUBE to give a fully ecological economic cycle model. As can be seen from the model, the entire ecosystem forms a complete closed loop around blockchain nodes, CUBE and exchanges.

\begin{figure}[htbp]
\centering
\includegraphics*[width=4.70in, height=2.9in, keepaspectratio=true]{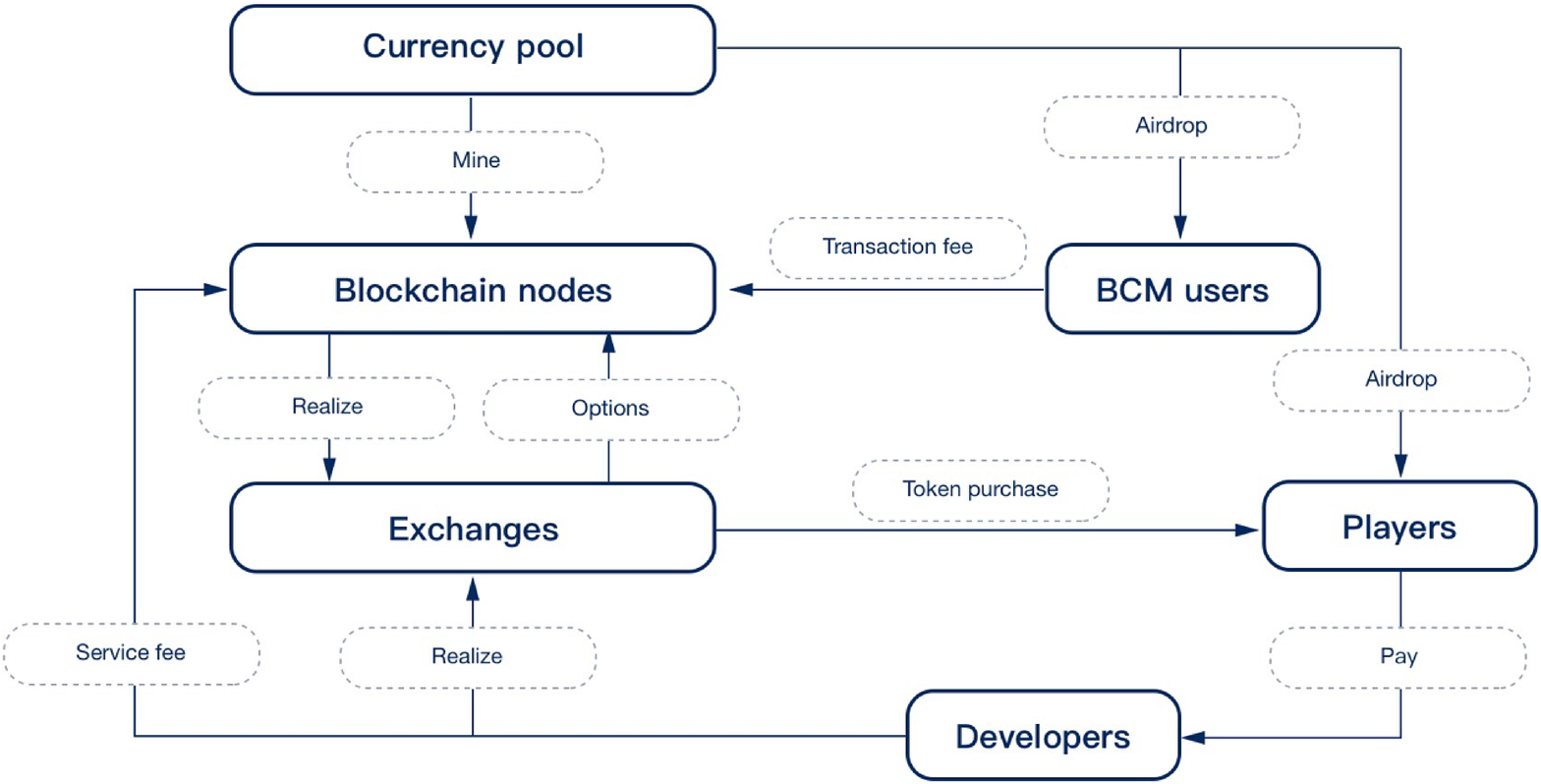}
\caption{Full Ecological Economic Cycle Model}
\end{figure}

As shown in \textbf{Figure 7.8}, the ABC Ecosystem builds a complete closed-loop economic model around blockchain nodes, CUBE and exchanges, providing reasonable access and exit for each economy participant. Compared with the open economic models of Bitcoin, Ethereum and EOS, ABC's economic model can enhance the liquidity and stability of tokens.

\subsubsection{Foundation Mechanism}

Throughout the ecology, participants, especially CUBE participants, are extremely sensitive to the market price of tokens. Stable currency price is particularly important in ABC's ecology. We utilize the foundation reserve pool pricing and market pricing to ensure that the token price is stable. The source of the Foundation Reserve Pool consists of two parts, namely a portion of the legal currency cash reserve for initial financing and a portion of the token reserved for the issuance of the token (about 20\% of the total circulation).

When there are n ABCs in circulation in the market, if the market price of ABC is higher than the expected theoretical price, the market will purchase ABC from the foundation to achieve a price equilibrium; When the ABC price is lower than the theoretical price, the demand for all tokens will be completed through open market transactions. The system will not sell. When the market price is less than 0.5 of the expected theoretical price, the foundation will buy back some of the ABCs, reduce the number of ABCs circulating in the market, and make the market price return to the expected price.

\paragraph*{The link among currency, inflation, and currency-policy}\mbox{}

\begin{figure}[htbp]
\centering
\includegraphics*[width=4.70in, height=1.4in, keepaspectratio=true]{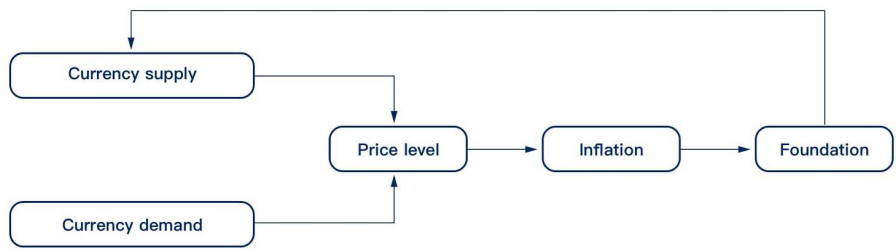}
\caption{Relationship among Currency, Inflation, and The Foundation Mechanism}
\end{figure}
\clearpage
According to the classical economic theory, money supply and demand jointly determine the current price level, and the price change determines the inflation rate\cite{7_7}. In the ABC ecosystem, the currency supply side is supplemented by 4\% of the additional currency each year, in addition to the currency that was first issued and circulated in the market. The currency demand side relies on the option deposit of blockchain nodes, the needs of CUBE game developers and gamers. This supply and demand relationship determines the purchasing power of the token, which in turn affects the price level.

The price level determines the ecological inflation rate. In the blockchain ecosystem, the money supply side can be regarded as a basically constant value. If there is a situation of excessive demand or weakness on the demand side of the currency, then the hyperinflation or deflation happens resulted in a serious impact on the ecology of the system. Therefore, it is necessary to establish a foundation mechanism and use a reasonable monetary policy to influence the amount of money on the money supply side when necessary, so that the entire ecology maintains a moderate inflation rate.
 
\subsection{Analysis}

A healthy blockchain economic operation model needs to include at least the following 4 aspects: transaction fee, network security, price stability mechanisms, and economic cycles. Next, we will use the method of comparison to illustrate the specifics of the AME blockchain ecology in these four aspects.

\subsubsection{Transaction Fee}

The transaction fee is the fee that the blockchain user needs to pay to the blockchain nodes or the system when the transaction is initiated. The fee may be transferred to the blockchain nodes and may also be consumed as fuel. The original intention of transaction-fee design is to allow users to use blockchain resources as a paid service, prevent the abuse of resources, and maintain the stability of the system. Typical representatives of this type of design are the Bitcoin network and Ethereum. Users must pay a certain transaction fee or fuel for the transaction. However, the problem is that charging the transaction fee will cause the blockchain usage threshold to be pulled higher, causing many potential users to be rejected. Next, EOS tries to solve the problem by cancelling the user's transaction fee. However, if you want to use the blockchain resources, you need to freeze a certain amount of tokens in the account. The amount of resources available and the amount of frozen tokens are positively related\cite{7_8}. This approach seems not requiring users to spend transaction fee. However, if users want to use blockchain resources, they still need to have a certain amount of tokens, which is a threshold itself. Therefore, EOS did not solve this problem very well.

In the ABC ecosystem, BCM is truly free to users. The system issues ``ABITs'' to users, and users chatting paying ``ABITs'' to nodes. The amount of ``ABITs'' received by nodes measures the workload of nodes. Finally, the system issues the corresponding amount of tokens to nodes. For resource abuse issues, we use a secure resource allocation pool and a game-like human-machine identification scheme to circumvent, separating the transaction fee from the use of resources and making the user layer truly free.

\subsubsection{Network Security}

The challenge of blockchain network security mainly comes from the blockchain node itself. In the absence of a clear penalty mechanism and sufficient positive incentive stimulus, blockchain nodes tend to behave maliciously for their own interests. There are usually two solutions to this problem: first, the deposit mechanism. That is, when the server is registered as a node, it needs to pay a certain amount of deposit to the blockchain system. If the node has a malicious error such as a packaging error, the deposit will be all or partially deducted. For example, both EOS and Steemit use this mechanism. Second, income turns into option partially. That is, the revenue of the blockchain node, whether it comes from mining or transaction fee, some of which are issued in the form of options that needed to be frozen for a period of time before exercising. We adopt the second method. Compared with the first one, there are two advantages of income partially turning into option.

\begin{enumerate}
\item  Reduce node access costs and thresholds. Contrary to the design of the EOS super node, the node does not need to pay the deposit at the time of registration. The node can concentrate on access and service without worrying about how many tokens are self-owned, further ensuring the decentralization of the blockchain system.

\item  Slowing down the rate of inflation. Under the ABC Ecology, the transaction fee and mining fee issued to the nodes are all new empty coins. Setting a part of the option currency not only guarantees the same economic benefits among nodes, but also slows down the inflation rate, which is conducive to firm currency prices.
\end{enumerate}
 
\subsubsection{Price Stability Mechanism}

When electronic currency is issued, if it is simply adjusted by the market itself, it will often lead to instability of the currency price, and the magnitude is large. This is not a good thing for providers and users of blockchain resources, because the use of resources and becoming nodes often require consumption or possession of tokens. Therefore, Steemit proposed a fiscal policy similar to bonds\cite{7_9}. Steemit issues bonds to investors. Bonds can be exchanged for equivalent tokens. Community leaders are responsible for maintaining the 1:1 equivalence between bonds and the US dollar. By means of bonds, Steem is linked to the US dollar. In this way, the interest rate of the bond can be used as a financial means to adjust the amount of tokens in the market, thereby affecting the price of the tokens. In addition, the foundation mechanism is also a common method. The foundation mechanism refers to a pool of funds and tokens that are artificially established by the blockchain development team to ensure the health of the economic order. The entire ecological economy is regulated by issuing pricing and repurchasing tokens. According to the theory of sticky information\cite{7_10}, a transparency and simple fiscal policy can improve the efficiency of monetary policy. The bond system is obviously more complicated than the foundation mechanism. For the foundation mechanism, ordinary investors only need to know the value of the token and that there are real gold and silver backing it up. The foundation mechanism enhances the public's confidence in the tokens, and stabilize the currency. Therefore, we first use the foundation mechanism as a price stability tool for ABC Ecology.

\paragraph*{Sticky Information Theory}\mbox{}
\vspace{\itemsep}

\noindent The core idea of the sticky information theory is that economic entities need to bear certain costs in the process of collecting, understanding and absorbing information\cite{7_11}. This has caused the economic entity not to continuously update the decision information set, still using the original plan and expired information. A simple understanding is the lag in information on economic decisions. Therefore, high transparency is conducive to enhancing the stabilizing effect of monetary policy. The central bank should communicate high-quality information to the public, and the central bank's intentions should be fully transparent and generally beneficial to social welfare. However, the increase in the transparency of sensitive information will have a negative impact on social welfare.

In the sticky information model, the transparency of monetary policy can improve the efficiency of monetary policy. If policy information is complex, people will ignore it because they think it is not worth spending too much effort on this information.

In the sticky information model, central bank announcements are influential\cite{7_12}. For example, because the central bank made an announcement, before the central bank adopted a tightening policy, the public might already have adjusted the plans. As a result, the central bank announced a reduction in the money supply growth rate before taking action, which would lead to a faster inflation response and a smaller loss of output comparing to a sudden reduction in the money supply growth rate.

As to the CFE token system, many designs are conducted under the guidance of the sticky information theory. For example, the reason of a foundation system rather than a bond mechanism is used to stabilize monetary value is, in the sticky information model, the transparency of monetary policy can improve the efficiency of monetary policy. If policy information is complex, people will ignore it because they think it is not worth spending too much effort on this information.

\subsection{Closed-Loop Token System}

This chapter proposes an economic plan based on the AME technology platform. The goal is to enable all participants in the ecology to perform their duties under reasonable economic incentives and fiscal governance to ensure the healthy operation of the entire ecosystem of the AME blockchain. The AME blockchain platform, BCM instant messenger and the BCM-based CUBE community are closely linked and functioning, by the mechanisms of the free economic market and macro-control policies. Finally, through the comparative economic analysis of AME blockchain and other well-known blockchain projects, the completeness and liquidity of the ABC economic model is crystal clear.

\clearpage 
\addcontentsline{toc}{section}{Conclusion}
\section*{Conclusion}
In this paper, we provide a three-layer pyramid-shaped blockchain ecosystem, consisting of the underlying level AME blockchain, the middle level BCM and the top-level CUBE.

The AME blockchain is the foundation of the entire system. To extend scalability, AME blockchain proposes a ``Double-Ring'' network topology design ABNN. It decouples network management logic from business logic. Thus, a flexible and scalable decentralized P2P system solution is realized. In terms of consensus mechanism, AME blockchain proposes a novel next-generation consensus protocol ACP. It uses randomness beacon and Verifiable Random Function to fairly select a random committee and then achieves agreement among the committee efficiently based on the $\text{PBFT}^*$ protocol. Beside the outstanding throughput and latency performance, the rigorous four-phase consensus protocol offers high security guarantees for AME blockchain.

As a decentralized application platform, we offer an AME blockchain based IM application and the blockchain-oriented gaming platform CUBE, particularly for gambling games, on top of the IM. To ensure the healthy operation of the ecosystem and stabilize the value of the tokens, we integrate CFE economic system into free markets including token issuance mechanism, consensus economics, economic cycle model. Powered with the economic incentive of CFE model, AME blockchain is able to link the participants in the three different layers and promote the healthy development of ABC ecology.

In addition, we offer a novel artificial intelligent based built-in governance mechanism in AME. The AME AI Governance Module is an intelligent module for privilege management, risk analysis and anomaly detection on the AME. It comprehensively utilizes multiple AI technologies and deeply adapts the characteristics of AME blockchain to automatically control the access, elimination and the role conversion of nodes on AME; It effectively improves the overall efficiency of AME systems. Moreover, it can reduce the risk of AME being subject to external surveillance, attacks, and large-scale node failures.

In spite of all these, there are many challenges faced by the enhancement of AME blockchain ecosystem. For instance, in order to make the AME AI Governance module more secure and intelligent, we need to develop more efficient privacy preserving decentralized learning algorithms. In the aspect of ecological construction, we planned to combine AI self-management technology with AME economic model to make token airdrops, option unlocking and the supply and demand adjustment more effectively and intelligent. These cutting-edge research into the fundamental problems of blockchain is the focus of our future work, which is also an inspiring exploration for the future of blockchain industry.

\end{document}